\documentclass{egpubl}

\JournalSubmission    
\usepackage[T1]{fontenc}
\usepackage{dfadobe}  
\usepackage{enumitem}
\usepackage{calc}
\usepackage{subfigure}
\usepackage{wrapfig}
\usepackage{array}
\usepackage{tabularx}
\usepackage{gensymb} 

\biberVersion
\BibtexOrBiblatex
\usepackage[backend=biber,bibstyle=EG,citestyle=alphabetic,backref=true]{biblatex} 
\electronicVersion
\PrintedOrElectronic
\ifpdf \usepackage[pdftex]{graphicx} \pdfcompresslevel=9
\else \usepackage[dvips]{graphicx} \fi

\usepackage{egweblnk}

\usepackage[draft, todonotes={disable}]{changes} 

\usepackage{setspace}
\title[Geospatial Network Visualization]%
      {Visualizing and Interacting with Geospatial Networks:\\A Survey and Design Space}

\author[S. Schöttler, 
Y. Yang, 
H. Pfister
\& 
B. Bach]
{\parbox{\textwidth}{\centering 
S. Schöttler$^1$,
Y. Yang$^2$, 
H. Pfister$^2$,
and B. Bach$^1$ 
}
        \\
{\parbox{\textwidth}{\centering 
$^1$University of Edinburgh, UK\\
$^2$Harvard University, MA
}
}
}

\newcommand{\fgeo}{\texttt{\textbf{GEO}}}
\newcommand{\ftopo}{\texttt{\textbf{NET}}}
\newcommand{\finteract}{\texttt{\textbf{INTERACT}}}
\newcommand{\fcomp}{\texttt{\textbf{COMP}}}

\newcommand{\fnode}{\texttt{\textbf{NODE}}}
\newcommand{\flink}{\texttt{\textbf{LINK}}}

\newcommand{\ncount}{95}

\newcounter{z}
\newcommand{\perc}[1]{\setcounter{z}{#1 * 100 / 95}\arabic{z}\%}

\newcommand{\iconfig}[1]{
    \setlength\intextsep{0pt}
    \begin{wrapfigure}{l}{0.08\columnwidth}
    \includegraphics[width=0.17\columnwidth]{#1}
    \end{wrapfigure}
}

\begin{document}


\maketitle
\begin{abstract}

This paper surveys visualization and interaction techniques for geospatial networks from a total of \ncount\ papers.
Geospatial networks are graphs where nodes and links can be associated with geographic locations. 
Examples can include social networks, trade and migration, as well as traffic and transport networks. 
Visualizing geospatial networks poses numerous challenges around the integration of both network and geographical information as well as additional information such as node and link attributes, time, and uncertainty.
Our overview analyzes existing techniques along four dimensions: 
i) the representation of geographical information, 
ii) the representation of network information, 
iii) the visual integration of both, and 
iv) the use of interaction. 
These four dimensions allow us to discuss techniques with respect to the trade-offs they make between showing information across all these dimensions and how they solve the problem of showing as much information as necessary while maintaining readability of the visualization. \url{https://geonetworks.github.io}.



\end{abstract}  

\vspace{-0.25em}
\section{Introduction}

\begin{figure}[t]
    \includegraphics[width=1\columnwidth]{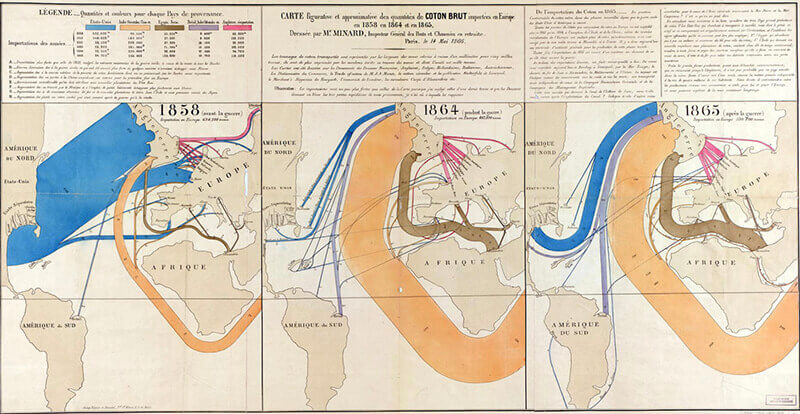}
    \caption{Charles Joseph Minard (1781-1870) depicting cotton imported into Europe in 1858 (left), 1864 (center), and 1865 (right). Color indicates the origin of flows: blue = United States; yellow = India, Orient, China, etc.; brown = Egypt, Syria; violet = Brazil, Oriental India; red = England, re-exportation.}
    \label{fig:minard}
\end{figure}

\begin{figure}[t]
    \subfigure[1920]{
        \includegraphics[height=3cm]{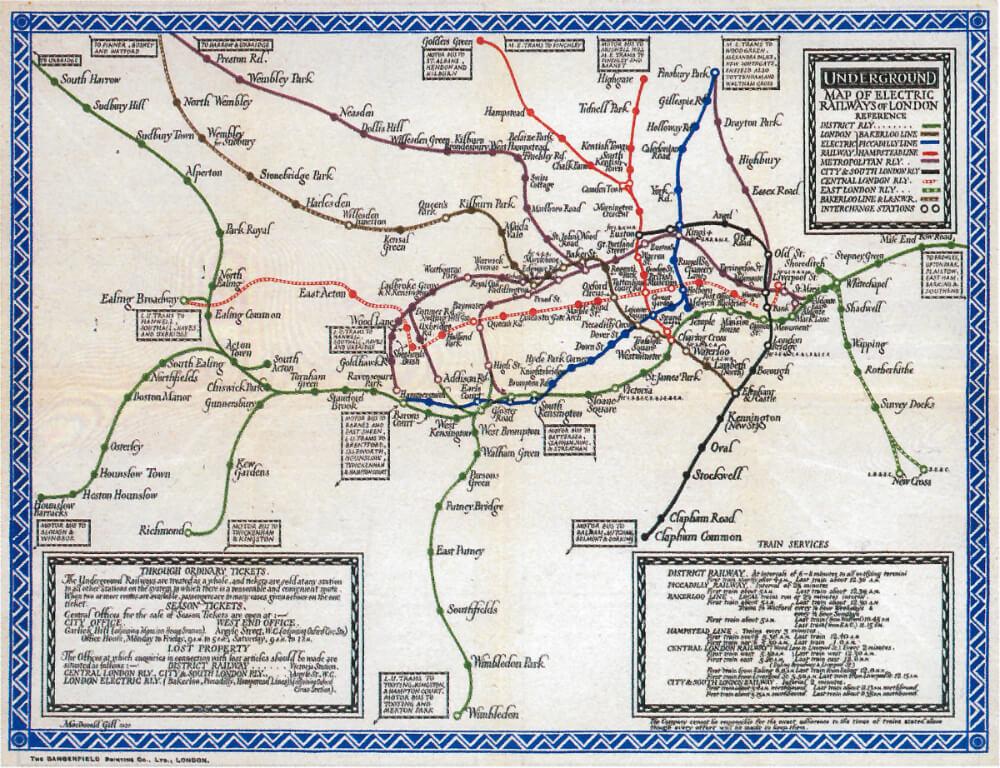}
        \label{fig:beck-1}
    }
    \hfill
    \subfigure[1933]{
        \includegraphics[height=3cm]{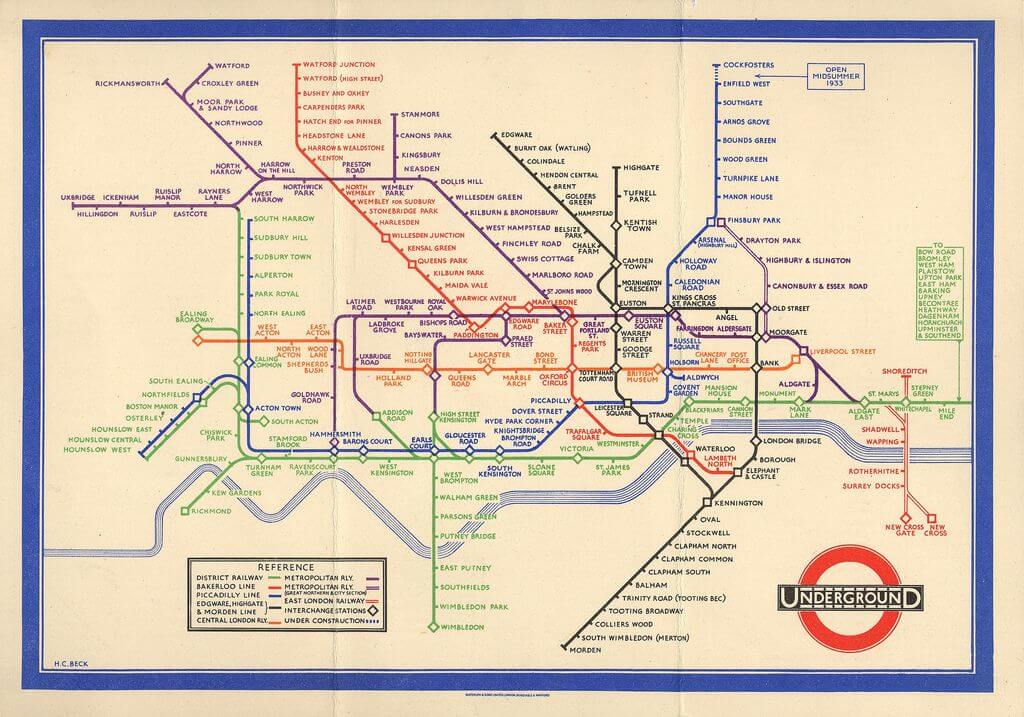}
        \label{fig:beck-2}
    }
    \caption{Metro map designs in London: 1920 (left) showing accurate spatial positions of stations; 1933 (right) showing a distorted version by Harry Beck (1902-1974).}
    \label{fig:beck}
\end{figure}

Geospatial networks are graphs whose nodes and links can be associated with geographic locations.
Examples of geospatial networks include social networks where social actors are found at specific locations, trade between countries (Fig. \ref{fig:minard}), or transport links between defined locations (Fig. \ref{fig:beck}). In all these networks, nodes are associated with individual geographic locations such as a city, a country, or a set of geographic coordinates, and are connected by links. 

The visualization of geospatial networks goes back to at least French civil engineer \textit{Charles Joseph Minard} (1781–1870), famously known for his depiction of Napoleon's March to Moscow and numerous other flow map visualizations~\cite{rendgen_minard_2018}.
Fig. \ref{fig:minard} shows one of Minard's graphics, visualizing the origin and amount of cotton imported into Europe in 1858, 1864, and 1865. The width of the flows shows the quantity of imported cotton and their color shows the country of origin. Minard cleverly distorts geographic shapes, positions, and sizes of countries, islands, and continents to provide space for these links. Showing all three years juxtaposed allows for understanding and exploring changes over time.

Another notable example of geospatial network visualization is \textit{Harry Beck}'s schematic map of the London Underground. Designed in 1933, Beck created his map (Fig. \ref{fig:beck-2}) to solve the problem of increasing complexity of the network. The growth in lines and stations had made the traditional approach to transport maps (Fig. \ref{fig:beck-1}), which was based on precise geographic locations, harder to read and therefore unfit for public display. 
Beck noticed that a lot of the geographic information in this map was unnecessary in the context it was to be used in---for tasks such as finding the fastest route between two stations. This insight led Beck to distort the underlying map to display the network of transport lines and stations more clearly; 
inspired by electronic circuit boards, he straightened lines and only used angles of 45 and 90 degrees. With some exceptions and numerous extensions, Beck's design has become the standard solution for public transport maps, providing an effective trade-off by abstracting geography to emphasize network topology.

Today, numerous techniques exist to offer solutions to the inherent complexity in geospatial networks, posed by the combination of geographic with network and potentially other information. Most of this work has been focusing on `traditional' \textit{node-link} diagrams superimposed onto geographic maps. For example, sometimes, links are drawn in a straight manner between source and destination~\cite[e.g.,][]{becker_visualizing_1995}, sometimes these links are slightly curved~\cite[e.g.,][]{jenny_force-directed_2017}---perhaps with the intention to communicate that the link does not have a specific geographic location. More recent work has suggested automated routing approaches~\cite{buchin_flow_2011}. Virtual reality is also offering new ways for rendering and interacting with three-dimensional globes~\cite{yang_origin-destination_2019}.

Another set of techniques continues Minard's and Beck's work and automatically deforms geographic space~\cite{bouts_visual_2016,otten_shifted_2018}. In its most extreme form, geographic space gets abstracted almost entirely and is instead represented as spatially grouped and ordered, colored segments on a circle in a chord diagram~\cite{hennemann_information-rich_2013,abel_quantifying_2014} (Figure~\ref{fig:abel2014}). Alternatively to node-link diagrams, adjacency matrices can solve the problem of dense networks side-by-side with geographic maps~\cite{yang_many--many_2017,guo_visual_2007} and can group and order nodes by geographic location if necessary~\cite{bach_ontotrix_2011,yang_many--many_2017}. Finally, there is a range of purely interactive techniques, e.g., to allow for navigating between distant nodes~\cite{moscovich_topology-aware_2009} or interactive lenses to reduce local link clutter in node-link diagrams~\cite{wong_edgelens:_2003}. 

In summary, the range of techniques is rich, and contributions have come from many different communities: visualization, graph drawing, geography. Moreover, many challenges are still unsolved.
Examples include moving nodes and dynamic geospatial networks and uncertainty in network topology (e.g., missing nodes and links), geographic locations (e.g., different granularities, identical positions), and their combination (e.g., uncertain, multiple, or missing node and link positions).

While many surveys, books, and articles have been written about visualizing networks, geographic visualization, and spatio-temporal data visualization (see Section~\ref{sec:relatedwork} for an overview), there is so far no structural approach to categorizing types of geospatial networks and the respective techniques. Different and inconsistent terminology, such as \textit{flow maps}, \textit{origin-destination maps}, \textit{geospatial networks}, etc., makes it hard to navigate the jungle of techniques and to inform 
\textit{i)} the application of existing techniques to specific (domain) problems, 
\textit{ii)} the design of novel techniques to address open challenges, and 
\textit{iii)} the comparison and study of the effectiveness of a given set of techniques for a set of analysis tasks.

The goal of this survey is to provide a structured review and to propose a design space for visualizations of geospatial networks, as well as to inform a discussion about current challenges. While our discussion of challenges focuses on how to practically address specific attributes of geospatial networks, our design space is informed by the trade-off each visualization design is confronted with: emphasizing some information while abstracting and aggregating other information in order to obtain a task-specific and clearly readable visualization. This trade-off, nicely illustrated in the works of Minard and Beck, is, in fact, common in visualization design and is demonstrated by numerous studies that show the complementariness of visualization designs.

In our design space (Section \ref{fig:designspace}), we describe existing techniques along four dimensions: 
\textit{i)} \fgeo: how explicitly geographic information is shown, 
\textit{ii)} \ftopo: how explicitly network information is shown, 
\textit{iii)} \fcomp: how geographic and network information are composed and integrated visually, and
\textit{iv)} \finteract: if and how interactivity is used to facilitate integration and exploration.

This design space aims to capture the dimensions along which a designer or analyst can make choices to create a balanced, purposeful visualization design and to address specific visualization challenges. This survey is addressed to readers in any discipline, including students new to visualization or geography and their applications as well as to experts in any of these domains as a reference and design space.

This survey is structured as follows: After discussing related work in Section \ref{sec:relatedwork}, we provide definitions and terminology for geospatial networks and their different types in Section \ref{sec:definition}. Section \ref{sec:methodology} then details our methodology for finding, selecting, and coding papers and how our design space evolved until its final state, while Sections \ref{sec:d-geography} to \ref{sec:d-interactivity} explain our design space's five dimensions and examples of visualization techniques. In Section \ref{sec:challenges}, we discuss specific challenges and how they may be addressed. We conclude the survey 
with a discussion and list of open problems in Section \ref{sec:discussion}. 

\vspace{-0.25em}
\section{Related Work}
\label{sec:relatedwork}

Surveys on \textbf{network visualization} have been compiled for many aspects in networks such as techniques for large graphs ~\cite{von_landesberger_visual_2011}, 
group structures in graphs~\cite{vehlow_state_2015},
dynamic graphs~\cite{beck_state_2014}, 
multivariate networks~\cite{nobre_state_2019}, temporal multivariate networks~\cite{kerren_temporal_2014}, multilayer networks~\cite{mcgee_state_2019}, and graph visualization in general~\cite{herman_graph_2000}. A variety of these surveys include visualizations for geospatial networks but do so for reasons other than surveying visualizations of geospatial networks as a whole. 
As a consequence, features and challenges specific to representing geospatial data are not discussed in detail, and do not play a significant role in any taxonomies or classifications introduced in these surveys.
For example, surveys on edge bundling techniques~\cite{zhou_edge_2013,lhuillier_ffteb:_2017} frequently include techniques with demonstrated applications to geospatial networks, or even specifically designed for this purpose, but they lack a wider discussion on visualization of geospatial networks in general.

Furthermore, a variety of surveys and textbooks on \textbf{geographic and spatio-temporal visualization} have been published. Bertin discusses maps as well as networks in his `Semiology of Graphics'~\cite{bertin_semiology_1983}, but not the combination of the two. Cartography, thematic mapping, and map design are discussed in numerous works by cartographers~\cite[e.g.,][]{robinson_elements_1995, dent_cartography:_2009, slocum_thematic_2009, field_cartography_2018}, but networks play only a small role, if any at all, in these books. A survey on `map-like' visualization was presented by Hogräfer~et~al.~\cite{hografer_state_2020}, describing techniques that either imitate or schematize cartographic maps in terms of their primary design elements: points, lines, areas, and fields. For spatio-temporal data, Andrienko and Andrienko offer a systematic approach for exploratory analysis~\cite{andrienko_exploratory_2006}, Andrienko~et~al. discuss visual analytics of movement data~\cite{andrienko_visual_2013}, and Bach~et~al. propose a descriptive framework for spatio-temporal visualizations based on generalized space-time cubes~\cite{bach_descriptive_2017}.

\textbf{Geospatial networks} have been discussed in a set of smaller surveys, focusing mostly on node-link diagrams, graph drawing, flow maps, trajectories~\cite{he_diverse_2019} or specific applications such as crime~\cite{wheeler_visualization_2015} or climate~\cite{nocke_review_2015}. 
Surveys on automatically drawing schematic transit maps, a topic that has received considerable attention in the graph drawing community, were published in 2007 and 2020~\cite{wolff_drawing_2007,wu_survey_2020}.
Rodgers~\cite{rodgers_graph_2005} provided a smaller overview of only node-link representations and graph drawing techniques. 
Similarly, Wolff discussed the use of graph drawing, node-link visualizations, and flow maps in cartography~\cite{wolff_graph_2013}.
However, neither of these present a full survey or comprehensive typology of geospatial network visualizations. 
The application of visual analytics methods to geographic networks is discussed by Rozenblat and Melançon \cite{rozenblat_methods_2013}, but their focus is not on visualization methods as such, although they include an overview of edge bundling methods. 
Jenny et al.~\cite{jenny_design_2018} have established design principles for flow networks.
Finally, a variety of geospatial network visualizations have been created by practitioners, compiled in the online resource \textit{Visual Complexity} (\href{http://www.visualcomplexity.com/}{visualcomplexity.com}).

Closest to our work, Hadlak et al.~\cite{hadlak_survey_2015} presented a survey on the visualization of multi-faceted graph data, in which spatial data is discussed as one possible facet of a multi-faceted graph. 
The classes of techniques in our \textit{Composition} dimension (Section \ref{sec:d-composition}) were informed by Hadlak~et~al.'s classification. However, Hadlak~et~al. do not provide a deeper discussion specifically on geospatial networks and their underlying visual representations. To the best of our knowledge, ours is the most comprehensive survey on the visualization of geospatial networks.

\vspace{-0.25em}
\section{Scope and Definitions}
\label{sec:definition}

Geospatial networks and their visualizations are used in many fields such as information visualization, geovisualization, and graph drawing and can include different types of data and terms.
This section aims to give an overview over the most common terms and to define the scope of this survey. 

\vspace{-0.5em}
\subsection{Terminology for Data Types}

\textbf{Networks} are considered synonymous to \textit{graphs} in this survey. A graph is formally defined as a pair $G = (V, E)$, where $V$ is a set of nodes (or \textit{vertices}) and $E$ is a set of links (or \textit{edges}), with each link either being a set of two nodes (undirected graphs) or an \textit{ordered} set of two nodes (directed graphs). Both nodes and links can have an arbitrary set of attributes $A_V$ and $A_E$ associated with them.

In a \textbf{spatial network}, nodes are associated with inherent and semantically meaningful spatial positions. For example, in brain connectivity networks, nodes are distinct regions, and their  position information is essential for understanding brain activities. Networks with arbitrarily determined positions such as those generated by network layout algorithms are \textit{not} considered spatial networks. Fixed node positions increase the difficulties when designing visualizations for spatial network data. 

\textbf{Geospatial networks} are a subgroup of spatial networks where the node locations are of a geographic nature, i.e., these nodes represent locations on the Earth or other planets. As locations on the surface of approximately (but not exactly) spherical bodies, \textbf{geospatial locations} have characteristics that differentiate them from other spatial data, making the distinction between \textit{spatial} and \textit{geospatial} essential. 
Besides precise geographic coordinates, locations can be defined more semantically and come with their own set of challenges. For example, areas can be well-defined and non-overlapping (e.g., \textit{Germany}, \textit{France}), roughly defined with fuzzy boundaries (\textit{Sahara desert}), nested (\textit{Scotland}, \textit{UK}), and overlapping (\textit{Schengen Area}, \textit{European Union}). Geographic information could point to multiple possible geographic positions of either certain value (\textit{Brest (France)} or \textit{Brest (Belarus)}) or uncertain value (\textit{`Mum's house'} or \textit{`the forest'}), or a geographic location could point to entirely fictional places (\textit{`Atlantis'}). Some of these cases introduce specific challenges to visualization. Reviewing the existing visualization and interaction techniques for geospatial network data is the focus of our survey.

\vspace{-0.5em}
\subsection{Related Concepts}

There are several concepts adjacent to geospatial networks in that they describe relationships or movements between different geographic locations. However, we do not consider all types of such data geospatial networks. 

Firstly, it is useful to think about these concepts in terms of \textit{link continuity}, which describes to what extent data about the trajectory of each link is available.
In some cases, like GPS tracking or air traffic data, continuous physical routes between origins and destinations are available. Such data are called \textbf{trajectories}. Trajectories have very high link continuity. Other data might be structured such that full trajectories are not available but \textbf{intermediate stops} between the origins and destinations are recorded, such as parcel tracking data. \textbf{Origin-Destination (or OD) data} consists of only origin-destination pairs, with no information on the trajectory between the two locations. OD data has the lowest link continuity.

Secondly, it is necessary to consider to what extent the locations in the data (i.e., origins and destinations) represent \textit{network nodes}. If locations are essentially arbitrary, for example in the case of ride-sharing data where a trip can start and end at any given location (as opposed to e.g., bus travel being limited to bus stops), each `node' would only be connected to a single link. Here, further abstraction in the form of aggregating individual origins and destinations into areas would be necessary to form a meaningful network with nodes that have more than one link each. Data types with higher link continuity can be abstracted to data types with lower link continuity; for example, trajectories can be treated as OD data by ignoring the information about trajectories between origins and destinations.

In summary, OD, trajectory, and other geospatial data can often be interpreted as or abstracted to geospatial network data---which is why our survey contains several techniques intended for these data types---but not all OD or trajectory data should automatically be considered geospatial network data. 

\vspace{-0.25em}
\section{Methodology and Design Space}
\label{sec:methodology}

Having defined the scope of our survey, this section details how we gathered papers from scientific venues, removed irrelevant papers, explored different approaches for classifying and discussing papers, and describes our final design space. 

\vspace{-0.5em}
\subsection{Collecting Papers}

Contributions to geospatial network visualization have come from many different fields including information visualization, graph drawing, and cartography. To account for the diverse range of publication venues for potential visualizations of geospatial networks, we followed a two-step approach. First, we looked at the proceedings and collections of major venues where work on geospatial and network visualization would naturally be published:

\begin{itemize}
    \item IEEE VIS (InfoVis) (Accessed via \cite{isenberg_vispubdata.org:_2017})
    \item ACM Conference on Human Factors in Computing Systems (CHI) (Accessed via \cite{acm_sigchi_acm_2019})
    \item IEEE/CGF EuroVis (Accessed via \cite{wiley_online_library_computer_2019})
    \item IEEE PacificVis / Asia-Pacific VIS (Accessed via \cite{pacificvis/apvis_visualization_nodate})
    \item Symposium on Graph Drawing (GD) (Accessed via \cite{springerlink_international_2019})
    \item ACM SIGSPATIAL conferences (Accessed via \cite{acm_sigspatial_acm_nodate})
\end{itemize}

For each of these venues, we manually scanned the proceedings and retrieved candidate papers based on their title. Papers were included in this initial collection if their titles contained references to geospatial networks and their visualization. In an effort to obtain as many relevant papers as possible, we also included many papers where the title only mentioned either networks or geovisualization. We chose this approach because `geospatial networks' are not a universally recognized concept, and many papers in our collection use different terms. The resulting variety in terms being used for similar concepts makes it problematic to identify relevant papers based on title or keywords alone. This resulted in a set of 191 candidate papers.

In a second pass we manually selected the most relevant papers out of the 191. To that end, we read the abstract and checked figures in each paper. This reduced our collection to 41 papers. 
From these 41 papers, we then extracted keywords, which were used to perform automated searches in the following online libraries which cover all major visualization and geographic journals:

\begin{itemize}
    \item ACM Digital Library
    \item IEEE Xplore
    \item Taylor \& Francis Online (publisher of several geography-related journals, e.g., \textit{Cartography and Geographic Information Science} \& \textit{International Journal of Geographical Information Science})
    \item Google Scholar
\end{itemize}

The search terms (keywords) are a combination of different terms to describe geospatial networks and terms used to describe the visualization aspect. Terms were combined in a Boolean search query as follows:

\begin{center}
\texttt{(geographical network(s) | geographic network(s) | geospatial network(s) | spatial network(s) | spatial interaction data | origin-destination)}

\texttt{\&}

\texttt{(visualization | visualisation | graph drawing | flow map)}
\end{center}

This yielded a large number of possible papers. Again, we manually narrowed down the results of the search by examining at least the abstract and figures in each of the retrieved papers. In addition, a number of papers were discovered through following references of some of the already retrieved papers as well as recommendations from reviewers. This selection step yielded another 52 papers, raising the final number of papers to \ncount. A paper was included if both of the following criteria were fulfilled:

\begin{itemize}
    \item \textbf{C1)} A technique must be \textbf{motivated by and designed for geospatial networks} and must visualize both geospatial and network information. If a technique is \textit{not} explicitly designed for geospatial networks, its \textbf{application must be demonstrated} \textit{and} address a challenge in visualizing geospatial networks.
    \item \textbf{C2)} A paper must contain a \textbf{novel and representative technique}, rather than iterating or adapting existing techniques. For example, we found many papers placing nodes at geographic positions and connecting them. Our survey does not list \textit{all} these papers but a manually chosen representative sample.
\end{itemize}

For example, we excluded an algorithm for clustering trajectories \cite{andrienko_clustering_2018} or a technique for visualizing vessel movements \cite{willems_visualization_2009} because while trajectory or movement data can often be abstracted to network data, it was not in the context of these techniques. Further, we excluded many edge and trail bundling techniques and instead selected a sample intended to represent the different possible types of edge bundling, which will allow for discussing implications for visualizing geospatial networks. In this survey, the term \textit{edge bundling} encompasses both edge and trail bundling techniques, since the differences between the two categories are negligible in our context~\cite{lhuillier_state_2017}.

\vspace{-0.5em}
\subsection{Creating a Design Space}
\label{sec:designspace}

Before arriving at our final design space, we went through several iterations of taxonomies and typologies, each one informed by the techniques themselves as well as alternative higher-level objectives.
For all iterations, a structured coding of our collection of papers was performed by one of the authors, informing our assessment of how useful each approach is. For the final design space, the full collection was independently coded by two people and disagreements resolved through discussion afterwards. The full, coded collection is available on the website (\href{https://geonetworks.github.io}{geonetworks.github.io}).

\textbf{Version \#1: Data-driven}---Our first approach was to structure techniques by the data types a given technique can be applied to, e.g., directed or undirected networks, additional link attributes, or dynamic geospatial networks. This approach is informative for describing how specific visualization challenges are addressed by the literature. However, this approach did not provide informative insight into the visual characteristics of different techniques and help discuss conceptual ideas behind techniques' design. We decided to discuss our grouping for \#1 later in Section \ref{sec:challenges}, complementing our design space.

\textbf{Version \#2: Technique-driven}---Our second approach involved coding papers by grouping common visualization techniques for networks such as node-link diagrams, adjacency matrices, flow maps, etc.
This is a common approach to classify techniques across the visualization community as it often leads to a high-level overview of major classes and `ideas' of techniques. However, it did not yield insightful results as most techniques (\perc{63}) used node-link diagrams, most of which apply edge bundling. We were not able to derive a meaningful discussion about how a specific challenge motivates a specific design and what problems it is solving. Moreover, we felt we would fail to capture the richness of all the different visualization approaches and to list meaningful directions for future design. 

\textbf{Version \#3: Challenge-driven}---Our third intention was to create a taxonomy around which problems and challenges a technique addresses. 
We were hoping this would result in a practical resource of solutions to common challenges.
However, we found that such an approach would suffer from three major drawbacks. First, a list of challenges is necessarily incomplete if not derived from a systematic schema. In other words, without a systematic approach to understand challenges in geospatial networks, any list would purely capture the state-of-the art and our `taxonomy' would be outdated with the next technique proposing a solution to an unsolved (and hence not appearing in our taxonomy) problem.
Second, we would potentially not be able to agree on the definition of a \textit{challenge}. For example, visual link clutter is a problem of node-link diagrams, but not of geospatial networks themselves. For example, using adjacency matrices \textit{avoids} this problem, rather than \textit{solving} it.
Lastly, a taxonomy based on challenges would not help to discuss and understand \textit{design solutions}, decisions, and to potentially inform new designs. 
Together with Version \#1, this informs our discussion of challenges, and techniques addressing them, in Section \ref{sec:challenges}.

\textbf{Version \#4: Representation-driven}---We eventually decided to code visualization techniques according to how they balance their visual representations at the tension between \textit{i)} explicitly showing all possible information in a geospatial network, i.e., all links, all nodes, all geographic information and places of interest, and 
\textit{ii)} managing visual clutter and information overload to provide for efficient task-oriented visual representation. 
We found our design space, dimensions, and classifications to best capture the trade-offs required in designing geospatial network visualizations and to provide a conceptual framework perhaps similar to the design space described by space-time cubes~\cite{bach_descriptive_2017} or map-like visualizations~\cite{hografer_state_2020}. The complete rationale is given in Section \ref{sec:finaldesignspace}.

\begin{figure}[t!]
\subfigure[D1: Geography representation (\fgeo)]{
    \includegraphics[width=\columnwidth]{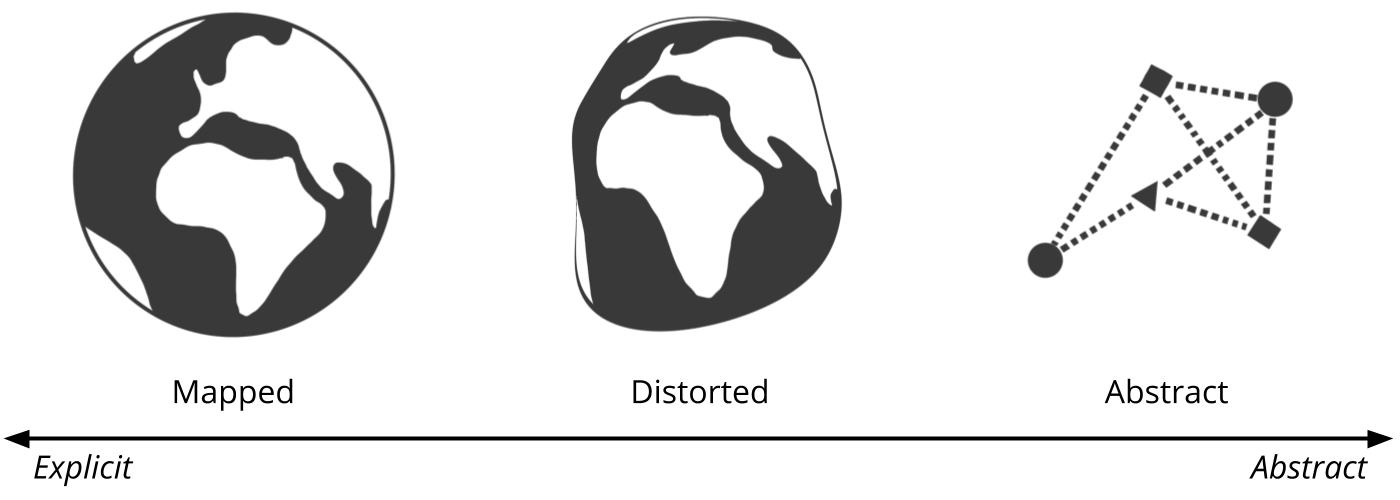}
    \label{fig:d-geo}
}
\subfigure[D2.1: Node representation (\ftopo: \fnode)]{
    \includegraphics[width=\columnwidth]{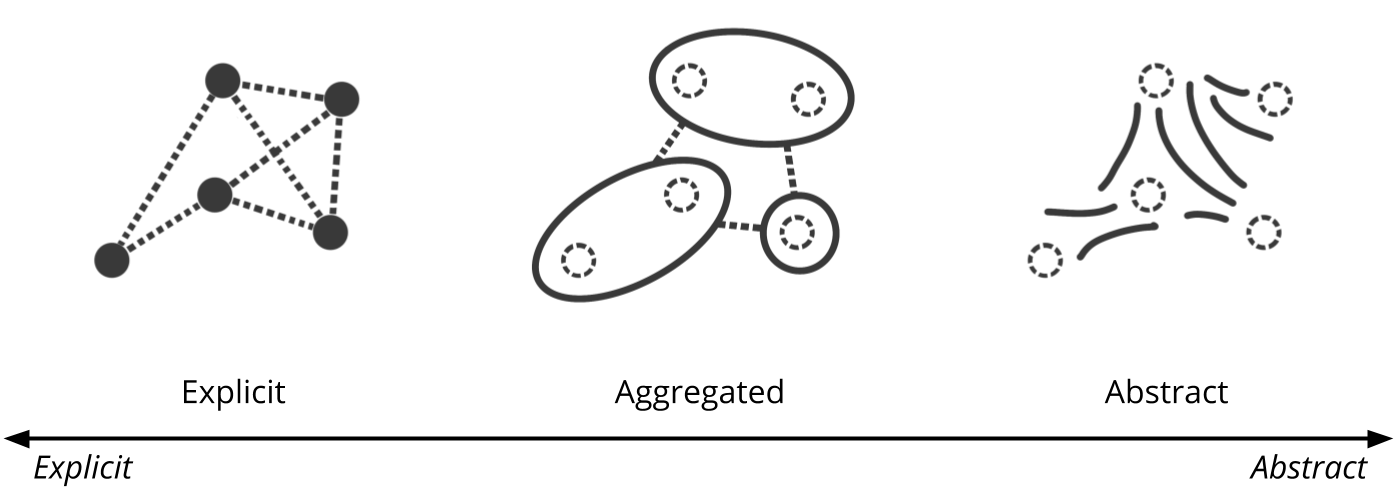}
    \label{fig:d-node}
}
\subfigure[D2.2: Link representation (\ftopo: \flink)]{
    \includegraphics[width=\columnwidth]{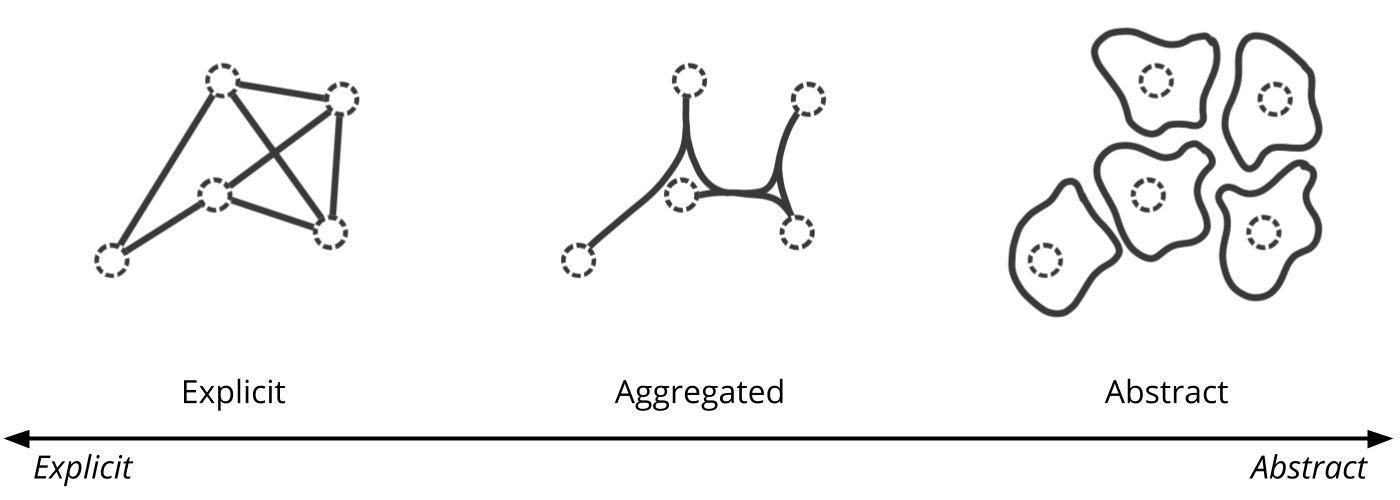}
    \label{fig:d-link}
}
\subfigure[D3: Composition (\fcomp)]{
    \includegraphics[width=\columnwidth]{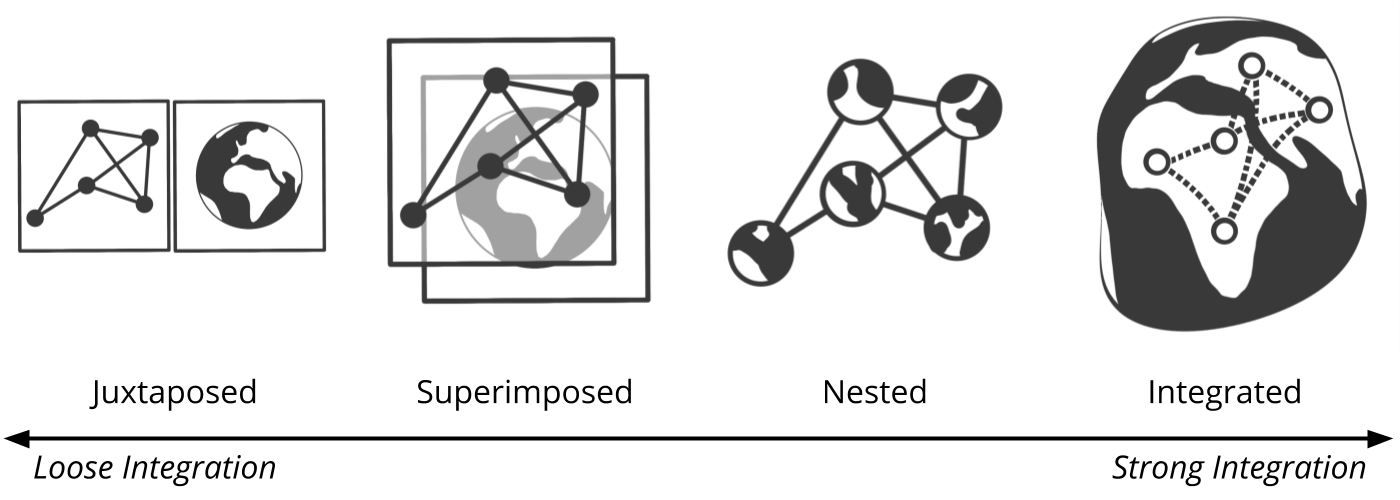}
    \label{fig:d-comp}
}
\subfigure[D4: Interactivity (\finteract)]{
    \includegraphics[width=\columnwidth]{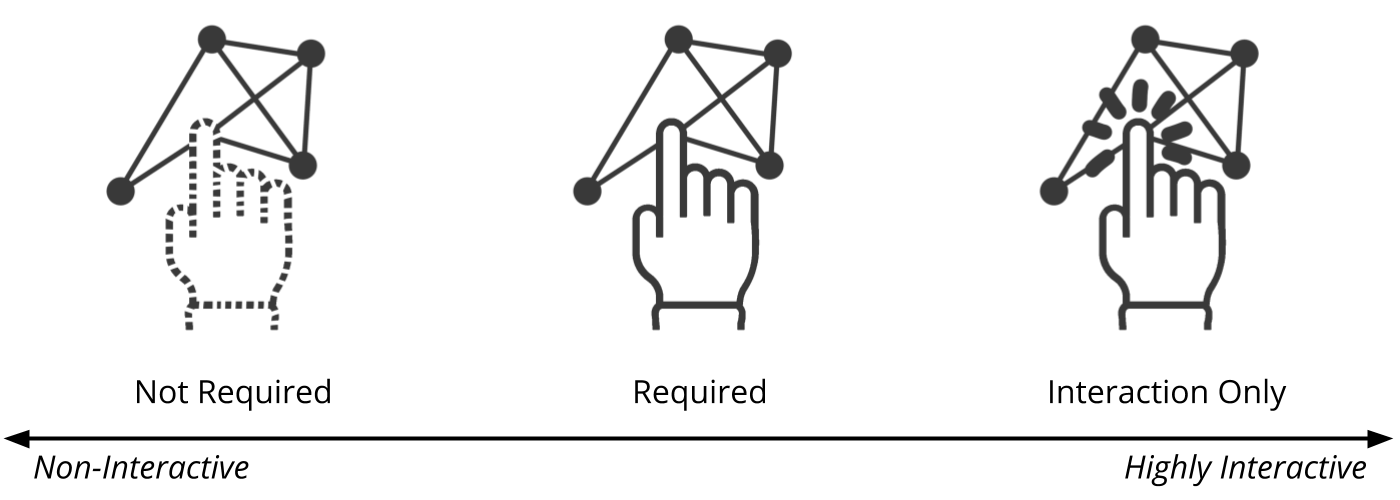}
    \label{fig:d-interact}
}
\caption{Dimensions and categories in our design space.}
\label{fig:designspace}
\end{figure}

\vspace{-0.5em}
\subsection{Final Design Space}
\label{sec:finaldesignspace}

Our final design space consists of four dimensions and two sub-dimensions (Figure \ref{fig:designspace}), each representing an essential aspect of a geospatial network visualization:

\begin{itemize}[noitemsep]
 \item \textbf{D1: Geography Representation} (\fgeo, Figure \ref{fig:d-geo}) describes how geographic information is visually represented.
 This dimension ranges from \textit{explicit} to \textit{abstract} visual encodings, where \textit{explicit} implies a representation that uses a (cartographic) map projection. Such \textit{mapped} (explicit) visual encodings follow the underlying principle of geographic maps: mapping geospatial locations, which are naturally three-dimensional, onto a 2D plane by using a map projection. Alternatively, locations may be displayed on a 3D globe.
 \textit{Abstract} encodings use visual encodings not based on map projections. These encodings use alternative ways of encoding geographic information, thus potentially leaving more visual space and visualization options (position, visual marks, etc.) for encoding network and other information in the visualization.
 Between mapped and abstract techniques, we can find techniques that distort geography (\textit{distorted}).
\end{itemize}

\begin{itemize} 
 \item \textbf{D2: Network Representation} (\ftopo, Figure \ref{fig:designspace}(b+c)) describes how topological information of the network is represented. Like \fgeo, this dimension describes visualization techniques along a spectrum ranging from \textit{explicit} to \textit{abstract}. 
 An \textit{explicit} encoding uses a one-to-one mapping, where each node and link is encoded as a separate visual element. This allows for precise topological tasks such as assessing if two nodes are connected. \textit{Aggregated} encodings use one-to-many mappings, where multiple nodes or links are represented as one visual element. 
 An \textit{abstract} encoding uses a more complex mapping of topological data to visual elements. Decoding the visualization to extract low-level topological information is often not possible,
 but an abstract network encoding can make more visual space and encodings available for the geographic aspect of the data.

 \end{itemize}
 
 As we found techniques that abstract only one of both \textit{nodes} or \textit{links}, we classify techniques separately for nodes and links. 
 
 \begin{itemize}
     \item \textbf{D2.1: Node representation} (\fnode, Figure \ref{fig:d-node}) describes how nodes are represented: \textit{explicit}, \textit{aggregated}, or \textit{abstract}.
     \item \textbf{D2.2: Link representation} (\flink, Figure \ref{fig:d-link}) describes how links are represented: \textit{explicit}, \textit{aggregated}, or \textit{abstract}.
 \end{itemize}

The dimensions \fgeo\ and \ftopo\ can be seen as two sides of the same coin: making decisions about the visual representation of network topology implies a decision on how to represent geography. For example, a visualization might choose to explicitly represent geography and maintain spatial distances between locations. This will most necessarily result in issues with node overlap if nodes are placed at similar or nearby locations. Or links overlap for the same reason and can stretch far over the geographical representation. On the other hand, a designer could choose to abstract geographic information, e.g., by distorting geographical distances to provide for better perception and understanding of a network's topology. Different visualization techniques propose different solutions to overcome this tension and abstract information in the network as we discuss in Sections \ref{sec:d-geography} to \ref{sec:d-link}.

Besides \fgeo\ and \ftopo, we include two further dimensions.

\begin{itemize}
\item \textbf{D3: Composition} (\fcomp, Figure \ref{fig:d-comp}) describes how network and geographical information are integrated visually. Inspired by Hadlak~et~al.'s composition mechanisms for multi-faceted networks~\cite{hadlak_survey_2015}, this dimension runs from a \textit{loose} integration (e.g., juxtaposition) to a \textit{strong} integration.

\item \textbf{D4: Interactivity} (\finteract, Figure \ref{fig:d-interact}) describes to what extent a technique \textit{requires} user interaction for exploring and connecting geography and network data of a geospatial network or whether a technique is an interaction technique in its own right (\textit{Interaction Only}).
\end{itemize} 

With this dimension-driven approach, we can capture the richness as well as some of the design decisions in existing techniques, while at the same time providing a design space to locate and compare existing techniques as well as inform discussions about missing approaches. Each dimension classifies techniques according to how much of that information (geography, nodes, links) they show explicitly, and how much of that information they abstract and aggregate. Along each dimension, techniques and designs are roughly grouped into categories, although transitions between these categories are fluent. The following Sections \ref{sec:d-geography} to \ref{sec:d-interactivity} detail each of these dimensions and discuss the types of representations, compositions, and interactions we found. A discussion on the limitations of this approach is provided in Section~\ref{sec:discussion}.


\section{D1: Geography Representation (\fgeo)}
\label{sec:d-geography}

The \fgeo\ dimension describes how geographic information is represented visually. Geographic information includes geographic locations (or node locations) as well as more general information related to spatial distances, regions, landmarks, relations between these locations (hierarchical, distances, etc.), and any additional geographic information important for the visualization. 
For \fgeo, we define three major categories along a continuous \textit{explicit---abstract} spectrum (Figure \ref{fig:d-geo}):
\textit{mapped}, \textit{distorted}, and \textit{abstract}. Note that these `categories' are not discrete sets, but rather steps along a continuous spectrum. 

\vspace{-0.25em}
\subsection{D1---\fgeo: Mapped}
\label{sec:mapped}
\iconfig{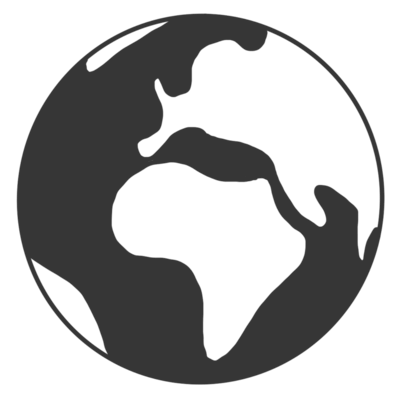}

\begin{figure}[b]
\subfigure[]{
  \includegraphics[height=2.4cm]{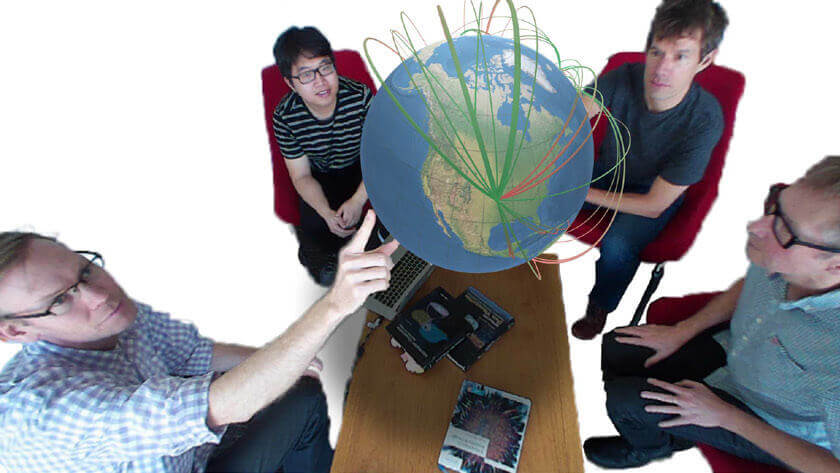}
  \label{fig:yang2019}
}
\hfill
\subfigure[]{
  \includegraphics[height=2.4cm]{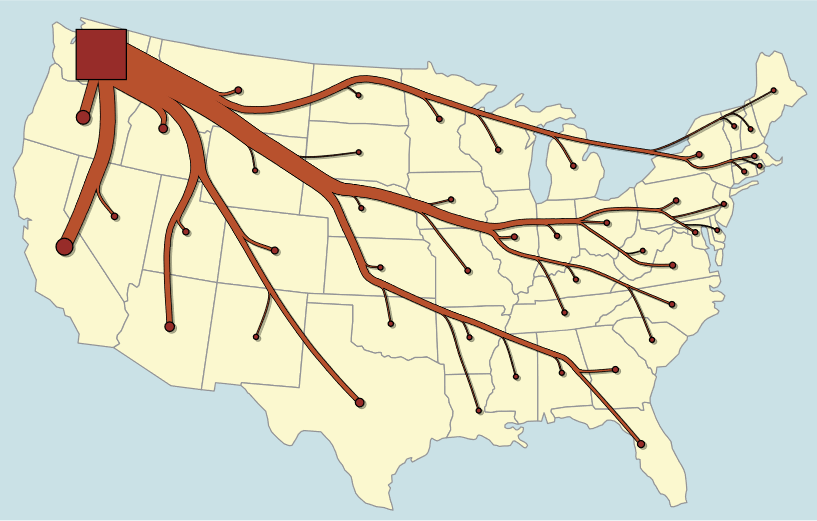}  
  \label{fig:buchin2011}
}
\caption{Examples for \fgeo--Mapped: 
(a) A globe in VR~\cite{yang_origin-destination_2019}, 
(b) A flow map superimposed on a map; it uses an automatic layout based on spiral trees~\cite{buchin_flow_2011, buchin_angle-restricted_2011} 
(Reprinted by permission from Springer Nature: Springer. Lecture Notes in Computer Science. ``Angle-Restricted  Steiner  Arborescences  for Flow Map Layout'', Buchin, K., Speckmann, B., and Verbeek, K.  \copyright\ 2011). 
}
\label{fig:geo_mapped}
\end{figure}

Mapped techniques are the most explicit geographic representations.
In mapped techniques, geospatial locations are visually represented by positioning visual elements using a geographic map projection. Any map projection introduces distortion as it is impossible to flatten a three-dimensional surface into a two-dimensional map without any distortion~\cite[][p.~3]{snyder_map_1987}. 
However, different map projections preserve different features of the geography, e.g., angles, areas or some distances, and as such the choice of projection is always a trade-off between different kinds of distortion. 
In addition, the \textit{mapped} category includes three-dimensional globes.
We chose not to differentiate between two-dimensional and three-dimensional representations as separate categories in this design space because the distinction is not about 2D or 3D but about a consistent presentation of a geographical space. 
Distortion, as introduced in the next section, distorts a geographic space based on the network data and thus introduces a data-driven distortion. 
For example, 3D globe representations can be distorted in the same way as 2D maps, based on the network's topology~\cite{alper_dynamic_2007}.
The majority of the surveyed techniques use mapped representations (\perc{71}).

\begin{figure}[b]
\centering
\subfigure[]{
  \includegraphics[width=0.9\linewidth]{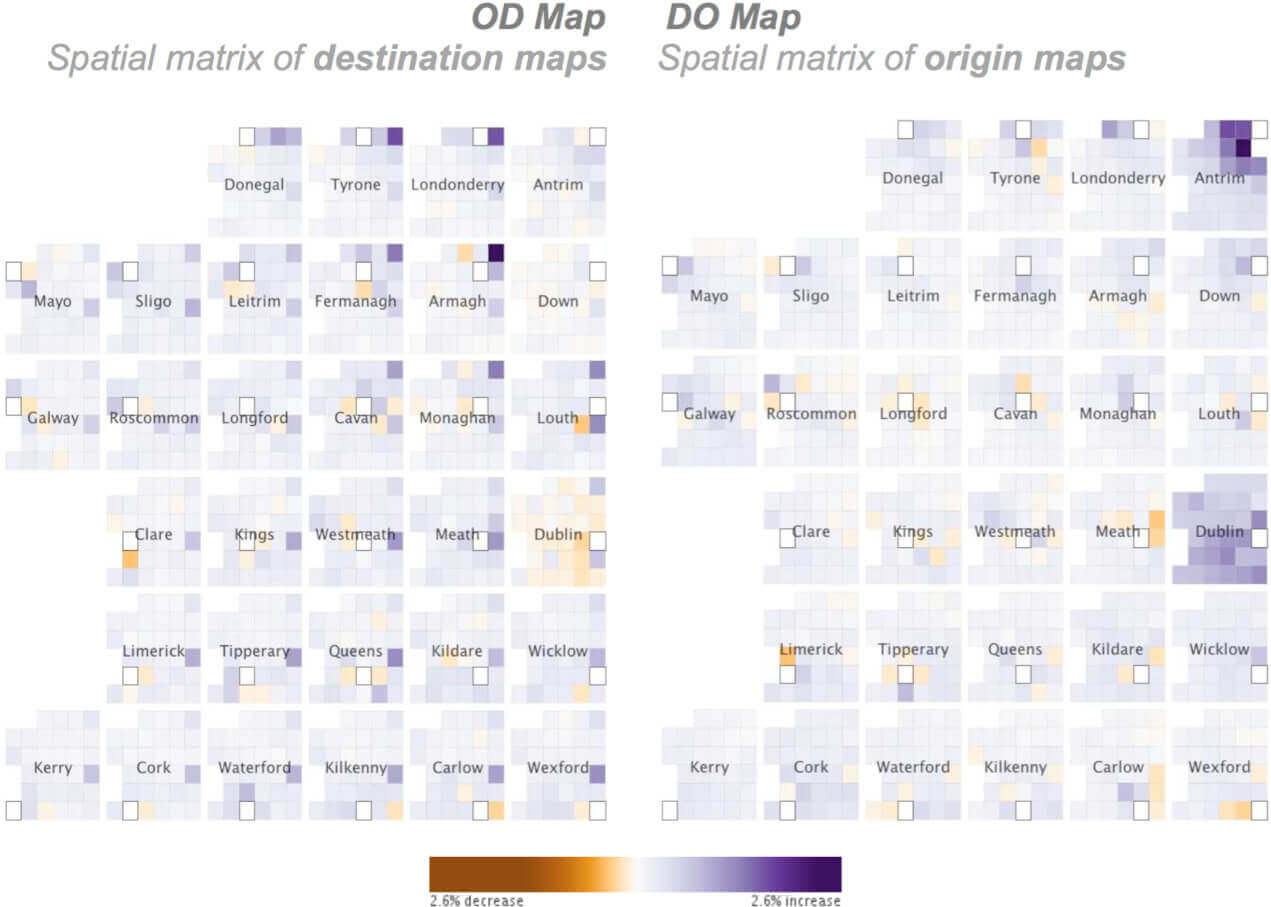}  
   \label{fig:wood2010-new}
}
\subfigure[]{
  \includegraphics[height=4.5cm]{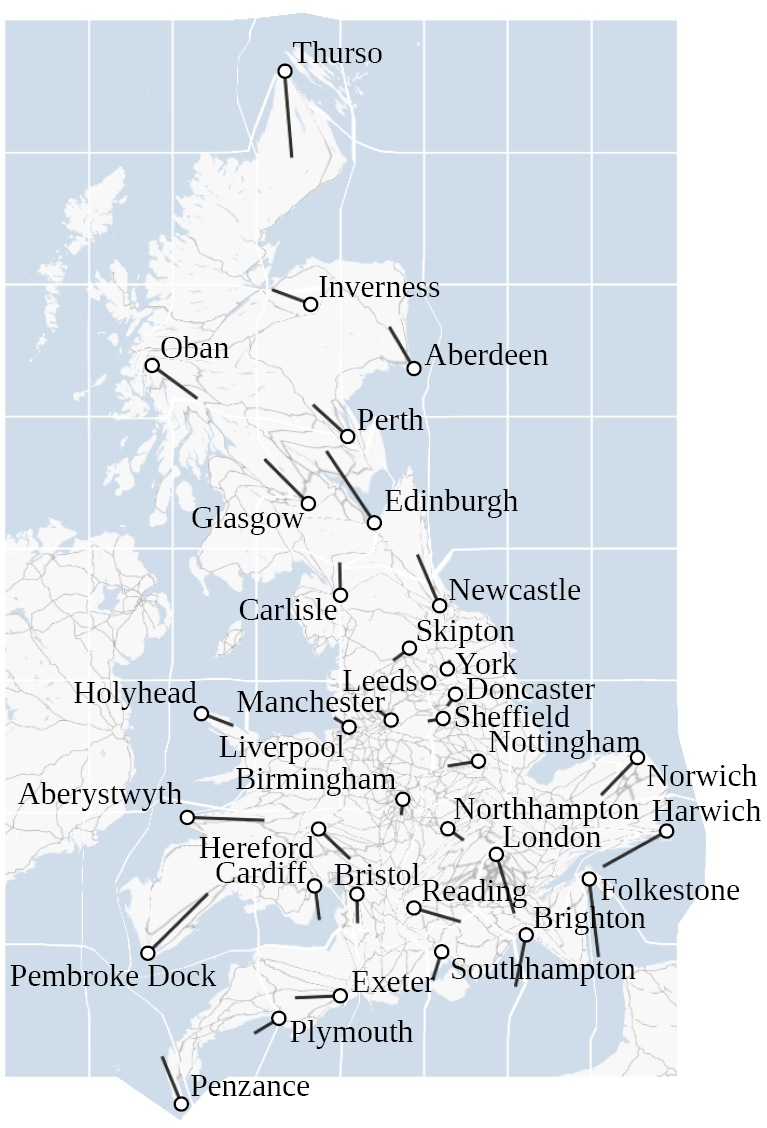}  
   \label{fig:bouts2016}
}
\subfigure[]{
  \includegraphics[height=4.5cm]{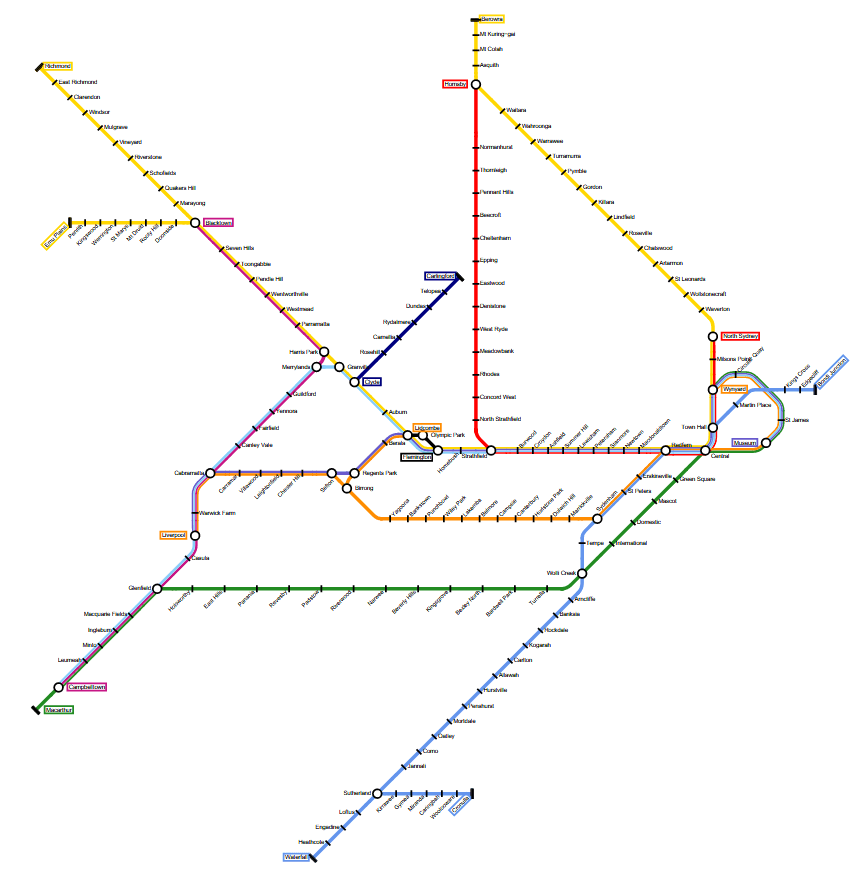}  
  \label{fig:nollenburg2011}
}
\caption{Examples for \fgeo--Distorted: 
(a) OD map~\cite{wood_visualisation_2010} (reprinted from \cite{kelly_historical_2013}), 
(b) A deformed map, visualizing travel times by train in the UK~\cite{bouts_visual_2016} (reprinted from \cite{bouts_thesis}), 
(c) An automatically created and labelled metro map~\cite{nollenburg_drawing_2011} (reprinted from \cite{nollenburg_network_2009}).
}
\label{fig:geo_distorted}
\end{figure}

\textbf{Three-dimensional (3D) globe} representations offer the most precise geographic information with respect to geographic area sizes and distances. 3D globes, either displayed on screens or in virtual reality (VR)~\cite{yang_origin-destination_2019} (Figure~\ref{fig:yang2019}), have been used to display node-link diagrams in various forms, 
e.g., using straight lines on a globe~\cite{cox_3d_1996}, arcs on a globe \cite{munzner_visualizing_1996,yang_origin-destination_2019, kaya_multi-resolution_2016}, 
edge bundling around the globe~\cite{lambert_3d_2010, zhang_visual_2018}, 
or flow maps spanning a globe~\cite{debiasi_force_2014}. 
3D globes preserve global distances and sizes of areas. Also, link crossings can be reduced since links can be drawn along their shortest path around the globe, naturally routing links around each other. The main shortcoming of globes is that, unless the entire network is located on one half of the globe, interaction is required for full exploration.

However, most mapped techniques use \textbf{2D maps}, despite the different types of distortions introduced by different map projections (such as Mercator or others)~\cite{battersby_effect_2009}. 
Still, 2D maps are highly usable on 2D screens and in print media. Many visualizations show geographic details such as roads~\cite{andrienko_revealing_2017} and country borders~\cite{ibarra_visualization_2016}. Other techniques reduce geographic detail to country shapes~\cite{buchin_flow_2011} (Figure~\ref{fig:buchin2011}) or remove any geographic detail except locating nodes at their respective geographic position on the screen~\cite{gansner_multilevel_2011, brandes_using_1998}. 

The major advantage of \textit{mapped} representations is that they \textbf{support tasks} related to 
a) purely geographic information such as \textit{Which regions are close? How far are these locations apart? Which country is this? Is there a mountain?} and 
b) geographic information about the network topology (if topology information is provided properly by the network representation): \textit{Is this region well-connected? How far apart are these two nodes? Which one is the longest link in my network?} A second advantage of mapped techniques is that people are familiar with geographic data being displayed on maps. Maps have been shown to have cognitive benefits when interpreting geographic data~\cite{hografer_state_2020}.

The \textbf{drawback} of mapped techniques is that they cause problems when nodes are close or at the same position, or when links span large distances. Depending on zoom level, nodes will often be displayed off screen while at the same time being highly related to the currently visible nodes. Abstracting the geographic information through distortion, aggregation, or removing information can offer some solutions. 
Another issue present in all \textit{mapped} node-link diagrams is the visual dominance of long links. A connection across continents is not necessarily more important than a local link (often quite the opposite), yet takes up much more space simply due to the geographic context, potentially covering shorter links as a side effect. This is further aggravated through the distorted distances caused by most map projections.
Finally, mapped representations result in positional variables of nodes not being available for other information such as connectivity or node type.

\vspace{-0.25em}
\subsection{D1---\fgeo: Distorted}
\label{sec:geo-distorted}
\iconfig{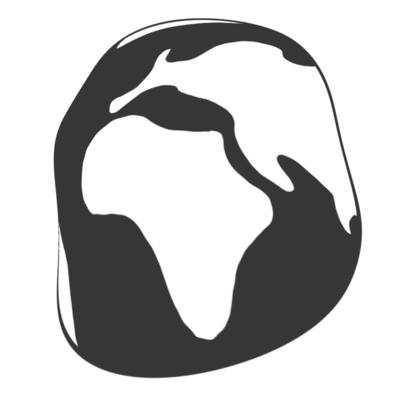}

\textit{Distorted} describes techniques where geospatial locations are displaced with respect to their original position in the initial map projection---both 2D and 3D. 
As discussed in Section~\ref{sec:mapped}, any map projection introduces some distortion as a consequence of projecting 3D space onto a 2D plane, but the \textit{Distorted} category deals with distortions based on the network data, either directly through data-driven algorithmic distortions, or indirectly through interaction techniques that let users distort the visualization to explore the network.
Generally, distorting the geography is used as a way to show the network topology more clearly and avoid difficulties such as described in the last paragraph of the previous section.

For non-spatial networks, laying out nodes based on their connectivity is common practice. For example, force-directed layouts place nodes with strong connectivity (many connections and many common neighbors) closer together. For geospatial networks, there is an inherent conflict between using node positions to represent geographic locations and using them to expose the topology. Distorting the underlying map or displacing individual nodes has been explored as a compromise to address this conflict.
Among the \perc{19} of techniques that use distortion, we identified five main approaches, including both continuous and discontinuous types of distortion: 

First, there are techniques that use continuous distortion to \textbf{show alternative measures} of distance instead of the geographic distance between two nodes~\cite{alper_dynamic_2007,bouts_visual_2016} (Figure~\ref{fig:bouts2016}). 
Alternative measures of distance in this context could be for example travel times or dissimilarity measures. 

Second, \textbf{tile maps} transform geographic regions into a grid of identical tiles, a form of discontinuous distortion. The tiles usually cannot be placed at their original geographic locations~\cite{mcneill_generating_2017}. For example, OD maps~\cite{wood_visualisation_2010} use nested tile maps to represent geospatial networks. As illustrated in Figure~\ref{fig:wood2010-new}, each cell contains a small version of the larger map. The color of each small cell indicates the flow volume into that cell from the larger cell it is nested into.

Third, \textbf{map insets} are a form of discontinuous distortion. Map insets either contain smaller (undistorted) sub-maps at different scales from the main map~\cite{brodkorb_overview_2016,otten_shifted_2018} (Figure~\ref{fig:comp:nested}), or individual nodes~\cite{ghani_dynamic_2011}. 
Insets scale parts of the map up or down, or move locations away from their original position, which we consider a form of distortion since it results in distances, areas and angles no longer being consistent across the map.

Fourth, there are representations where node positions are computed as a \textbf{trade-off} between showing the true geospatial location and clearly showing the network topology, usually using continuous distortions. Essentially, nodes are shifted to increase the legibility of the network representation. 
An example of this is centrality-based scaling \cite{merrick_increasing_2006}, where the underlying geography is distorted such that dense areas in the network are enlarged compared to sparser areas, while preserving link orientation as much as possible. A similar method specifically for road networks is proposed by Haunert and Sering~\cite{haunert_drawing_2011}.
A notable application of this type of distortion are metro map layouts~\cite[e.g.,][]{hong_automatic_2006, wang_focus+context_2011, bottger_map_2008, van_dijk_practical_2019}) (Figure~\ref{fig:nollenburg2011}), a classic application of graph drawing. The common place of such transit maps is that users do not need to know the precise geography of the transport network---it is more important that they can clearly see how different lines connect so that they can plan their route accordingly. As such, a map that distorts the geography to the extent necessary to create an easily legible network map is an ideal trade-off. 

Lastly, continuous as well as discontinuous distortion can be used as an \textbf{interaction technique}, scaling different parts of the network up and down based on user interaction. A fisheye lens \cite{brown_browsing_1993} is a classic implementation of this. For metro maps, a custom scaling method specifically for this purpose has been proposed \cite{wang_focus+context_2011}. Interaction techniques are discussed separately in Section~\ref{sec:interactiontech}.

\vspace{-0.25em}
\subsection{D1---\fgeo: Abstract}
\iconfig{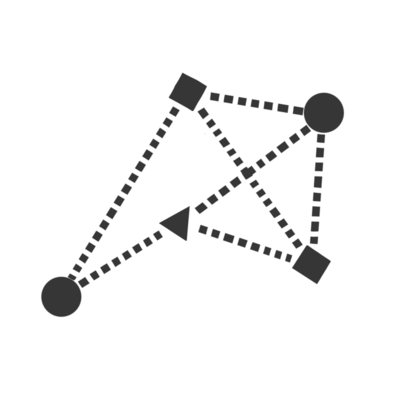}
\label{sec:geo-abstract}

\begin{figure}[t]
\centering
\subfigure[]{
  \includegraphics[height=3.7cm]{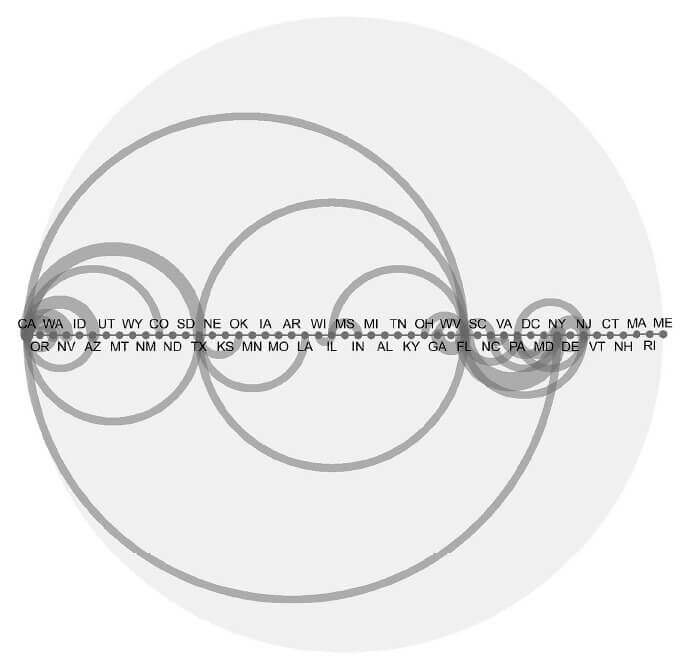}  
  \label{fig:xiao2009}
}
\subfigure[]{
\includegraphics[height=3.7cm]{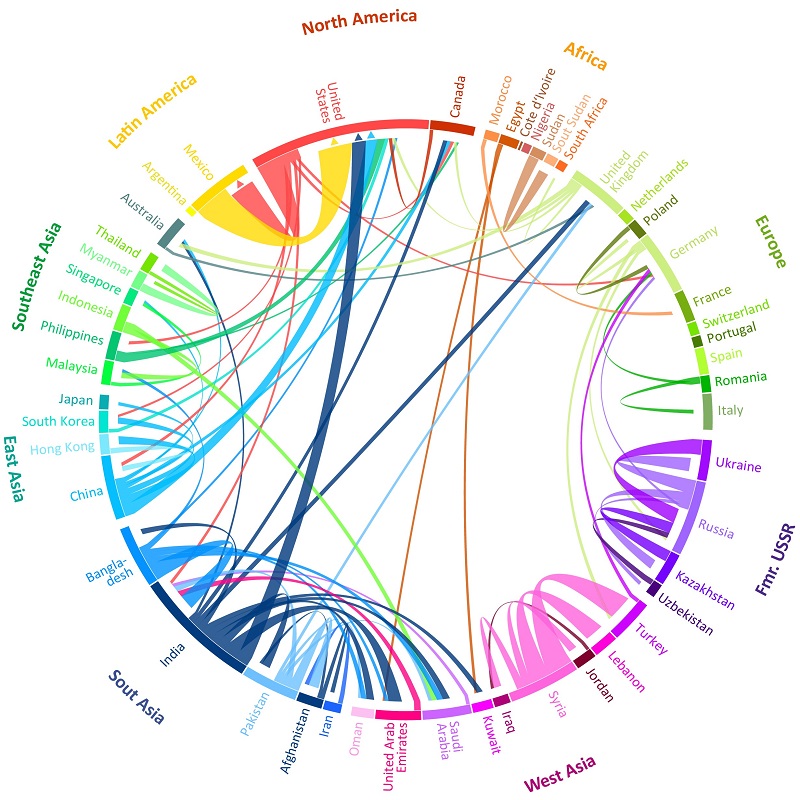}  
  \label{fig:abel2014}
}
\subfigure[]{
\includegraphics[width=1\columnwidth]{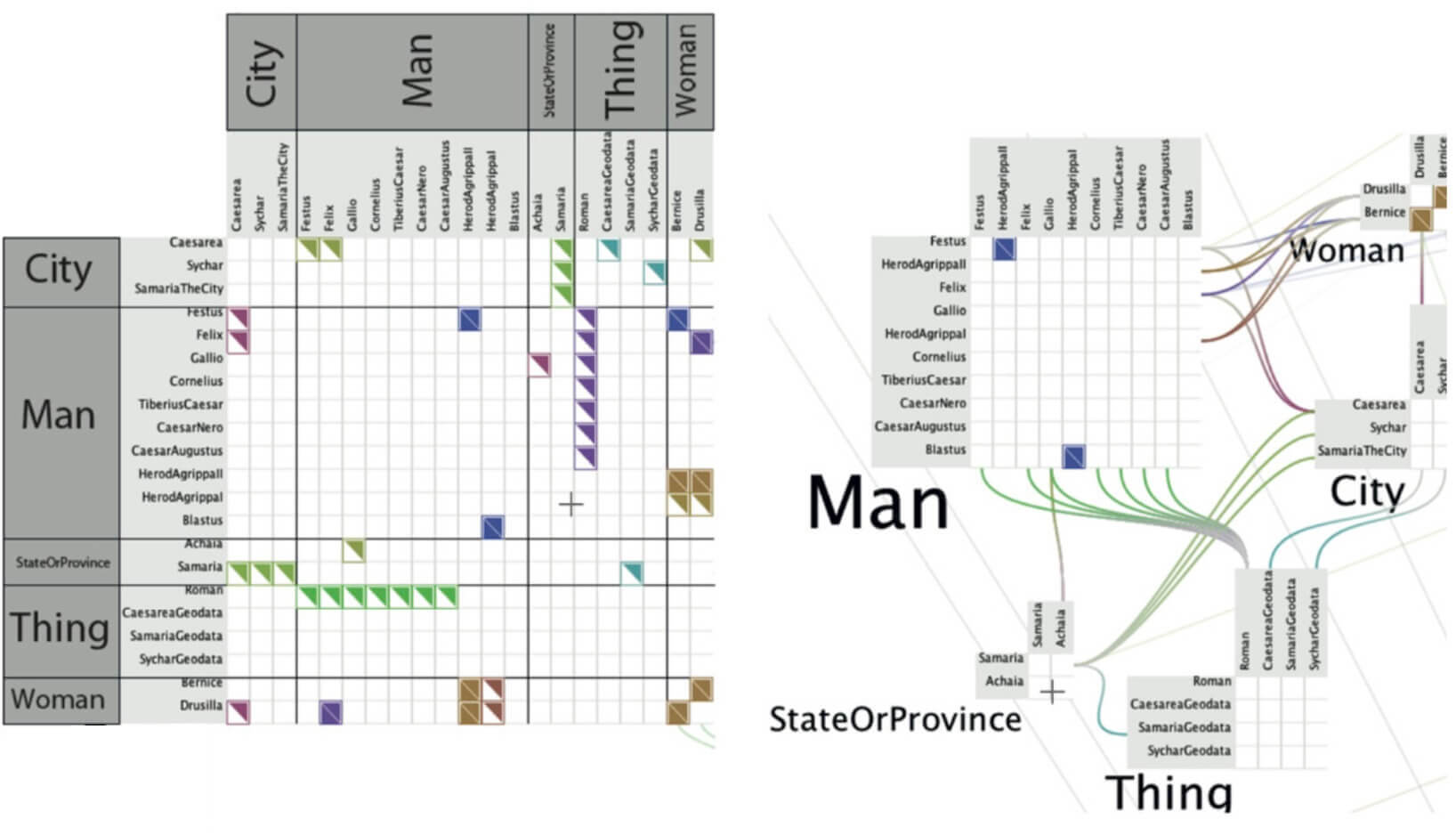}
  \label{fig:bach_ontotrix_2011}
}
\caption{Examples for \fgeo--Abstract:
(a) `Kriskograms', the locations are projected to positions on a one-dimensional straight line~\cite{xiao_visualizing_2009}  (reprinted by permission of the publisher, Taylor \& Francis Ltd, \href{http://www.tandfonline.com}{www.tandfonline.com}),
(b) Global international migration flows shown in a chord diagram, the locations are grouped and color-coded~\cite{abel_quantifying_2014} (Figure courtesy of Federal Institute for Population Research (BiB), Germany),
(c) `OntoTrix', geographic nodes are grouped (`City' node)~\cite{bach_ontotrix_2011}.
}
\label{fig:geo_abstract}
\end{figure}

An abstract geography representation encodes geographic information without the use of map projections. 
This usually includes non-spatial variables such as \textbf{color} or \textbf{shape} (Figure \ref{fig:geo_abstract}) and can be seen as the exact opposite of explicit geographic representations. 
Often, geographic locations are aggregated into coarser groups. 
Note that a visualization can still use position to place visual marks on the picture plane, but in \textit{abstract} geographic representations there is no natural mapping between an element's position on the screen and its geospatial location. 
Displaying locations along a line could be considered a \textit{projection} from 3D to 1D, but it is not a \textit{map projection}, which is defined as a projection onto a 2D plane~\cite[][p.~3]{snyder_map_1987}.

Abstract representations allow for a great variety of designs and, we believe, offer many unexplored solutions as we only found five papers with abstract geographic encodings (\perc{5}). 
Two papers use circular chord diagrams~\cite{hennemann_information-rich_2013,abel_quantifying_2014} (Figure~\ref{fig:geo_abstract}~(b)) in which geographic information is encoded as groups of nodes along the circle (technically along a single spatial dimension, i.e., the circle's circumference), using color as redundant encoding for geographic regions. 
A third technique uses a variation of arc diagrams termed~\textit{Kriskograms}~\cite{xiao_visualizing_2009} (Figure~\ref{fig:xiao2009}), which orders nodes on a horizontal 1-dimensional straight line. 
Our specific example orders nodes according to their position from west to east but other encodings, e.g., grouped by country as in the chord diagrams are easily imaginable. 
In both examples, geospatial locations can be approximated with greater or less detail, e.g., locations on the Northern and Southern as well as Eastern and Western Hemisphere. 
A fourth technique is an \textbf{adjacency matrix}, in which nodes can be grouped by geographic location or region~\cite{bach_ontotrix_2011} (Figure~\ref{fig:bach_ontotrix_2011}) or geographic regions being integrated into the network topology as individual nodes, e.g., a node for every location having links to the nodes related to these locations. 
Further examples of abstract encodings could include coloring nodes in a force-directed node-link layout based on their geographic location, e.g., country.

The \fgeo\ dimension represents a continuous spectrum and provides the designer with many choices and opportunities. Decisions depend on the level of precision required for the geospatial aspect of the data. For example, if a task requires geographic fidelity (e.g., estimating spatial distribution, densities, and distances of nodes), then mapped representations should be naturally considered. If, however, a task can neglect certain geographic fidelity (e.g., estimating number of nodes in a given region), geographic information could be abstracted or distorted to provide space for visualizing information relevant to the task (e.g., distorting space to remove overlap between nodes to support estimating the number of nodes in a given region).


\section{D2.1: Node Representation (\fnode)}
\label{sec:d-node}

The \fnode\ dimension is the first subdimension of \ftopo\ and describes the visual representation of the nodes of the network along an \textit{explicit---abstract} spectrum. In a geospatial network, nodes represent locations or geolocated entities, and they are related to each other by links.  

The most \textit{explicit} representation is one where each node is individually visually represented, either through displaying a visual element such as a dot, or by otherwise clearly indicating its position. For example, node symbols are often omitted in node-link diagrams, but the start and end points of links clearly indicate node positions (e.g., Figure~\ref{fig:holten2009}).
The most \textit{abstract} representation is one where individual nodes are not visually indicated at all, making it hard to reconstruct the overall network topology. In between these two extremes, we find that nodes can be \textit{aggregated} into groups---each node is assigned to a specific group, but it is not displayed as a separate element. The node representation is entirely independent from the link representation, and any of the three \fnode\ classes can be combined with any of the three \flink\ classes. 

\vspace{-0.25em}
\subsection{D2.1---\fnode: Explicit} 
\iconfig{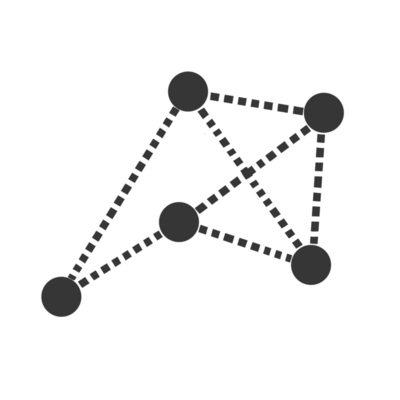}

\begin{figure}
\vspace{-0.5em}
\centering
\subfigure[]
{
  \includegraphics[height=3.85cm]{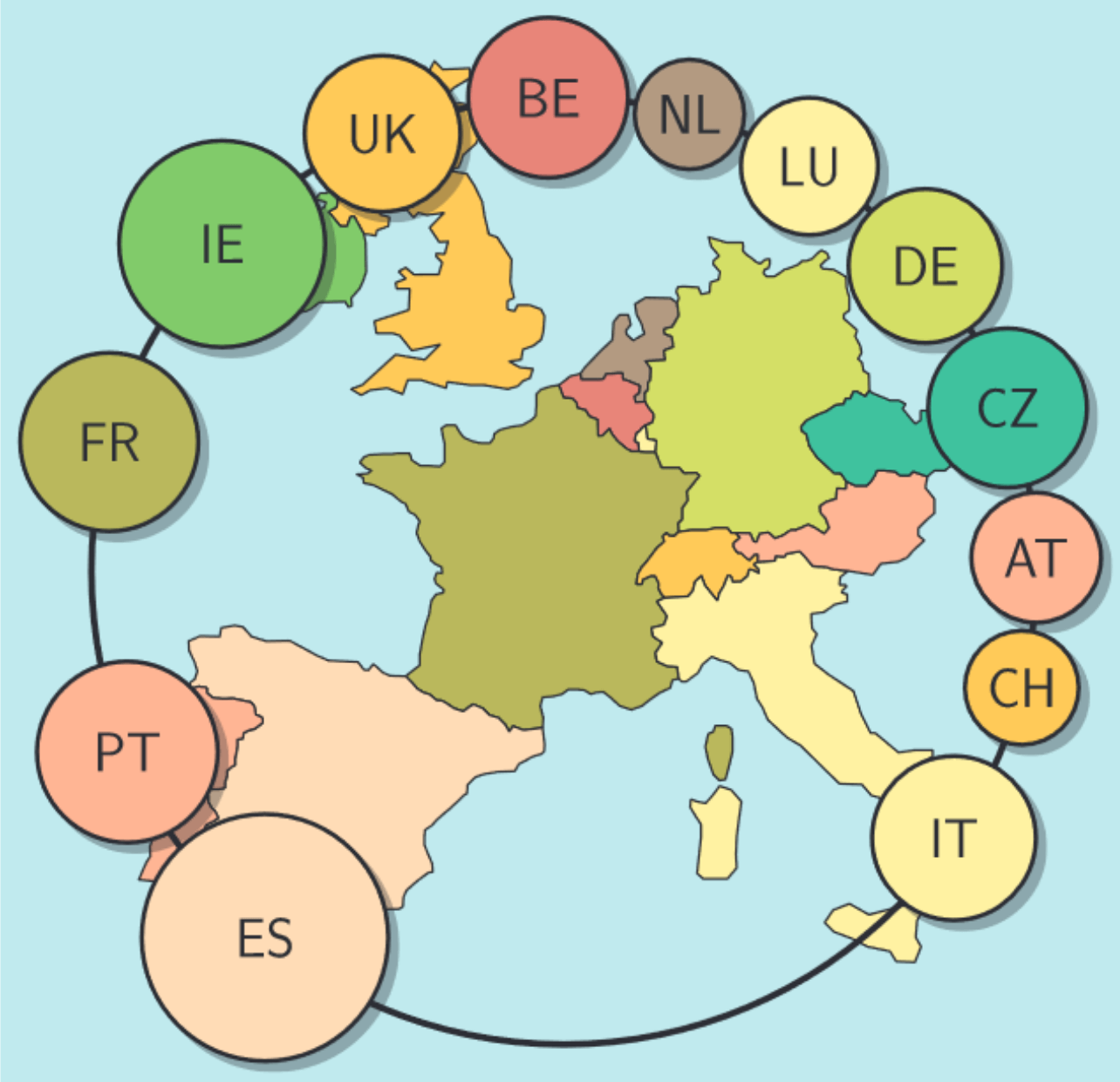}  
  \label{fig:speckmann2010}
}
\subfigure[]{
    \includegraphics[height=4cm]{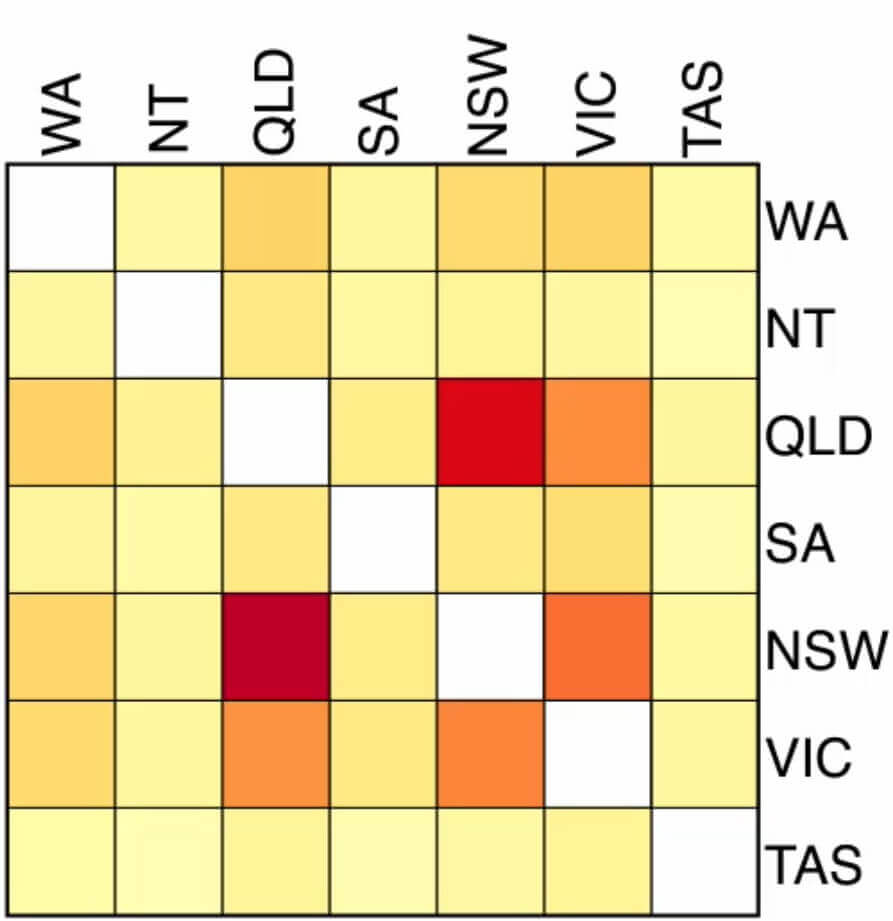}  
  \label{fig:od-matrix}
} 
\caption{Examples for \fnode--Explicit:
(a) A `Necklace map'; the space in the center can be used for displaying links between the nodes on the outer circle~\cite{speckmann_necklace_2010, speckmann_algorithms_2015} (reprinted from \cite{speckmann_algorithms_2015}),
(b) An Origin-Destination (OD) Matrix.}
\label{fig:node_explicit}
\end{figure}

With \perc{83} of the surveyed techniques, the majority of techniques show nodes explicitly, i.e., show each individual node in the network. In any of these representations, nodes can be shown explicitly as points (node-link diagrams, arc diagrams) or rows and columns (adjacency matrix) (Figure~\ref{fig:od-matrix}).  Sometimes, especially in techniques using edge bundling, explicit visual markers for nodes are omitted but  node locations are still clearly identifiable from the endings of links.

Most techniques with explicit node representations are variations of node-link diagrams, differentiated only by the types of links they use (such as straight lines, arcs, bundled edges, etc.). These types of visualizations are typically \textit{superimposed} on a \textit{mapped} geography representation. 
 
However, explicit node representations have also been used in combinations with both \textit{explicit} and \textit{distorted} geography. For example, \textit{Necklace maps}~\cite{speckmann_necklace_2010} (Figure~\ref{fig:speckmann2010}) display nodes on a circle surrounding the relevant part of the map. Nodes are placed on the circle according to their geospatial position, resulting in a \textit{distorted} display of the geography. A similar method is proposed by Stephen and Jenny~\cite{stephen_automated_2017}, where nodes that are out of view are laid out around a target area in a circular arrangement and connected to nodes in the target area (Figure~\ref{fig:stephen2017}). The distorted globe by Alper~et~al.~\cite{alper_dynamic_2007} (Figure~\ref{fig:alper2007}) and deformed map by Bouts~et~al.~\cite{bouts_visual_2016} (Figure~\ref{fig:bouts2016}) use explicit nodes on a globe/map, combined with an abstract link representation.

Combined with an \textit{abstract} geography representation, we have found techniques that display nodes in a circular, spatially ordered layout, or on a single spatially ordered axis, with the network topology shown as a chord diagram~\cite{abel_quantifying_2014, hennemann_information-rich_2013} (Figures~\ref{fig:abel2014} \& \ref{fig:hennemann2013}) or arc diagram~\cite{xiao_visualizing_2009} (Figure~\ref{fig:xiao2009}).

\vspace{-0.25em}
\subsection{D2.1---\fnode: Aggregated}
\iconfig{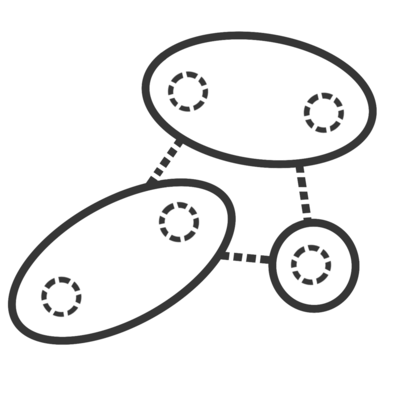}

\begin{figure}[b]
\centering
\subfigure[]{
  \includegraphics[width=0.95\columnwidth]{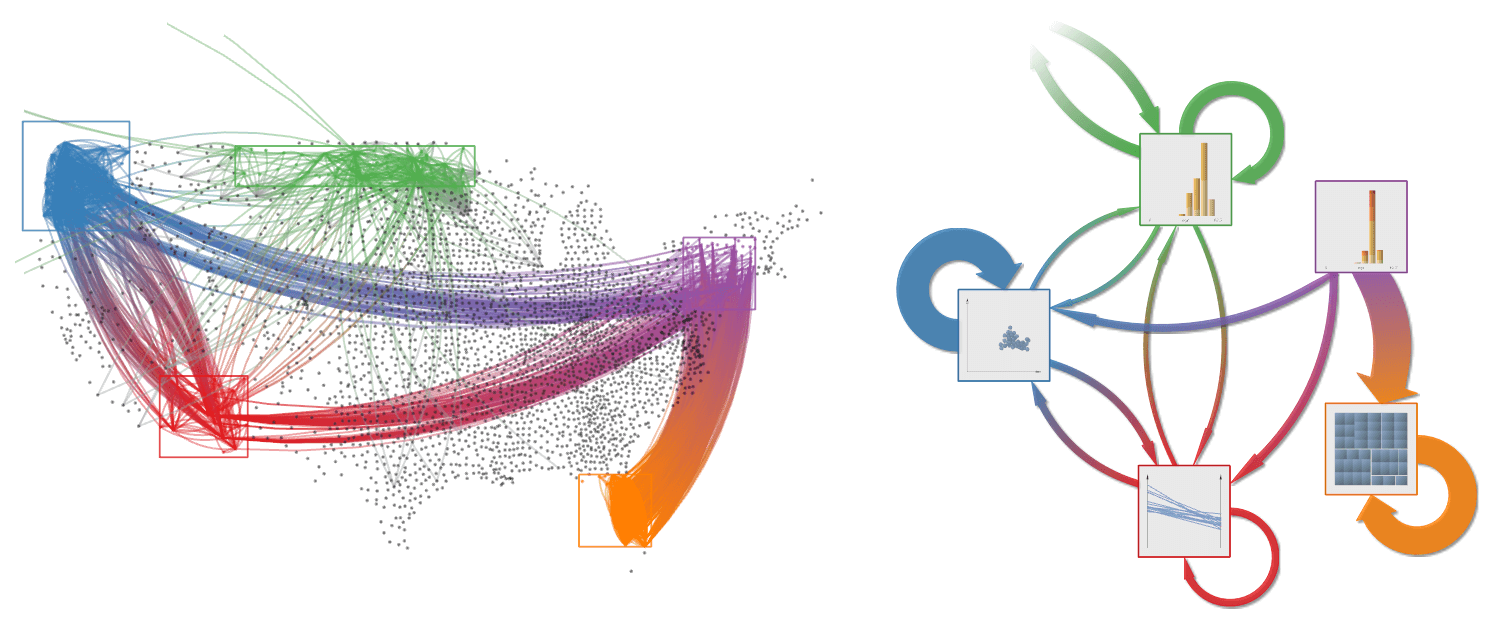}
  \label{fig:elzen2014-left}
}
\subfigure[]{
\includegraphics[width=0.95\columnwidth]{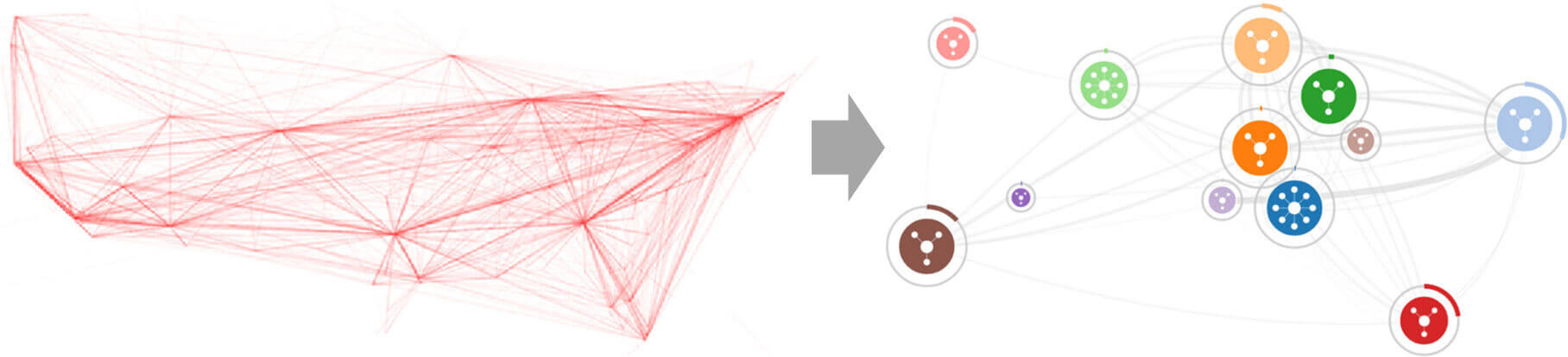}  
  \label{fig:li2017}
}
\caption{Examples for \fnode--Aggregated: 
(a) The user can interactively select regions to aggregate nodes~\cite{elzen_multivariate_2014},
(b) Module-based visualization: Nodes are automatically aggregated into clusters~\cite{li_module-based_2017}
(Reprinted by permission from Springer Nature: Springer. Journal of Visualization 20, 205–215. ``Module-based visualization of large-scale graph network data'', Li, C., Baciu, G., and Wang, Y. \copyright\ 2017).
}
\label{fig:node_aggregation}
\end{figure}

Aggregated node representations group individual nodes into metanodes and show only these metanodes explicitly.
Aggregation can happen through grouping nodes in predetermined areas~\cite{guo_flow_2009}, or more flexible partitions either through user interaction, e.g., specifying geographic regions \cite{elzen_multivariate_2014} (Figure~\ref{fig:elzen2014-left}), or by algorithmically identifying dense clusters and grouping nodes by these clusters \cite{li_module-based_2017} (Figure~\ref{fig:li2017}). The \textit{OD map}~\cite{wood_visualisation_2010} (Figure~\ref{fig:wood2010-new}) uses a regular spatial grid or tiling, overlaid onto a 2D map and groups nodes within each grid cell. The volume of flows (link weight) between each pair of cells is then further encoded inside the cell using a nested map and color.

Aggregating nodes reduces the number of visual elements displayed in the visualization. This can automatically reduce the number of links as they can be aggregated into metalinks, i.e., links between metanodes. In reducing detail, aggregation can highlight higher-level patterns and allows for showing summary statistics and data for each metanode (Figure~\ref{fig:elzen2014-left}-right).

\vspace{-0.25em}
\subsection{D2.1---\fnode: Abstract}
\iconfig{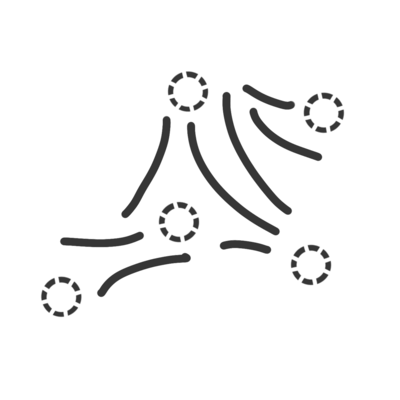}

While aggregated representations still visualize network topology to some degree, \textit{abstract} node representations do not show any visual marks that are identifiable as individual nodes or metanodes but rather communicate approximate locations and areas with fuzzy boundaries where individual nodes are situated.

\begin{figure}[b]
    \centering
    \includegraphics[height=4cm]{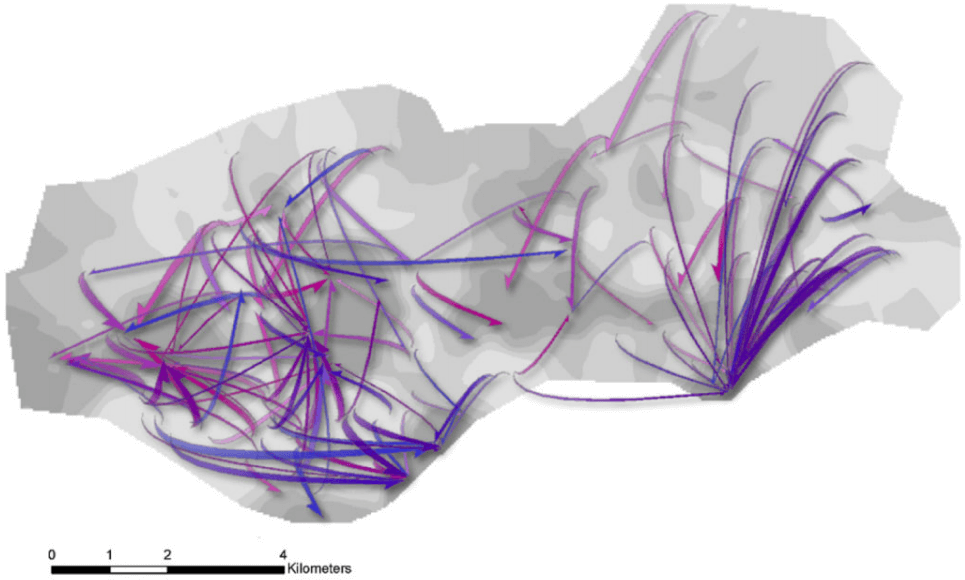}
\caption{Example for \fnode--Abstract: 
No nodes are shown in this technique by Zhu and Guo, the displayed `links' have been obtained by clustering together links of a very dense network~\cite{guo_origin-destination_2014}.}
\label{fig:node_abstract}
\end{figure}

We found five examples of abstract node representations (\perc{5}), all of which were created for flow data. 
Guo and Zhu use a flow-based density estimation method to extract patterns from the flow data~\cite{guo_origin-destination_2014} and then show these patterns as flows on a map. 
The same authors propose a second, similar method, where flows are clustered based on similarity, and each cluster displayed as a single representative flow~\cite{zhu_mapping_2014} (Figure~\ref{fig:node_abstract}).
Using these abstract techniques, the geographic representation can show location-specific information (attributes in color) such as bandwidth or unemployment, instead of overloading the visualization with explicit or aggregated nodes. 
This information can be visually correlated with the flows while providing detailed geographic information.
Similarly, Kim~et~al.~\cite{kim_data_2018} propose a combination of heatmaps, `field line' maps (Figure~\ref{fig:kim2018}), and arrows superimposed on a map to visualize flows, whereas Yao~et~al.~\cite{yao_visualizing_2019} use hexagonal \textit{pattern maps} to indicate directions (Figure \ref{fig:yao2019}). 

As these techniques show, abstracting nodes to fuzzy areas or glyphs can reveal overall patterns of how different areas are connected, especially for very large and dense networks. However, since any abstraction selectively hides some details of the data, the quality of the final visualization heavily relies on choosing methods appropriate to the data and task.

In summary, the \fnode\ dimension, similar to \fgeo, represents a continuous spectrum that provides a range of possible design choices: showing nodes explicitly and with full detail, aggregating nodes into clusters and other meaningful groups, and eventually fully abstracting the visual node representation, focusing only on connectivity. The tasks supported by this range depend on the required detail about nodes; the less individual nodes are important, the more nodes can be abstracted and the visualization be designed for tasks focusing on, e.g., density, geographic landmarks, etc.


\section{D2.2: Link Representation (\flink)}
\label{sec:d-link}

Link representation is the second sub-dimension of the network representation dimension (\ftopo). Similarly to \fnode, \flink\ ranges from \textit{explicit} to \textit{abstract}, and we classify techniques into one of the three classes \textit{explicit}, \textit{aggregated}, and \textit{abstract}. 

\vspace{2em}
\subsection{D2.2---\flink: Explicit}
\iconfig{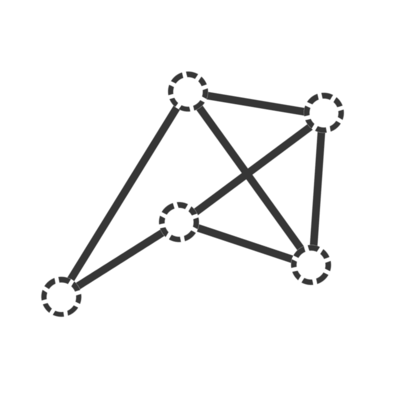}

The largest group of techniques in this class are techniques based on node-link diagrams, which are largely differentiated by the visual representation of their links. A flow map is essentially a special case of node-link diagram, in which all links are weighted and directed, represented by arrows of varying widths. The simplest form of flow map uses straight lines or arrows to connect nodes, superimposed on a regular map. Particularly early computer-generated visualizations use this approach \cite[e.g.,][]{tobler_experiments_1987,becker_visualizing_1995}. However, straight lines create cluttered maps even with comparably small data. As a consequence, a multitude of techniques have been developed to reduce clutter in node-link diagrams and flow maps, of which many are based on \textit{aggregation} and as such discussed in the following section.

\begin{figure}[t]
\centering
\subfigure[]{ 
  \includegraphics[height=2.6cm]{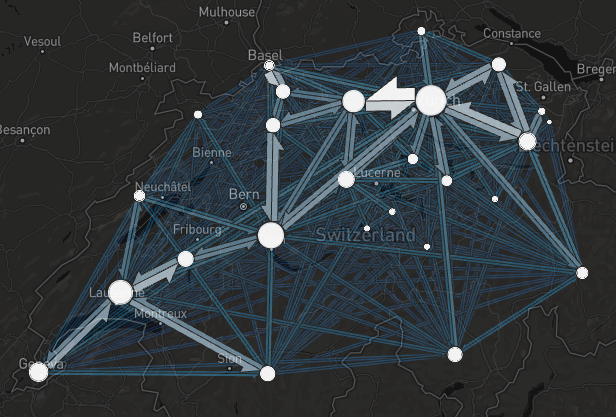}  
  \label{fig:riche2012}
}
\hfill
\subfigure[]{ 
  \includegraphics[height=2.6cm]{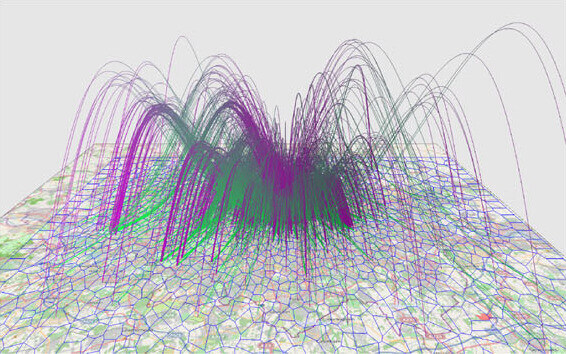}  
  \label{fig:vrotsou2017}
}
\subfigure[]{ 
  \includegraphics[height=5.5cm]{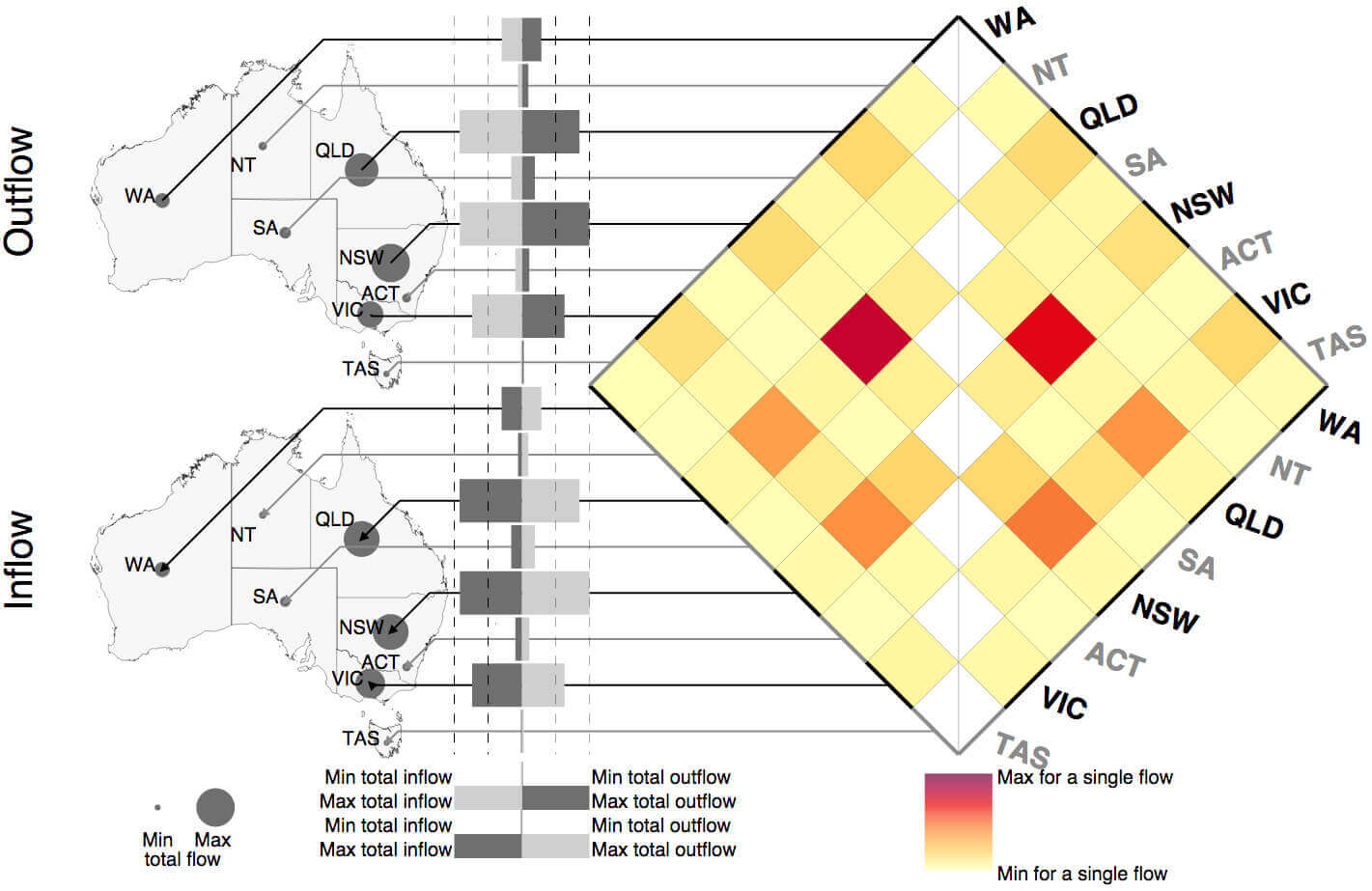}  
  \label{fig:yang2017}
}
\caption{Examples for \flink--Explicit:
(a) An interactive flowmap created with \href{https://flowmap.blue}{flowmap.blue}~\cite{flowmap_blue}, 
(b) A directed node-link flow map with 3D arcs~\cite{vrotsou_interactive_2017},
(c) `MapTrix', a matrix connected to two maps~\cite{yang_many--many_2017}.
}
\label{fig:edge_explicit}
\end{figure}

In smaller, less dense networks, clutter is less of an issue. Yet, links may overlap, or multiple links may connect the same nodes. To address this, several techniques have been developed to ensure links are clearly visible as separate lines. This can be achieved by bending them, for example using Bézier curves \cite{brandes_using_1998}, which can be further spread out using angle constraints where the links connect to the nodes \cite{brandes_improving_2000}. 
In three-dimensional visualizations, lifted arcs may reduce overlap compared to a two-dimensional display, particularly when the user can interactively navigate the visualization~\cite{yang_origin-destination_2019, vrotsou_interactive_2017} (Figures~\ref{fig:yang2019} and \ref{fig:vrotsou2017}).
Kaya~et~al.~\cite{kaya_multi-resolution_2016} introduce a technique to spread out links around a globe.

In addition to reducing overlap, another challenge in node-link diagrams is the visualization of link attributes, which can be as simple as direction or weight, but also take more complex forms. When using arcs, the arc height can be used to encode data attributes such as link weight or distance~\cite{vrotsou_interactive_2017} (Figure~\ref{fig:vrotsou2017}). For networks with additional link attributes more complex than weights or directions, different modifications to the links have been proposed, such as animated link textures \cite{romat_animated_2018}, where shape, size, color, and direction of particles moving along the links can be used to encode attributes. A static variant are patterned links \cite{cornel_composite_2016}. 

In a flow map, each link would typically be represented by a line of uniform thickness, determined by its weight. An alternative to this are tapered links. In addition, there are different shapes and sizes of arrowheads that can be used \cite{kim_developing_2012}.

Dynamic geospatial networks are networks that change over time. Changes can be limited to certain attributes, e.g., link weights, but can also affect the entire network, including new or deleted nodes or links. Dynamic networks can be visualized through animating node-link diagrams or displaying them as small multiples~\cite{boyandin_qualitative_2012}. \textit{Flowstrates} is an alternative for data that can be represented as a directed, bipartite graph. The flows are displayed using two maps, one for origins and one for destinations. The two maps are linked through a heatmap showing the flows between the linked locations over time. 
Figure~\ref{fig:flowstrates} shows refugee flows as an example, which change over time, and often go from one region of the world to another~\cite{boyandin_flowstrates:_2011}. Displaying the network in a space-time cube is another method~\cite{kapler_geotime_2005}.

Node-link diagrams have been adapted to visualize uncertainty in the underlying graph data. Schulz~et~al.~\cite{schulz_probabilistic_2017} introduce a technique that first decomposes the uncertain graph into its possible instances, then creates a visualization from these. Von~Landesberger~et~al.~\cite{landesberger_typology_2017} make further suggestions for visualizing uncertainty in geospatial network data.

A variety of interaction and navigation techniques have been developed for node-link diagrams. These are discussed in Section \ref{sec:interactiontech}.

A different form of explicit link representation is used by matrix-based techniques, where each node is represented by a row and column in the matrix, representing outgoing and incoming flows to other nodes. 
Representing the network as a matrix requires the geospatial aspect to be displayed separately. 
This can be done on a juxtaposed map where color-coding relates the locations to the matrix~\cite{guo_visual_2007}. 
For dense, directed, and weighted geospatial networks (often called \textit{many-to-many flow data}), 
Yang~et~al.~\cite{yang_many--many_2017} introduce \textit{MapTrix}, a technique that shows a matrix with a 45{\degree} rotation, rows and columns connected to their associated location on two juxtaposed maps (Figure~\ref{fig:yang2017}). 
As such, \textit{MapTrix} employs the same basic concept as \textit{Flowstrates}: inserting a graph representation in between two maps showing origins and destinations.

\vspace{-0.25em}
\subsection{D2.2---\flink: Aggregated}
\iconfig{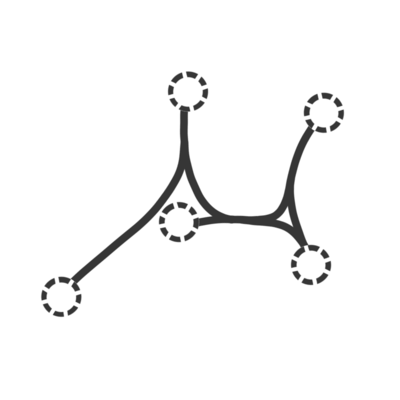}

\begin{figure}[t]
\centering
\subfigure[]{
\includegraphics[height=3cm]{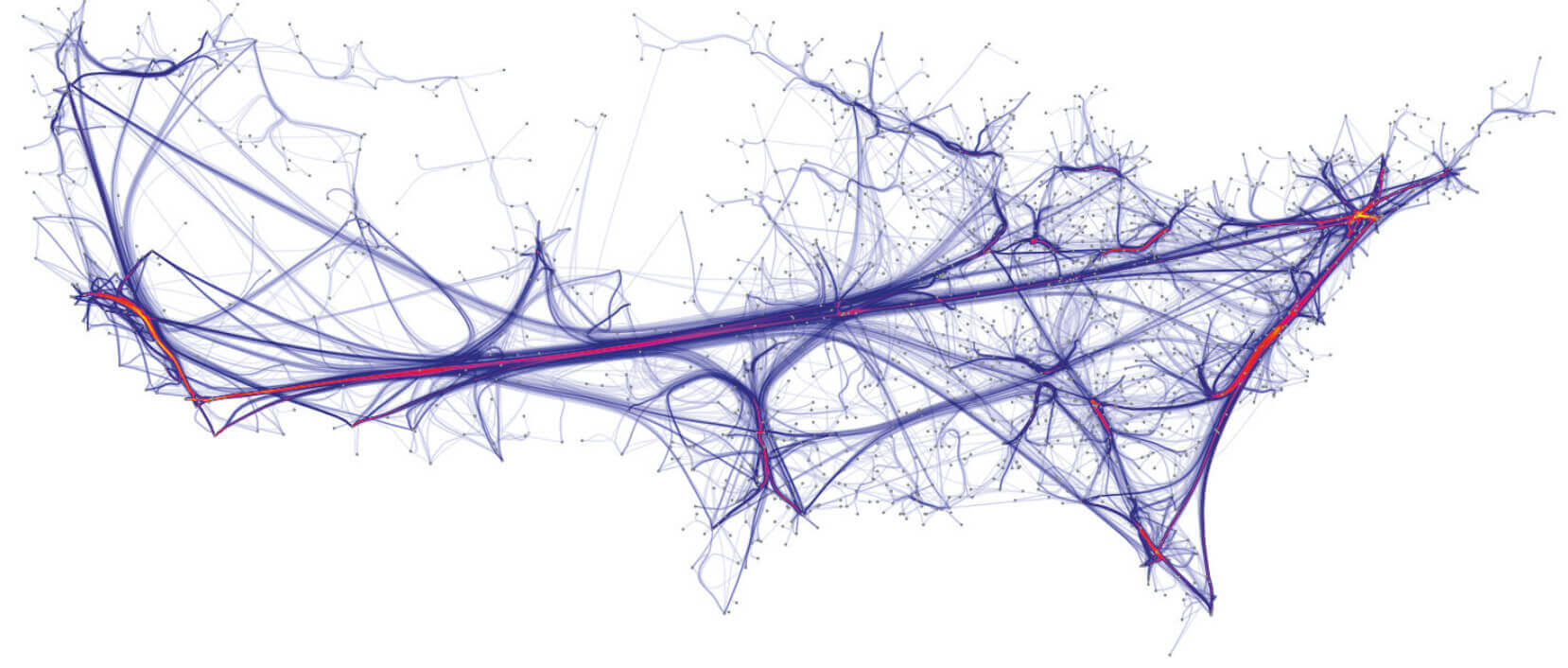}
  \label{fig:holten2009}
}
\subfigure[]{
  \includegraphics[height=3cm]{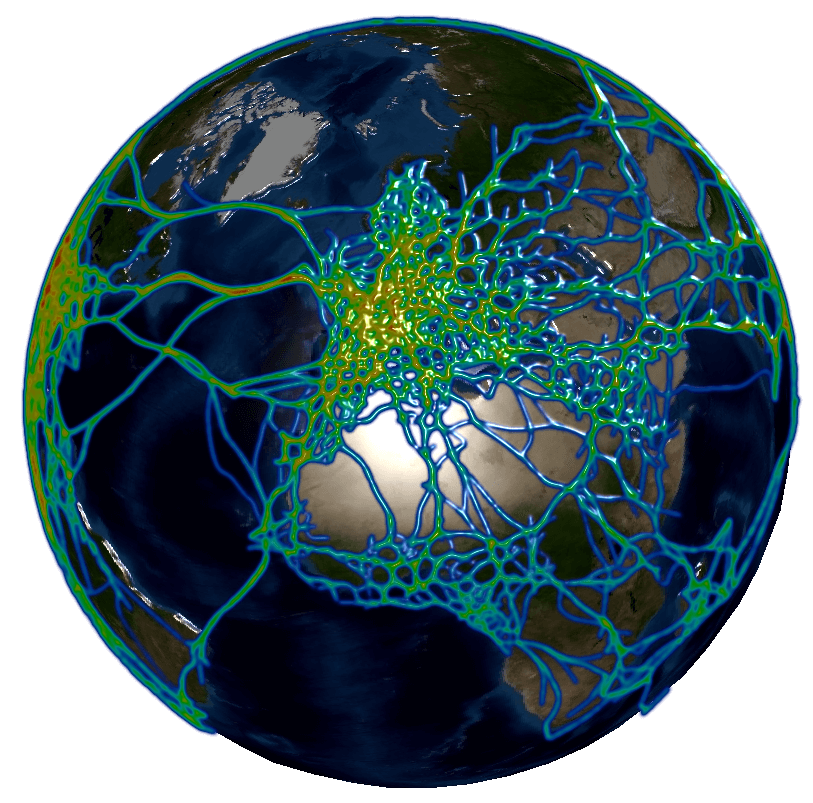}  
  \label{fig:lambert2010}
}
\hfill
\subfigure[]{
  \includegraphics[height=3cm]{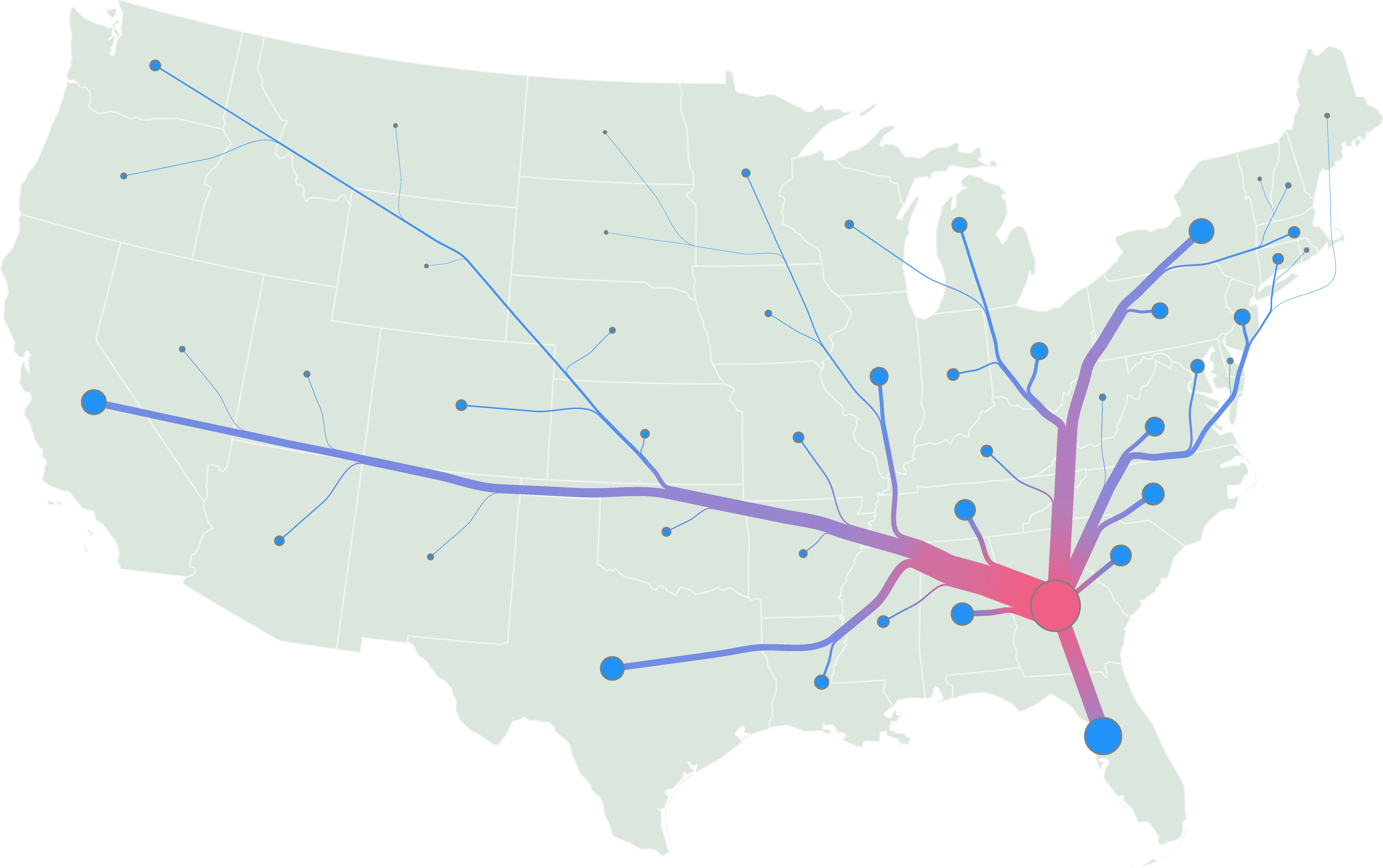}  
  \label{fig:sun2019-1}
}
\caption{Examples for \flink--Aggregated:
(a) Force-directed edge bundling~\cite{holten_force-directed_2009}, 
(b) Edge bundling and splatting on a 3D globe~\cite{lambert_winding_2010},
(c) A one-to-many flow map~\cite{sun_spatial_2019}.
}
\label{fig:edge_aggregated}
\end{figure}

Links are considered \textit{aggregated} if they cannot be individually identified anymore. Aggregated links can be a good solution to overlap and clutter on node-link diagrams, and as a result many (\perc{27}) of our surveyed techniques fall into this category.

When \textit{nodes} are aggregated, this nearly always leads to aggregated or abstract link representations because most visual representations remove, hide, or combine links between aggregated nodes. Examples are Van den Elzen and Van Wijk's technique for multivariate network exploration~\cite{elzen_multivariate_2014} or Li~et~al.'s \textit{module-based visualization}~\cite{li_module-based_2017}, both shown in Figure~\ref{fig:node_aggregation}. However, the majority of aggregated link techniques we found use explicit node representations, aggregating the links only. 
Note that nodes can be aggregated while links are explicit: in such a case, individual links would be shown between clusters of nodes. However, we did not find such a technique applied to geospatial networks.

Edge bundling, for which numerous algorithms have been proposed, is a technique that bends links to form bundles~\cite[e.g.,][]{holten_force-directed_2009, lhuillier_ffteb:_2017, peysakhovich_attribute-driven_2015, peng_sideknot:_2012, gansner_multilevel_2011, cui_geometry-based_2008} (Figure~\ref{fig:holten2009}). The level of aggregation depends on the algorithm and parameters that are chosen --- if the bundling is very `loose', with individual links clearly visible, the representation may in fact still be considered \textit{explicit}. However, most edge bundling techniques generate clearly aggregated results, for example the 3D edge bundling algorithm by Lambert~et~al.~\cite{lambert_winding_2010} (Figure~\ref{fig:lambert2010}). A more complete overview of edge bundling techniques is provided by Zhou~et~al.~\cite{zhou_edge_2013} and Lhuillier~et~al.~\cite{lhuillier_state_2017}. 
Edge routing is an alternative approach that defines fixed routes that links must follow, effectively also bundling edges~\cite{bouts_clustered_2015}.

\textit{Flow maps}, i.e., node-link diagrams with directed, weighted links, suffer from clutter problems even for relatively small data sets. Link weights are represented by line thickness, and arrow heads are added to each link to show its direction, resulting in thicker links easily covering up nodes and other links.
This can be somewhat mitigated by changing the styling of the links and drawing thinner links on top of thicker ones~\cite{jenny_design_2018},
but nonetheless, straight-line layouts are severely limited. To mitigate this, different types of bundled flow map layout algorithms have been proposed: a layout based on combining edge routing and spline interpolation~\cite{doantam_phan_flow_2005}, a layout based on spiral trees~\cite{buchin_flow_2011} (Figure~\ref{fig:buchin2011}), stub bundling \cite{nocaj_stub_2013}, and a force-directed layout~\cite{jenny_force-directed_2017}.
To create a layout similar to that of the original flow maps by Charles Minard \cite[reproduced in][]{rendgen_minard_2018}, Sun proposes a layout that routes links through oceans where possible~\cite{sun_spatial_2019} (Figure~\ref{fig:sun2019-1}).
Additionally, several `hybrid' techniques between edge bundling and bundled flow maps exist. These techniques take account of the link direction when bundling spatially proximal links, for example, divided edge bundling which bundles flows in different directions separately~\cite{selassie_divided_2011} or Graser~et~al.'s edge bundling-based flow map technique~\cite{graser_untangling_2019}.
There are also flow map techniques developed for three-dimensional globes \cite{debiasi_force_2014}.

\vspace{-0.25em}
\subsection{D2.2---\flink: Abstract} 
\iconfig{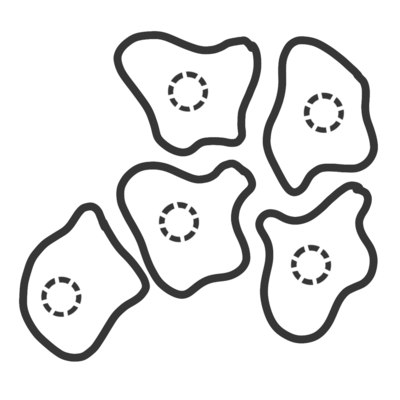}

Abstract link representations are those that do not display each link as an explicit visual element. In general, we observe that abstract link representations are often used to visualize a specific aspect of the network data rather than provide an overview of the network topology in general.

\begin{figure}[t]
\centering
\subfigure[]{
  \includegraphics[height=3.5cm]{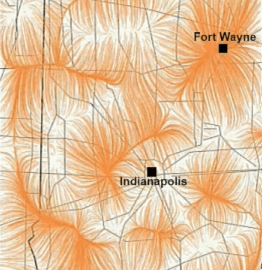}  
  \label{fig:kim2018}
}
\hfill
\subfigure[]{
\includegraphics[height=3.5cm]{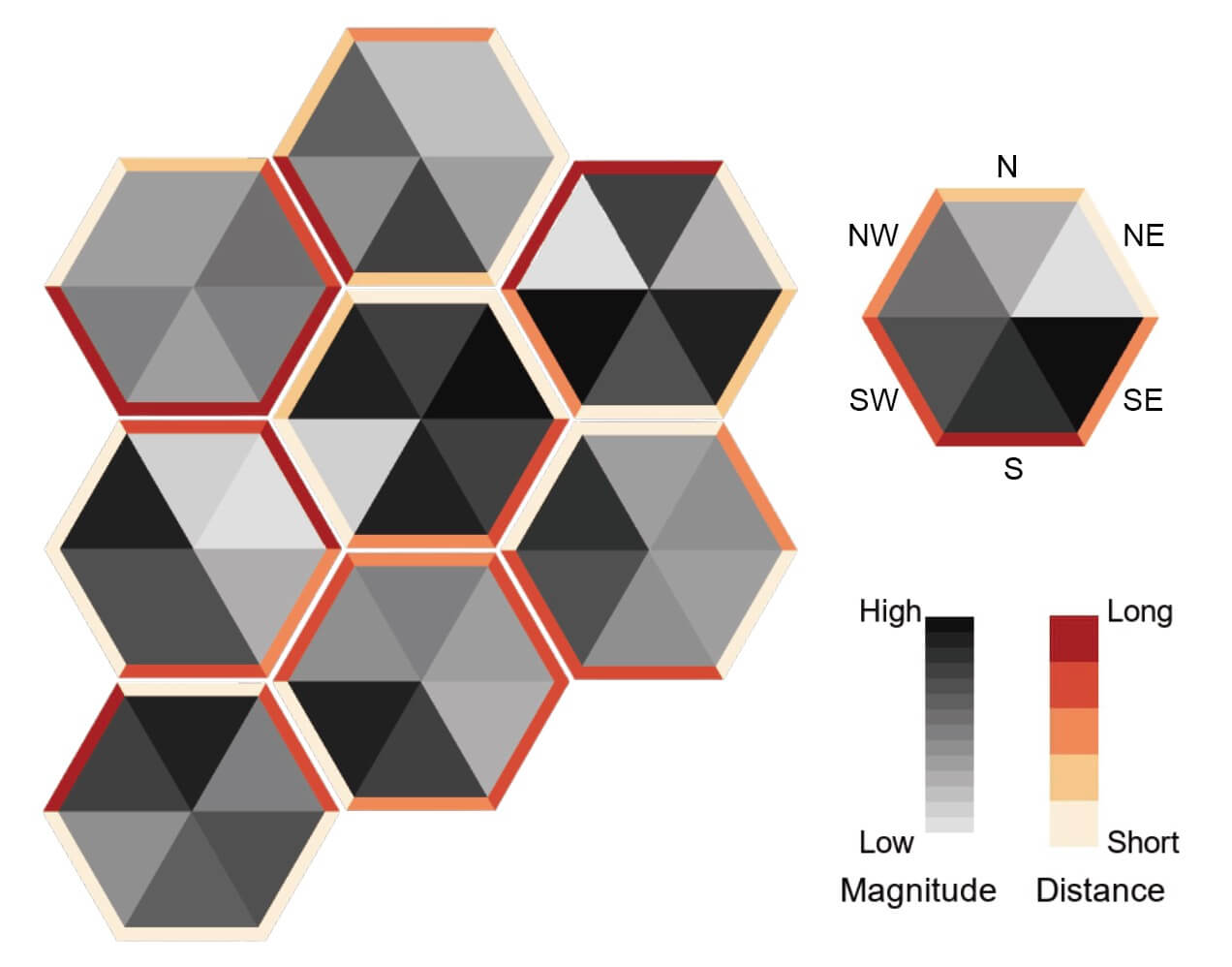} 
  \label{fig:yao2019}
}
\caption{Examples for \flink--Abstract:
(a) Showing patterns using `field lines'~\cite{kim_data_2018},
(b) Detail and legend of a `pattern map'~\cite{yao_visualizing_2019}.} 

\label{fig:edge_abstract}
\end{figure}

Several techniques focus on visualizing the directional aspect of origin-destination network data. To visualize the direction of flows, Andrienko~et~al.~\cite{andrienko_revealing_2017} propose a technique where node locations are marked by glyphs, which show the strength of flows in different directions over time. \textit{Pattern maps}~\cite{yao_visualizing_2019} visualize flows on a hexagonal grid which is superimposed on the geography. Each hexagon indicates the magnitude of the flow on its inside (grey), and the distance travelled on its boundary (red/orange) (Figure~\ref{fig:yao2019}). Kim~et~al.~\cite{kim_data_2018} propose a technique that generates `field lines' to show flow strength and direction (Figure~\ref{fig:kim2018}).

Other techniques focus on showing the `connectedness' or similarity of different nodes. 
The distorted globe technique by Alper~et~al.~\cite{alper_dynamic_2007} shows the node locations on a globe, distorting the globe such that nodes are closer together the more closely linked they are, without showing links as arcs or similar (Figure~\ref{fig:alper2007}). 
A similar technique on a flat map has been introduced by Bouts~et~al.~\cite{bouts_visual_2016} (Figure~\ref{fig:bouts2016}).

Finally, several techniques exist that extract high level patterns from the network data. Andrienko~et~al.~\cite{andrienko_spatial_2011} developed a technique that at first glance, seems to show a regular flow map, yet not only makes use of nodes aggregated into areas, but also of partitioned trajectories such that flows are only drawn between neighboring areas. The previously discussed techniques by Guo and Zhu~\cite{guo_origin-destination_2014, zhu_mapping_2014} follow this pattern as well  (Figure~\ref{fig:node_abstract}). 

In summary, the \flink\ dimension describes the spectrum of information shown for links. Explicit link representations support tasks about individual links (e.g., path following, link type, link direction), while aggregating links supports general connectivity tasks (e.g., high-level network structure, connectivity of regions). Abstracting links can help showing direction and strength of links without occluding any geographic features. This is helpful where geographic features or nodes are important,  where nodes are very closely placed on a map so that space for visualizing links is scarce, or where a specific network metric (e.g., density per region) is important and can best be shown by showing meta-information about links and thus choosing a more abstract representation rather than displaying individual links.


\section{D3: Composition (\fcomp)}
\label{sec:d-composition}

This dimension describes how geographic (\fgeo) and network topology information (\ftopo) are composed into a single visualization to provide for tasks that potentially involve information from both: \textit{Which region has the most nodes? Which regions are most connected to each other? What is the location of a particular node?}


For combining general views in visualization, Javed and Elmqvist~\cite{javed_exploring_2012} propose a taxonomy consisting of \textit{superimposition}, \textit{overloading}, \textit{juxtaposition}, \textit{integration}, and \textit{nesting}. In their survey on multi-faceted graphs, Hadlak~et~al.~\cite{hadlak_survey_2015} use similar categories: \textit{juxtaposition}, \textit{superimposition}, and \textit{nesting}.

Based on our collection of techniques and inspired by both Javed and Elmqvist as well as Hadlak~et~al. we define four types in which geographical and network information are combined in geospatial network visualizations, ranging from a loose integration (\textit{juxtaposed}) via \textit{superimposed} and \textit{nested} to a strong integration (\textit{integrated}).

\vspace{-0.25em}
\subsection{D3---\fcomp: Juxtaposed}
\iconfig{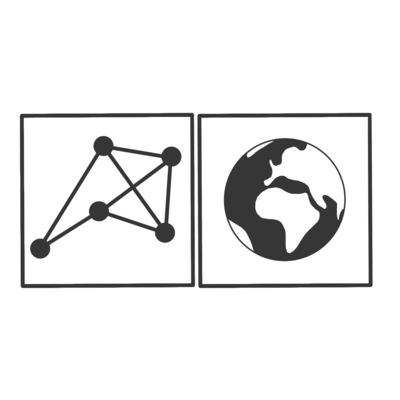}

In a juxtaposition, network representation and geography representation are placed side-by-side on the screen as commonly done in Coordinated Multiple Views (CMV) environments. In a CMV system~\cite{roberts_state_2007,wang_baldonado_guidelines_2000}, views are meant to show complementary information and each view can act as selection tool to filter or select information for other views. Brushing and linking are common interactions for these systems. Typical examples for juxtaposed visualizations for geographic networks include the \textit{OD Wheel}~\cite{lu_od-wheel:_2015} and Ibarra~et~al.'s juxtaposition of a map and an arc diagram~\cite{ibarra_visualization_2016}. Techniques that use multiple representations of either geography or network, e.g., a 3D globe and a 2D map~\cite{cox_3d_1996}) or multiple perspectives on network topology~\cite{guo_flow_2009} do not count as juxtaposed in our survey.

\begin{figure}
\centering
\includegraphics[width=0.9\columnwidth]{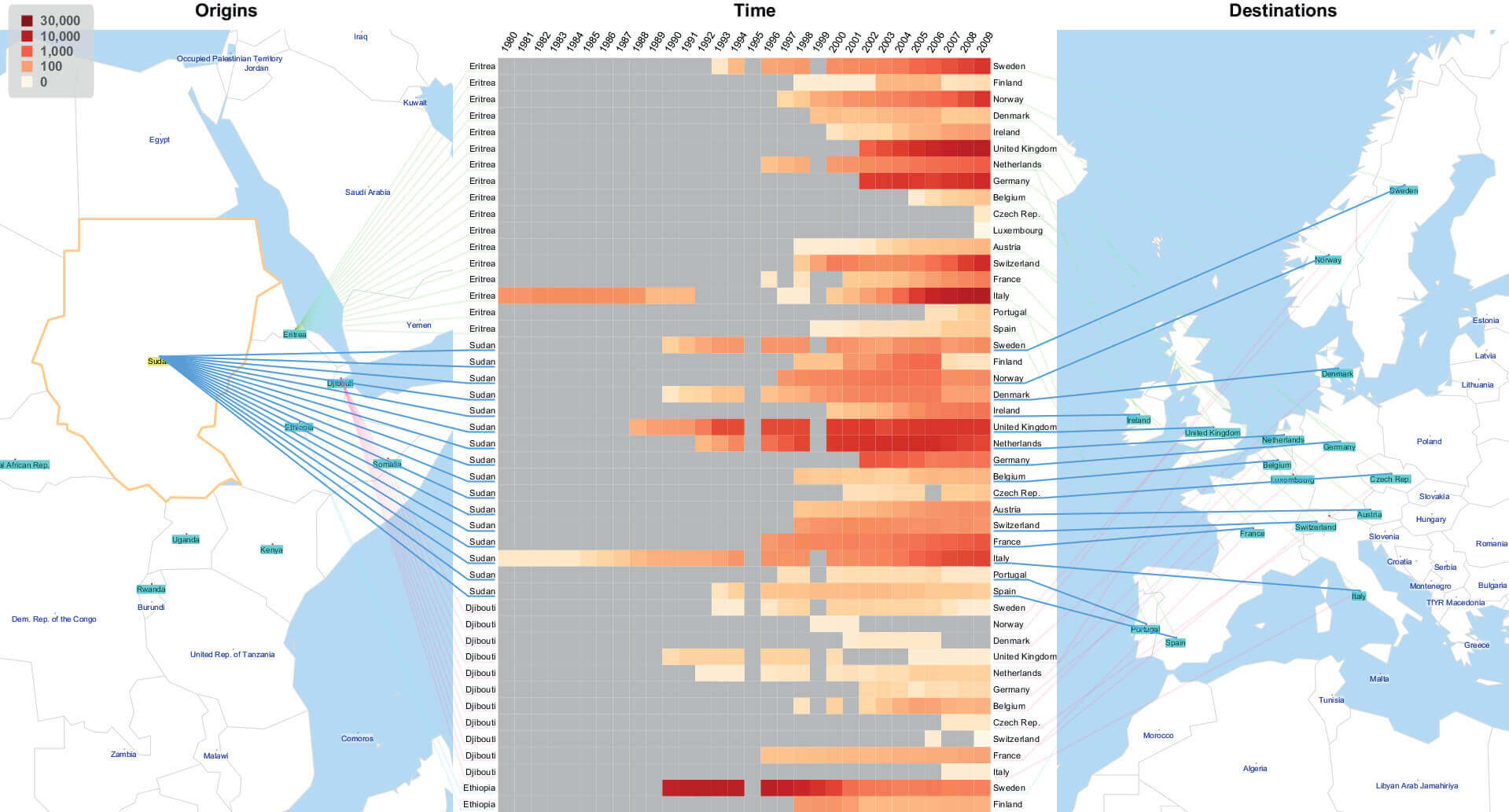}
\caption{Example for \fcomp--Juxtaposed:
`Flowstrates'~\cite{boyandin_flowstrates:_2011}.}\vspace{-0.5em}
\label{fig:flowstrates}
\end{figure}

To better relate elements across different views of network topology and geography, some techniques draw leader lines between locations and nodes, e.g., between regions on a map and rows and columns (the network's nodes) of an adjacency matrix~\cite{yang_many--many_2017} (Figure~\ref{fig:yang2017}). Another example of juxtaposition is Boyandin~et~al.'s \textit{Flowstrates}~\cite{boyandin_flowstrates:_2011} (Figure~\ref{fig:flowstrates}) which combines two geographic maps with a third visualization in between the two maps, showing changes in link weight over time. Each of the timelines in the central visualization is linked to the respective regions in both maps to provide for a network representation.

Juxtaposition strongly reduces visual clutter and allows for combining precise and mapped geographic representations with any type of network representation (node-link, matrix, etc.), at the same time allowing for explicit representation of nodes and links. However, as pointed out by Javed and Elmqvist~\cite{javed_exploring_2012}, users need to build a mental mapping between the two representations. This can require extra mental effort, especially when multiple objects are targeted. Elements need to be connected across visualizations using visual elements such as lines or interaction such as brushing and linking. Also, the visual space is divided, so each representation has less visual space for presentation.

\vspace{-0.25em}
\subsection{D3---\fcomp: Superimposed}
\label{sec:comp:superimposition}
\iconfig{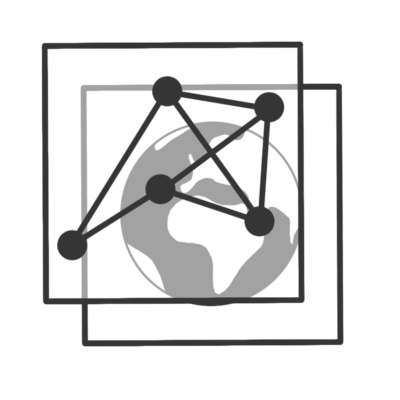}

\begin{figure}
\centering
\subfigure[]{
  \includegraphics[height=3.6cm]{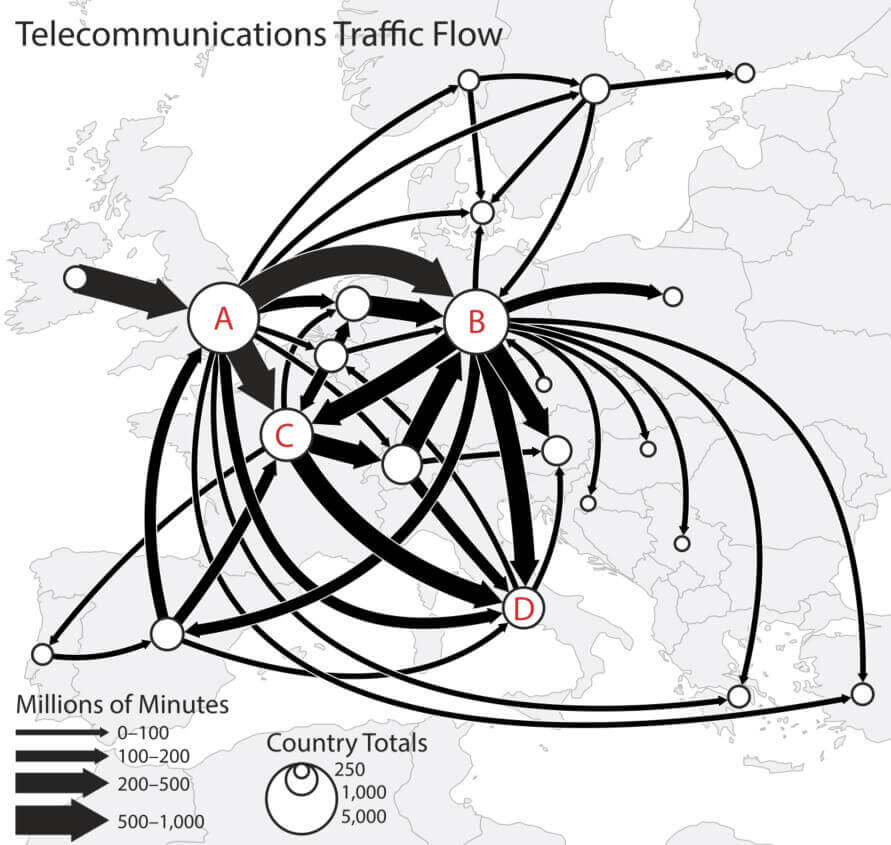}  
  \label{fig:jenny2018}
}
\subfigure[]{
  \includegraphics[height=3.6cm]{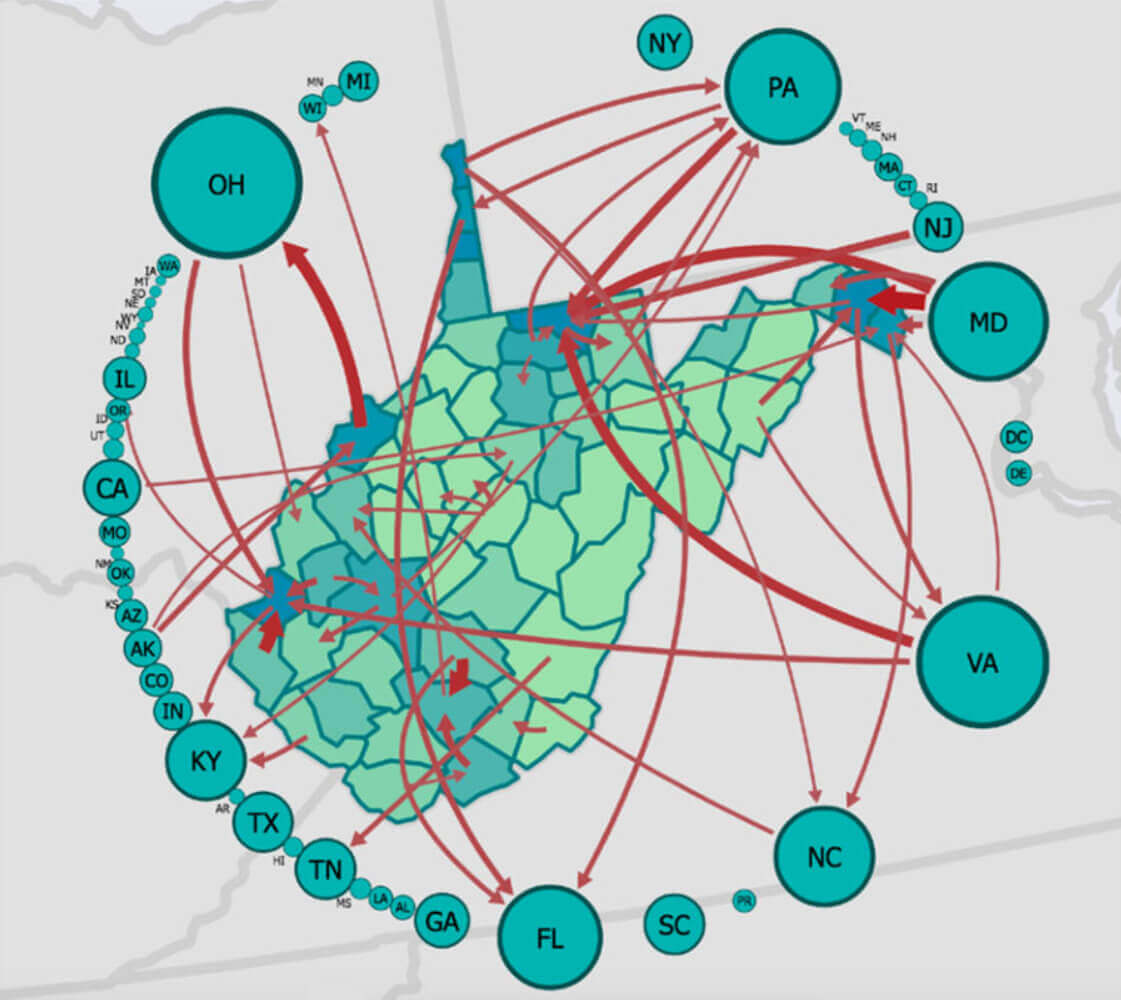}  
  \label{fig:stephen2017}
}
\caption{Examples for \fcomp--Superimposed:
(a) A flow map using curved arrows~\cite{jenny_design_2018} (reprinted by permission of the publisher, Taylor \& Francis Ltd, \href{http://www.tandfonline.com}{www.tandfonline.com}),
(b) Nodes laid out around a map~\cite{stephen_automated_2017}.}\vspace{-0.5em}
\label{fig:comp:superimposed}
\end{figure}

In superimposed compositions, the network and geography representation are visually overlaid, in most cases the network on top of the geography. Superimposition can happen in that geography is seen as a reference for placing nodes at their precise geographic positions (Figure~\ref{fig:jenny2018}) or as approximation for node positions (Figure~\ref{fig:stephen2017}). It also does not matter whether geography is mapped or distorted, or abstract. As discussed in Section \ref{sec:geo-abstract}, in abstract geographies, geography can still be shown along a single spatial dimension while links are overlaid (Figure~\ref{fig:xiao2009}).
Still, the most commonly found versions of superimposition (\perc{53}) overlay explicit nodes on mapped geographies
~\cite{doantam_phan_flow_2005,buchin_flow_2011,jenny_design_2018,sun_spatial_2019,yang_origin-destination_2019} (Figure~\ref{fig:buchin2011}, \ref{fig:sun2019-1} \& \ref{fig:jenny2018}). Some techniques also superimpose abstract node representations on a mapped geography representation~\cite{andrienko_revealing_2017,yao_visualizing_2019} (Figure~\ref{fig:yao2019}).

An interesting variation of this approach is \textit{In Situ Exploration}~\cite{hadlak_situ_2011}, where the user can draw rectangular `portals' onto the map and select a type of visualization, e.g., a matrix. The `portal' inset then displays a visualization of the data of the region which it covers on the underlying map.

Superimposition can often be easier to interpret and creates less cognitive load for the viewer. However, superimposition has visual elements of both geography and network in the same visual space, which may easily produce visual clutter and might make it difficult to visualize additional attributes.

\vspace{-0.75em}
\subsection{D3---\fcomp: Nested}
\iconfig{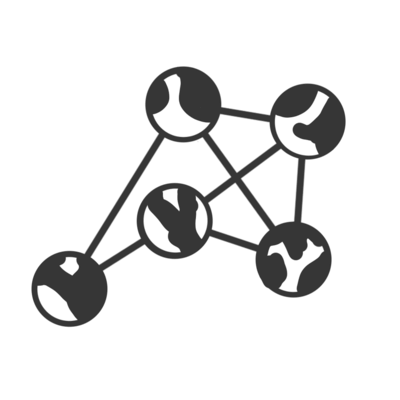}

In a nested representation, one of network or geography representation is \textit{nested} inside the visual elements of the respective other representation (geography or network).
One example of a nested representation is the \textit{OD Map} by Wood~et~al.~\cite{wood_visualisation_2010}. An \textit{OD Map} divides a geographic map into cells in a regular grid. Then, inside each cell, i.e., a confined geographic space, a small representation of the entire geographic space is nested (Figure \ref{fig:wood2010-new}). Geographic regions on each of the small maps can now be used to encode information about the links between the region represented by the cell and each region on the miniature map. 
Hadlak~et~al. and Schulz~et~al. propose techniques that nest point-based tree representations of hierarchical data into geographic regions on a map~\cite{hadlak_visualization_2010, schulz_point-based_2011}.

\begin{figure}[t]
    \centering
    \subfigure[]{
        \includegraphics[height=2.5cm]{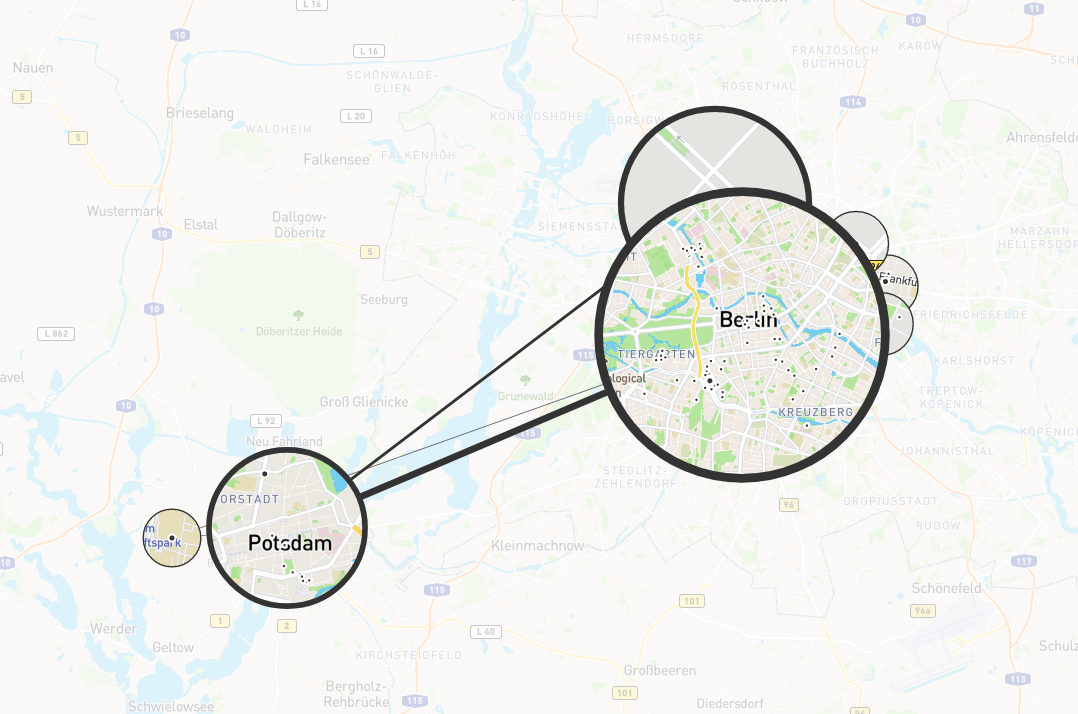}
    }
    \hfill
    \subfigure[]{
        \includegraphics[height=2.5cm]{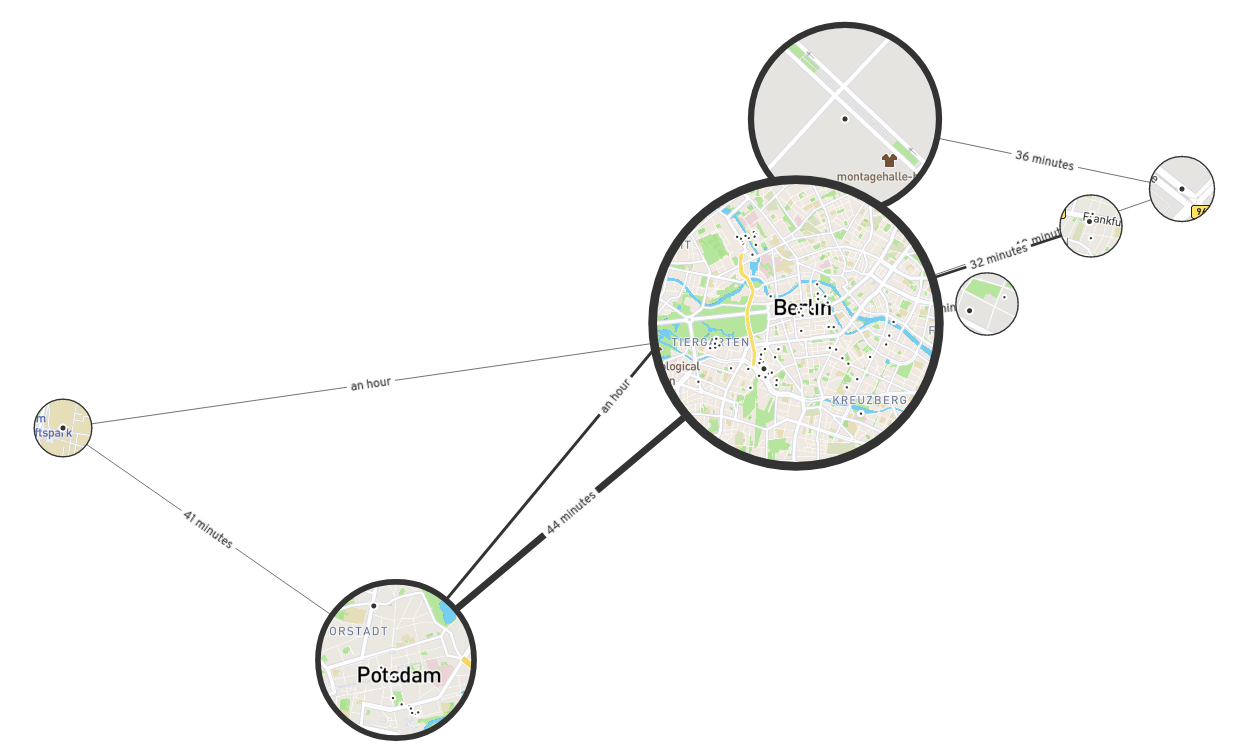}
    }
    \caption{Example for \fcomp--Nested:
    Shifted maps~\cite{otten_shifted_2018} nesting geography (mapped) into network nodes (explicit) and providing seamless transition between (a) mapped geography and (b) distorted geography showing travel time between locations (images via \href{https://shifted-maps.com/}{shifted-maps.com}). 
    }
    \label{fig:comp:nested}
\end{figure}

Another nested technique are \textit{shifted maps}, which show geographic maps inside the nodes of a node-link diagram~\cite{otten_shifted_2018} (Figure \ref{fig:comp:nested}). Nesting in \textit{shifted maps} allows this technique to provide for a seamless transition between a geographic placement for nodes, and a geographic distortion that renders the maps' visual (Euclidian) distances between the nodes on the screen depending on node connectivity using a force-directed layout. 

Nesting creates compact representations and shows specific information about specific regions or nodes. However, there is usually limited space for the nested representations as nesting requires space to render the nested views which can then occlude information in the background (in the case of geography). Alternatively, nested views can get quite small.
Moreover, extra efforts are required to understand the context, e.g., the distortion in \textit{OD Maps} or the different scales of geography in~\cite{brodkorb_overview_2016,otten_shifted_2018}, and to provide for a holistic view in the viewer's mind. 

\vspace{-0.25em}
\subsection{D3---\fcomp: Integrated}
\iconfig{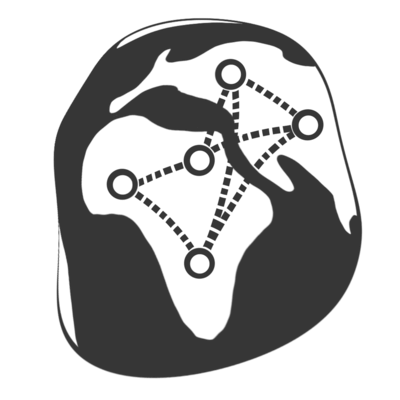}

A technique represents an integrated view of network topology and geography if it is not possible to separate network topology and geography into two clearly distinct visualizations. The visualizations of both geography and network topology depend on each other in terms of layout or other visual mapping decisions. In other words, an integrated technique cannot be shown in a juxtaposition or superimposition.

\begin{figure}[t]
\centering
\subfigure[]{
  \includegraphics[height=3.6cm]{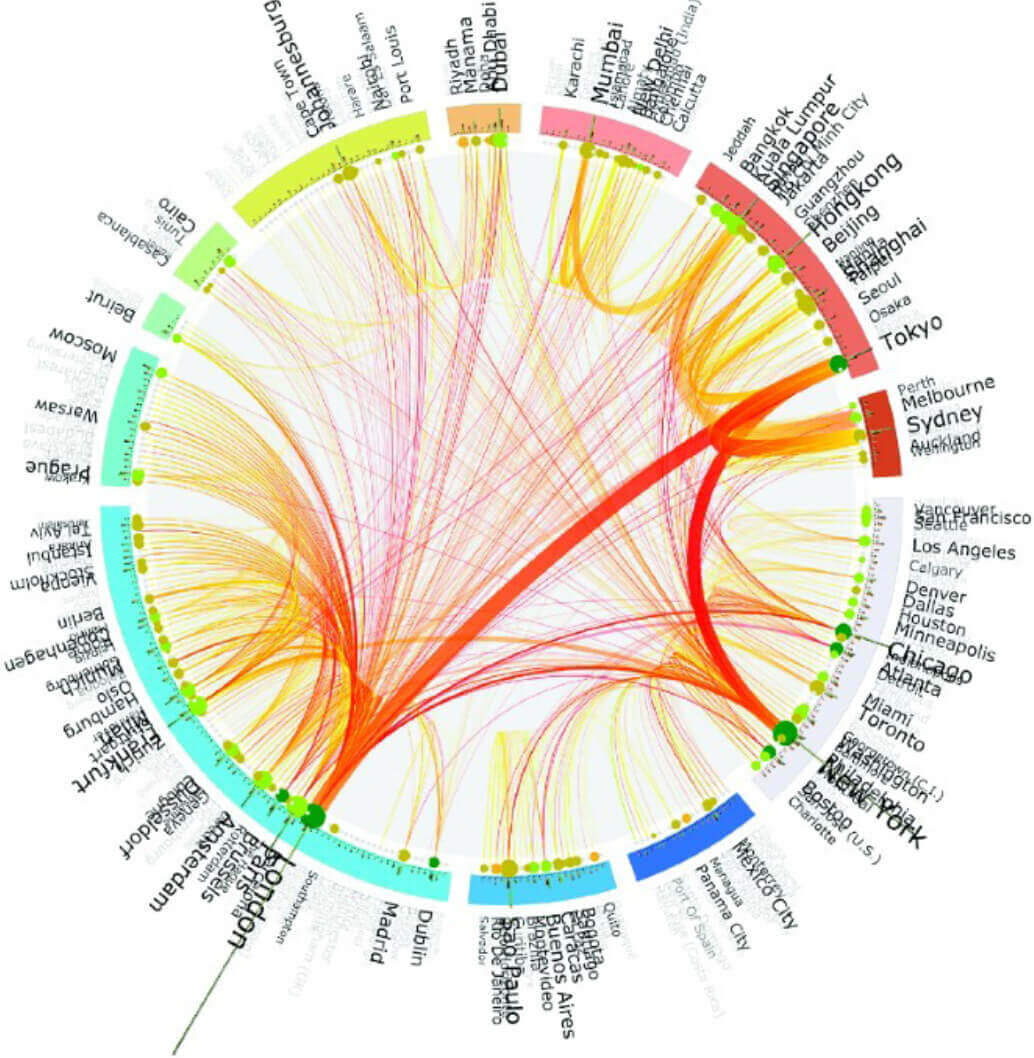}  
  \label{fig:hennemann2013}
}
\subfigure[]{
  \includegraphics[height=3.2cm]{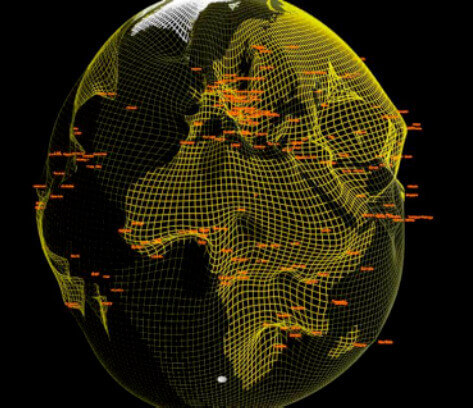}  
  \label{fig:alper2007}
}
\caption{Examples for \fcomp--Integrated: 
(a) Nodes are arranged in a circle for a clear network representation, but grouped geographically~\cite{hennemann_information-rich_2013},
(b) distortions of the globe are used to visualize the network connectivity~\cite{alper_dynamic_2007}.}
\label{fig:comp:integrated}
\end{figure}

For example, the chord diagram in Figure~\ref{fig:hennemann2013}~\cite{hennemann_information-rich_2013} can explicitly show the network topology, while nodes are grouped, placed, and abstractly color-coded according to their geographic locations. Geographic information is \textit{integrated} into an explicit network visualization. Another example for integration are Xiao~et~al.'s \textit{Kriskograms}~\cite{xiao_visualizing_2009} that show geographic information as node labels. The ordering of the nodes is a one-dimensional projection (not a \textit{map} projection) of their geographic locations. Network information can also be represented by modifying a geographic representation. For example, Alper~et~al.~\cite{alper_dynamic_2007} use a distorted geographic representation to represent the network information (Figure~\ref{fig:alper2007}).

Integration reduces visual clutter even more than juxtaposition, as it does not need to split the display space between two components. However, in all integrated techniques we found, one of the components (either \fgeo\ or \ftopo) is visually abstracted. Thus, people may find it difficult to interpret the abstracted component.

In summary, the \fcomp\ dimension describes a spectrum of combinations of visual representations for \fgeo\ and \ftopo\ (from loose to strong integration). 
\textit{Juxtaposition} reduces visual clutter by placing geographic representation and network representation side-by-side. But extra mental effort is needed to link them visually. 
\textit{Superimposition} overlays one representation on top of another (usually \ftopo\ on top of \fgeo). The two representations are intuitively linked, but the overlapping of visual elements usually introduces visual clutter.
\textit{Nesting} embeds one representation inside the visual elements of the other one, which results in compact designs. \textit{Integration} fully utilizes the display space by combining the visual representations of \fgeo\ and \ftopo. However, the abstraction introduced requires more effort to interpret the visual representation.

\section{D4: Interactivity (\finteract)}
\label{sec:d-interactivity}

In addition to the construction of the visual representation through geography, network and their composition as described in the previous sections, interactivity can be essential for visual exploration.
This dimension classifies techniques related to understanding geospatial networks, i.e., not including generic techniques such as pan and zoom, or highlighting visual marks for details. We classified techniques and papers along a scale from least to most interactive, with interaction being either \textit{not required}, \textit{required}, or the technique being \textit{interaction only}.

We decided to include such pure interaction techniques into this survey because they provide useful cases about how to address specific problems in geospatial network visualization; as any interaction technique requires a visualization to function on, the respective papers (or combined interaction$+$visualization technique) can still be classified under all of our other dimensions. 
Papers that did not explicitly mention any interaction were classified as interaction being \textit{not required}.

\subsection{D4---\finteract: Not Required}
\iconfig{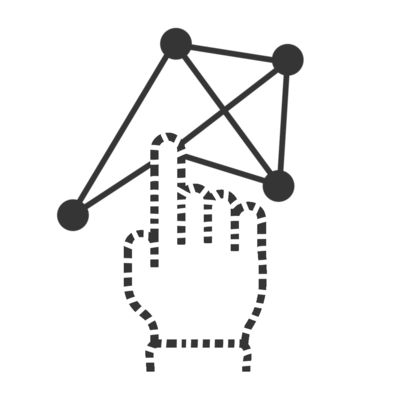}

Out of the surveyed techniques, \perc{62} do not require user interaction. In many cases, this means that the technique generates a static image, which can optionally be enhanced with basic interactions like pan and zoom, scaling (parts of) the visualization up or down, expanding and collapsing parts of the visualization, or filtering for a subset of the data. While any technique in this category is \textit{usable} without interaction, the extent to which it is \textit{useful} varies among the techniques and between different data sets. In general, techniques that do not require interaction have two major advantages: firstly, computational complexity is less of a concern if the image only needs to be rendered once. Secondly, non-interactive visualizations can be used in contexts where user interaction is not possible, such as in print materials or displayed on public screens.

Some spatial generalization techniques \cite[e.g.,][]{guo_flow_2009, andrienko_spatial_2011, cornel_composite_2016}, 
i.e., techniques that derive a less detailed visualization from a more detailed one,
can be used to create one-off generalizations for a display at a specific size and scale, but are also computationally efficient enough to be used as part of an interactively scalable map or other visualization that automatically updates the level of generalization based on the zoom level.

Edge bundling techniques also typically do not require interaction, though they can be enhanced with interactive features to explore bundles. The most common interactive tool for edge bundling techniques is \textit{relaxation}, which lets the user interpolate between the bundled and unbundled views~\cite{lhuillier_state_2017}. In addition to this generic technique, the creators of the \textit{SideKnot} edge bundling technique~\cite{peng_sideknot:_2012} explicitly discuss the possibility of integrating the technique with the \textit{EdgeLens} interaction technique~\cite{wong_edgelens:_2003}, which we include under \textit{Interaction Only}.

Not all non-interactive visualizations are necessarily static images; animated link textures as proposed by Romat~et~al.~\cite{romat_animated_2018} use animation to move particles along a network's links to indicate link weight (flow). Again, interaction is not required but could be added for playing and controlling the animations.

\subsection{D4---\finteract: Required}
\iconfig{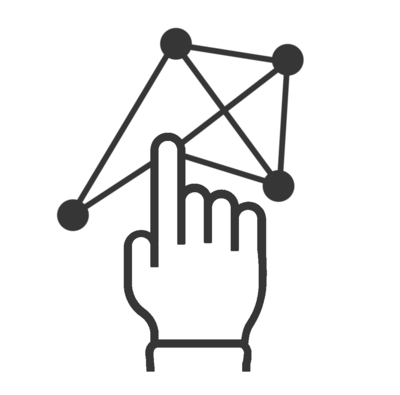}

Techniques are labelled as \textit{required} if the visualization requires user interaction to fully explore the data set because some parts of the data are hidden in the default view, e.g., to reduce visual complexity (\perc{23} of surveyed techniques).

For example, in three-dimensional globe representations~\cite{cox_3d_1996, alper_dynamic_2007, lambert_3d_2010} (Figure~\ref{fig:lambert2010}), or in the \textit{GeoTime} system~\cite{kapler_geotime_2005}, which uses a space-time cube, user interaction is required to navigate 3D space through rotation and view the complete visualization. 
Without interaction, the visualization hides important information (e.g., data on the other side of the globe). 
If the visualization requires configuration, it requires interaction.
Most interactive visualizations are displayed on a regular screen, limiting interaction modalities to using a mouse or touch screen. Virtual reality, on the other hand, can offer rich interaction methods~\cite{yang_origin-destination_2019} (Figure~\ref{fig:yang2019}).

Interaction is also often required in dealing with large data sets. Several techniques present the user with an abstract overview and let them selectively expand parts of the visualization. Li's \textit{module-based visualization} aggregates nodes into meta nodes, which can be expanded to view the underlying network structure~\cite{li_module-based_2017}. Hadlak~et~al. initially present an aggregated node-link diagram superimposed on a map, then let the user open up additional visualizations such as matrices to further explore subsets of the data~\cite{hadlak_situ_2011}. The opposite approach is to initially present the user with a detailed, possibly cluttered, visualization, from which they can select regions of interest that are then visualized in a more abstract, overview-type visualization~\cite{elzen_multivariate_2014} (Figure~\ref{fig:elzen2014-left}).

A second group of techniques rely on interaction as an exploration tool, e.g., through filtering the data~\cite{vrotsou_interactive_2017}, or selecting subsets of the data, which are then displayed in separate, superimposed or juxtaposed visualizations~\cite{hadlak_situ_2011, luo_spatial-social_2011}.
The \textit{Flowstrates} technique offers rich interaction capabilities, for example letting the user interactively select which origin and destination locations should be displayed in the central heatmap~\cite{boyandin_flowstrates:_2011}.
Some techniques create a visualization entirely based on user input, for example metro maps that are arranged with a focus on the user's travel route~\cite{wu_travel-route-centered_2012}.

A third application of interactivity is morphing between different network layouts, which can be useful in relating a network to its geographical context. An example is \textit{OD morphing}, which morphs a node-link diagram between an edge-bundled display and a geographically accurate one, where links are routed along their true route~\cite{lyu_od_2019}. Another example are \textit{shifted maps}~\cite{otten_shifted_2018} (Figure~\ref{fig:comp:nested}), where users can switch between different node-link layouts based on a mapped representation, and alternative layouts showing travel distance, travel time, and travel frequency.

\subsection{D4---\finteract: Interaction Only}
\label{sec:interactiontech}
\iconfig{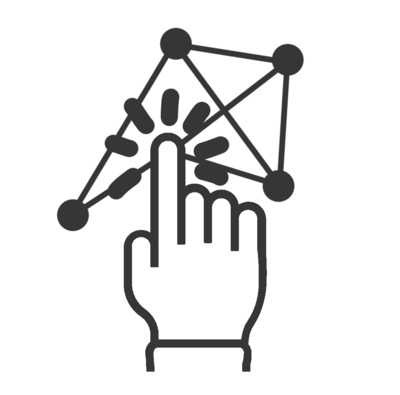}

\textit{Interaction Only} describes techniques that are not visualization techniques in their own right, but
pure interaction techniques. As such, they are intended to be applied on top of a visualization created with an existing technique.
The majority of \textit{Interaction Only} techniques are designed to be applied to \textit{mapped} node-link diagrams and mitigate issues node-link diagrams frequently suffer from, such as overlap, link clutter, and varying densities.

\begin{figure}[b]
\centering
\subfigure[]{
  \includegraphics[width=\linewidth]{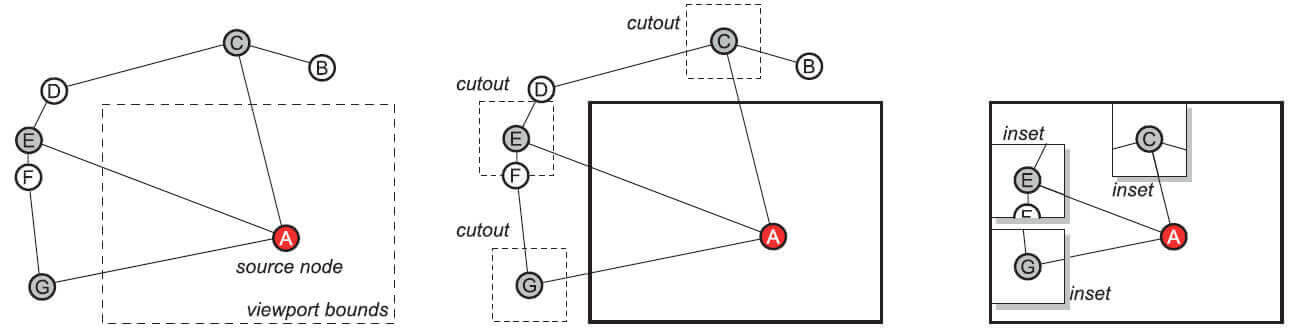}  
  \label{fig:ghani2011}
}
\subfigure[]{
  \includegraphics[width=0.3\linewidth]{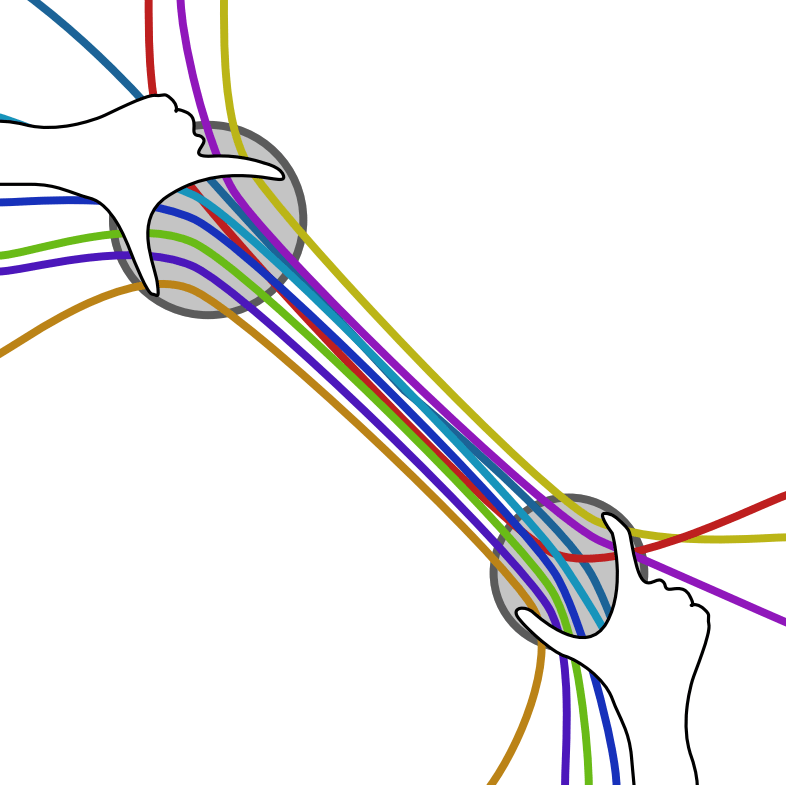}  
  \label{fig:riche2012-1}
}
\subfigure[]{
  \includegraphics[width=0.3\linewidth]{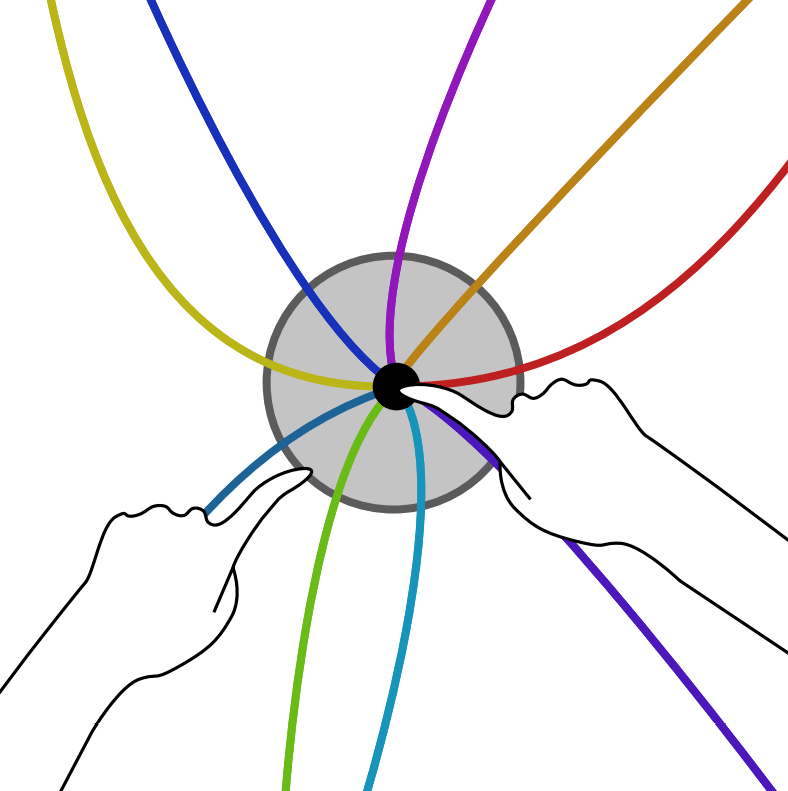}  
  \label{fig:riche2012-2}
}
\subfigure[]{
  \includegraphics[width=0.3\linewidth]{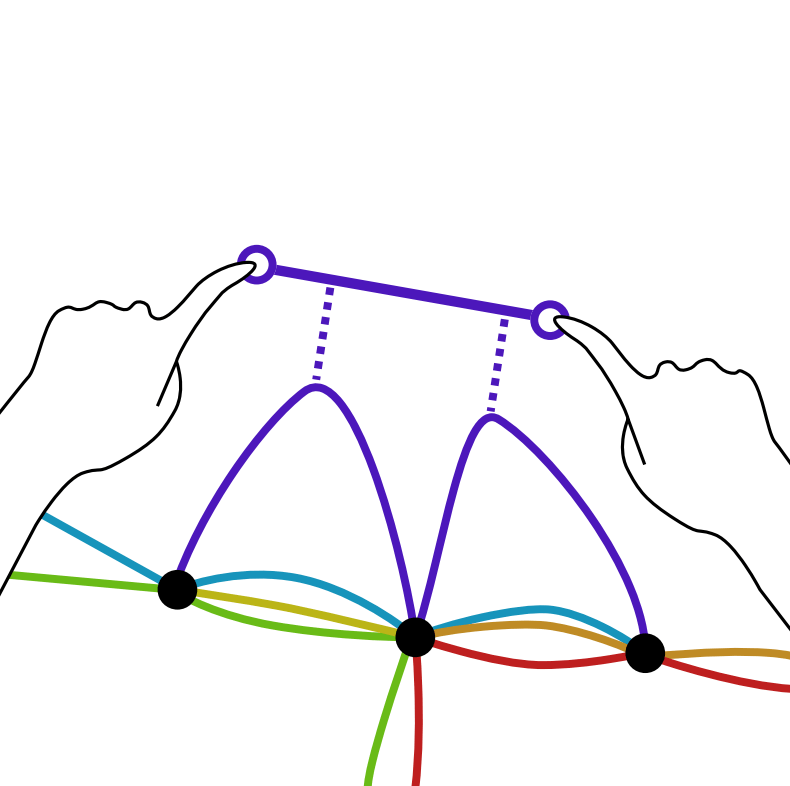}  
  \label{fig:riche2012-3}
}
\caption{Examples for \finteract--Interaction~Only:
(a) Map insets for navigation: the most relevant nodes are selected and displayed as insets~\cite{ghani_dynamic_2011},
(b-d) Illustrations of techniques by Riche et al.~\cite{riche_exploring_2012}: Interactive link bundling, fanning, and magnets.}
\label{fig:interaction}
\end{figure}

A number of different \textbf{lenses} have been proposed to help manage clutter in superimposed compositions. The \textit{Fisheye lens} distorts the entire visualization to allow for zooming into a region while still viewing the entire network~\cite{brown_browsing_1993}.
The \textit{EdgeLens}~\cite{wong_edgelens:_2003} and its 3D counterpart, the \textit{3DArcLens}~\cite{debiasi_3darclens:_2014}, push links away from the cursor by bending them. This makes it possible to see nodes and the underlying map more clearly. An application-specific lens is provided by the \textit{Focus+Context} metro map, which combines a detailed view of a small region of the metro map with an aggregated, simplified view of the larger network context~\cite{wang_focus+context_2011}.

There are techniques that give the user fine-grained control over moving and \textbf{displaying links}. \textit{Link plucking}~\cite{wong_supporting_2007} lets the user drag groups of links to the side to reveal what is underneath. Riche~et~al.~\cite{riche_exploring_2012} present three techniques: \textit{interactive bundling}, which lets the user select links to bundle together; \textit{link fanning}, which lets users select a node around which links are then spread out so the user can see individual connections; and \textit{link magnets}, to be placed by the user, which attract links in their vicinity (Figure~\ref{fig:interaction}(b-d)).

Lastly, there are techniques focused on \textbf{navigation}: \textit{Link sliding}~\cite{moscovich_topology-aware_2009} snaps the cursor to a link and while dragging the mouse, slides the field-of-view along the link until it reaches the other end of the link.
The same paper introduces another technique: \textit{Bring \& Go}, which, upon selecting a node in a network, brings all its direct neighbors into view, even if they are not located in the current zoomed-in view. The user can then navigate to any of the connected nodes by selecting it.
Ghani~et~al.~\cite{ghani_dynamic_2011} present a navigation technique based on small map insets (Figure~\ref{fig:ghani2011}). A \textit{degree-of-interest function} is used to determine which out-of-view nodes are most relevant; the selected nodes are then displayed in small map insets along the boundary of the map.

In summary, some visualization techniques propose solutions to reduce clutter in a static way, e.g., by bundling edges or distorting space to try to address a specific set of tasks. Such techniques are good for static visualizations (e.g., posters) or contexts where interaction is tedious or unlikely. However, they make decisions for the observer with regard to what information to show and in how much detail. Other techniques introduce interaction to allow for exploration of e.g., cluttered visualizations, or filter for relevant information. This aims to support a broader range of tasks than static visualizations. Eventually, a range of interaction techniques (\textit{Interaction Only}, Section \ref{sec:interactiontech}) can potentially be integrated into visualizations to provide for specific exploration tasks, e.g., \textit{Link Sliding} or the \textit{EdgeLens} can be applied to \textit{Kriskograms}, \textit{OntoTrix}, or flow maps on a 3D globe alike.

\vspace{-0.5em}
\section{Addressing Specific Challenges}
\label{sec:challenges}

While Sections \ref{sec:d-geography} to \ref{sec:d-interactivity} discussed techniques based on their design, i.e.,  how they visualize and abstract information (\fgeo, \ftopo), how these representations are composed (\fcomp), and to what extent a technique requires user interaction (\finteract), this section provides a complementary view on visualization techniques.

Based on earlier versions of our taxonomy (Section \ref{sec:methodology}, Versions \#1 and \#3), we discuss common challenges in visualizing specific types of geospatial networks and to what extent these challenges are addressed by existing techniques. For each of the techniques included in this survey, we assess whether it is suitable for thirteen \textbf{additional data attributes or characteristics}: 
\textit{directed links,
weighted links,
additional link attributes,
additional node attributes,
exact point locations,
area locations,
co-located nodes,
dense networks,
networks with varying density,
dynamic networks,
uncertain locations,
uncertain network topology,} and
\textit{uncertain additional attributes}.
As such, this section does not aim to provide a comprehensive list of challenges but to provide a practical resource for finding the most commonly encountered challenges and existing approaches to address these challenges when visualizing specific types of geospatial networks.

\vspace{-0.5em}
\subsection{Link and Node Attributes}

Geospatial networks can have a variety of attributes associated with nodes and links. We consider four specific additional data attributes here:  Most common are \textbf{directed links} and \textbf{weighted links}, which are usually visualized as node-link diagrams making use of arrows of different thickness (e.g., in flow maps). However, there can be node and link attributes beyond direction and weight.

\textbf{Directed links:}
Out of the additional data attributes, directed links have received the most attention, most often in the context of visualizing flows between locations. In node-link diagrams, common options include arrows, tapered links, and animated links, although there is no clear consensus on which design is preferable, especially in a geospatial context. 
Holten~et~al.~conducted a series of studies~\cite{holten_user_2009,holten_extended_2011} comparing different encodings for networks in general and recommend the use of tapered and animated compressed links.
Koylu~et~al.~\cite{koylu_design_2017} compared tapered links with arrows. They found that tapered links are only better for identifying long distance links.
Jenny~et~al.~\cite{jenny_design_2018} evaluated different designs based on tasks including reading flow magnitude and counting the degree of nodes. They found arrows outperforming tapered links significantly and provide three potential reasons: 1) the gradients of tapered links are inconsistent due to different link lengths, which is confusing; 2) links in flow maps are usually thin, resulting a very weak gradient; and 3) incoming flows are difficult to determine with tapered links.

An alternative to node-link diagrams are matrices. For example, \textit{MapTrix}~\cite{yang_origin-destination_2019} juxtaposes a matrix whose rows and columns are linked to two maps, one for incoming and one for outgoing flows. A matrix naturally visualizes direction since its rows and columns typically indicate origins and destinations, respectively.

\textbf{Weighted links:}
Another common attribute of links is \textit{weight}, i.e., a simple numerical value associated with each link (e.g., trade volume). Link weight in node-link diagrams is often visualized through width of the links. In geospatial networks, this can lead to obstructing important geographic and other information. Using color-gradient encoding (light to dark) compared to link width has been found more effective and efficient in a controlled user study~\cite{dong_using_2018} for small geospatial networks; participants commented that maps using color gradients are clearer than maps using line thicknesses.
As an alternative to static maps, Romat~et~al.~\cite{romat_animated_2018} use animated link textures to indicate link weight, while keeping all links to the same width. Their user study found that participants could discriminate up to six different values based on animated link textures, but that the use of particle patterns resulted a poor performance for a quick estimate yet was the best choice for accessing details. However, no comprehensive and comparative study exists so far. 

\textbf{Additional link attributes:}
In cases where links have additional attributes beyond simple weights and directions, colors are commonly used to encode categorical information. For example, Abel and Sander~\cite{abel_quantifying_2014} use color in links to represent regions of the countries in global migration. In transit maps, colors are usually used to represent different transit lines~\cite{wu_survey_2020}.

\textbf{Additional node attributes:}
In many cases, nodes are represented as identical shapes (e.g., circles). 
However, there are scenarios that require representing additional attributes on the nodes.
One common case is using the size of the node to encode a quantitative value. The quantitative value can be the net in/out flow into the geographic location the node represents, or a property associated with the geographic location but irrelevant to the network topology (e.g., the population of a country~\cite{speckmann_necklace_2010}). 
When it comes to the need of encoding more complex data on nodes, glyphs have been commonly used. 
For example, Yang~et~al.~\cite{yang_many--many_2017} place two semicircles on one location: one half represents net inflow and the other half represents the net outflow. 
Statistical charts can also be nested into nodes to visualize a variety of additional information~\cite{elzen_multivariate_2014}.
If node size is used to encode quantitative values on a mapped geography representation, small and close regions will pose a problem due to glyphs overlapping other glyphs, links, and potentially information on the underlying map.

\vspace{-0.5em}
\subsection{Geographic Locations}

Our definition of a \textit{geospatial network} includes networks associated with a variety of types of geolocations. We consider three different data type-related challenges. A rough division can be made into \textbf{point locations} (0D) and \textbf{area locations} (2D). For both of these location types, a multitude of techniques exist, with many techniques usable for both types. However, the distinction matters when addressing a common challenge related to locations: \textbf{co-located nodes}, i.e., multiple nodes at the same location (e.g., companies in the same building).

\begin{figure}
\includegraphics[height=3cm]{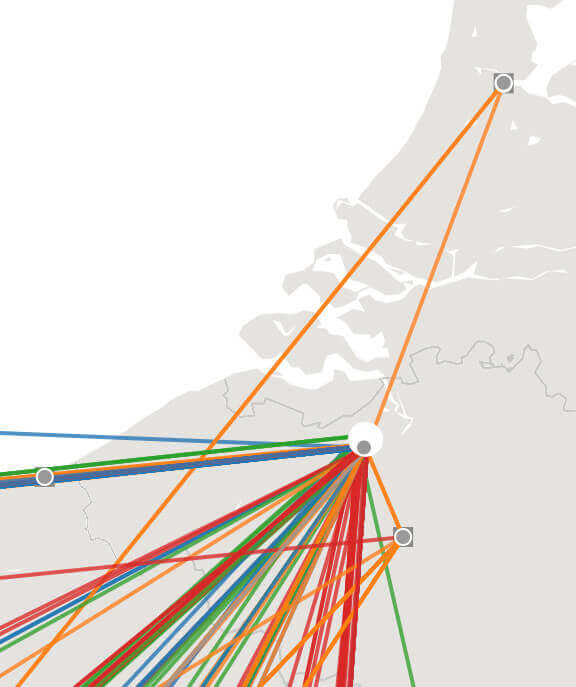}
\hfill{}
\includegraphics[height=3cm]{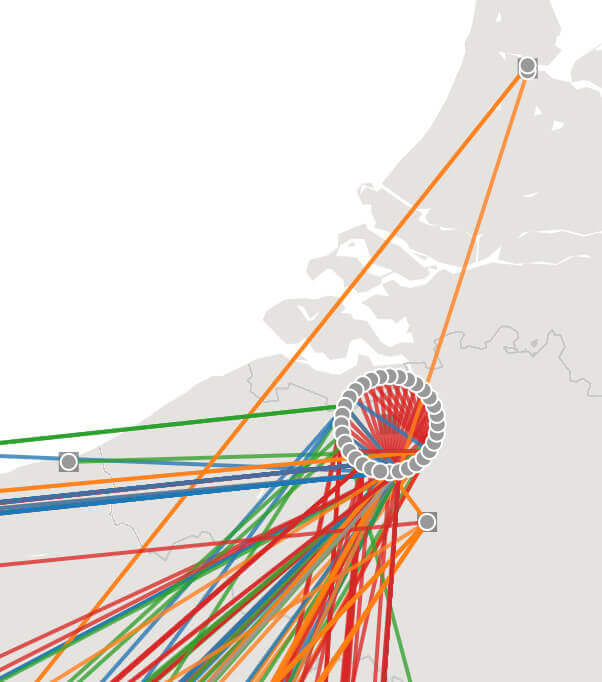}
\hfill{}
\includegraphics[height=3cm]{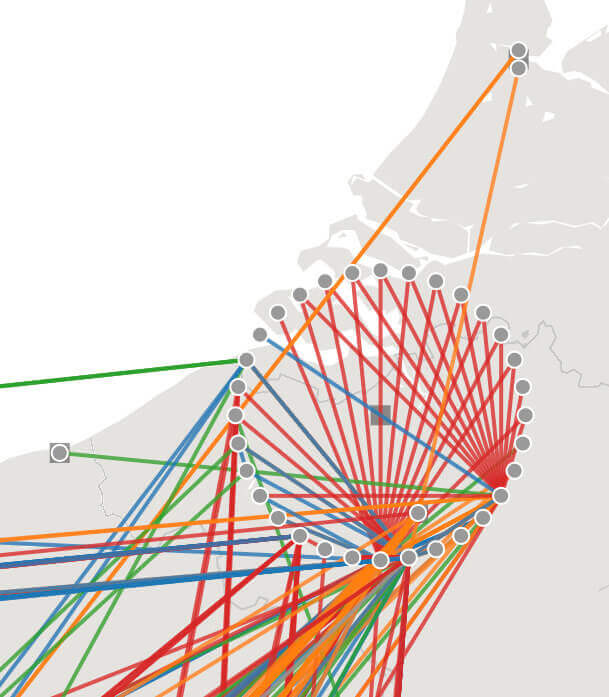}
    \caption{Node-overlap removal in the `Vistorian': Overlapping nodes are expanded circularly, using a simple range slider to set the expansion radius of the node circles. Links between nodes in the same position are shown inside the circle~\cite{bach_networkcube:_2015}.}
    \label{fig:vistorian}
\end{figure}

Co-located nodes are a challenge because placing multiple nodes in the same position leads to occlusion not only of the nodes, but also of potential links between them and surrounding geographical information. For near-identical positions, some \textit{distorted} geography representations (see Section~\ref{sec:geo-distorted}) can be used to mitigate this issue to an extent, though distortion cannot address \textit{exactly} identical positions. The \textit{Vistorian}~\cite{bach_networkcube:_2015} orders nodes from the same location on a circle around their position while the user interactively controls the circle radius (Figure \ref{fig:vistorian}).

When multiple nodes are located in the same area (e.g., a country), they could be juxtaposed inside this area, although we could not find a technique implementing this. Nodes might even be placed in an intelligent way that minimizes link crossings. 

An approach that can theoretically be used for both point and area locations that are co-located is to detach the network topology from the precise geography to an extent. This can be achieved by using \textit{abstract} or \textit{distorted} geography representations, e.g., showing locations by grouping nodes occupying the same space into segments along a circle~\cite{abel_quantifying_2014, hennemann_information-rich_2013}, a line~\cite{xiao_visualizing_2009} or a matrix~\cite{bach_ontotrix_2011} (\textit{abstract}) or compositions such as Necklace maps~\cite{speckmann_necklace_2010} (\textit{distorted}), which arrange nodes on a circle superimposed on a map, with node positions based on geolocations as shown on the map. Rather than abstracting the geography, an alternative approach can be the use of a \textit{juxtaposed} composition, which produces a looser integration of geography and network. An example is the \textit{MapTrix} technique~\cite{yang_many--many_2017}, which uses a matrix to display the network. Leader lines connecting the matrix to a juxtaposed map are used to link geography and network; these can easily link multiple matrix rows or columns to the same location.

The inverse challenge to co-located nodes would be nodes with multiple positions (e.g., if an organization has multiple offices in different locations). However, we are not aware of any techniques addressing this specifically for geospatial networks, or discussion of this challenge elsewhere. 
In non-geographic networks, this has been addressed by node-cloning~\cite{abello_ask-graphview_2006}, but this technique has not been adapted for geospatial networks and comes with its own set of challenges (e.g., showing which nodes are the same, deciding how to draw or duplicate links).

An additional potential challenge we see are node locations at different granularities, i.e., a network where some nodes are associated with geolocations at the level of countries, others with cities or precise street addresses. Similarly to the previous challenge, this is not a commonly discussed challenge, and we are not aware of any techniques addressing this.

\vspace{-0.5em}
\subsection{Link Density}

In the context of link density, we consider two main data characteristics: Firstly, \textbf{dense networks} have too many links that cause visual clutter when visualized in node-link and arc diagrams. Superimposing a node-link diagram on a \textit{mapped} or \textit{distorted} geography representation will clutter the display and can make geography as well as the network unreadable. A different, but related challenge is posed by \textbf{networks with varying density}, where some geographic locations feature significantly more nodes and links than others.

A common solution for the visualization of dense non-geospatial networks are \textbf{matrices}, which scale to complete networks, i.e., networks where each node is connected to all other nodes (density$=$1.0). 
Matrices have been extensively studied in visualization, starting with Ghoniem~et~al.~\cite{ghoniem_comparison_2004} who found that matrices scale well to larger network data. Matrices have subsequently been found more efficient for weighted and undirected networks~\cite{alper_weighted_2013}. In a crowdsourcing study, Okoe~et~al.~\cite{okoe_revisited_2018,okoe_node-link_2019} found that node-link diagrams worked better for topology and memorability tasks, while matrices were better for cluster-related tasks. 
However, it is challenging to integrate geographic information into a matrix. One solution is a \textit{juxtaposed} composition in which leader lines connect rows and columns of a matrix to their associated geographic locations, shown on two maps~\cite{yang_many--many_2017} or grouped by location~\cite{bach_ontotrix_2011}.
In a user study by Yang~et~al.~\cite{yang_many--many_2017}, these two designs had similar performance and scaled better than bundled node-link diagrams superimposed on maps. An alternative solution is the \textit{nested} composition of Wood~et~al.'s \textit{OD Maps}~\cite{wood_visualisation_2010}, in which the entire map is duplicated and nested into itself to resemble a map and thus represent the geographic context. 

For applications where superimposed node-link diagrams are required, different methods exist to make these visualizations more readable and navigable even at high link densities.
\textbf{Edge bundling}~\cite[e.g.,][]{holten_force-directed_2009, lhuillier_ffteb:_2017} or \textbf{edge routing}~\cite{bouts_clustered_2015} can free visual space by visually grouping links, though this results in a loss of accuracy when following individual links~\cite{bach_towards_2017}. Less problematic are techniques that keep links separate, such as manually routing links through `magnets' ~\cite{riche_exploring_2012} or automatically routing links around each other~\cite{brandes_improving_2000, brandes_using_1998, jenny_force-directed_2017}. Two comprehensive studies found that curved links performed better than straight ones when superimposed on maps~\cite{jenny_design_2018,dong_using_2018}. One of them, Jenny~et~al.~\cite{jenny_design_2018} conclude that well-designed curved links can \textit{i)} minimize the overlaps between links, \textit{ii)} avoid them intersecting at acute angles, and that \textit{iii)} curved links should be curved as little as possible. 

\textbf{Aggregating nodes} is another option as it automatically leads to link aggregation as well. Examples are Li~et~al.'s \textit{module-based visualization}~\cite{li_module-based_2017}, in which dense clusters of the network are visually represented by aggregate symbols, as well as the use of hierarchical clustering to aggregate nodes by Zhu and Guo~\cite{zhu_mapping_2014}.

Moreover, rather than aggregating nodes or links individually, the \textbf{entire network can be abstracted} such that higher-level features of the topology are shown.
Andrienko~et~al.~\cite{andrienko_spatial_2011} introduce an algorithm to generalize and aggregate directed geospatial networks such that a simpler version of the network is produced, which can then be visualized. A slightly different approach is Guo~and~Zhu's \textit{flow data smoothing}~\cite{guo_origin-destination_2014}, which displays link-like arrows, but they are not connected to fixed nodes. \textit{Pattern maps}~\cite{yao_visualizing_2019} indicate directions of flows in space through a hexagonal pattern; another technique by Kim~et~al.~\cite{kim_data_2018} uses a field line-like visualization for the same purpose.

While most of the above-mentioned techniques for reducing link clutter work globally on the entire network, other techniques have a \textit{local scope} in that they resolve clutter for \textit{some} nodes and links: dynamic edge lenses~\cite{zou_dynamic_2016}, temporarily pushing away links~\cite{debiasi_3darclens:_2014}, or topological fisheye techniques~\cite{brown_browsing_1993}.
Additionally, there are many generic graph interaction techniques which could be used for geospatial networks, such as the \textit{Local Edge Lens} and \textit{Bring Neighbors Lens}~\cite{tominski_fisheye_2006}.
These local \textit{interaction only} techniques are good for keeping the overall visualization stable while exploring subgraphs and specific geographic regions. Also, while globally applied techniques can help to deal with non-uniform densities, local approaches like these do not unnecessarily sacrifice detail in less dense areas.

Finally, there are several approaches specifically useful for dense areas in networks with non-uniform densities, i.e., networks that also contain much sparser areas.  One solution is using \textbf{distortion} to increase the size of dense areas, for example, through \textit{centrality-based scaling}~\cite{merrick_increasing_2006}. Metro map layouts~\cite{hong_automatic_2006} address this issue as well and could potentially be applied to non-transport networks as well. \textbf{Interaction techniques} that let the user view separate visualizations for subgraphs are also a possible solution \cite{hadlak_situ_2011}. The problem can be avoided using an abstract geographic representation~\cite{abel_quantifying_2014} or matrices to avoid positioning nodes based on their geographic location.
Another solution to varying densities can be \textbf{map insets}~\cite{ghani_dynamic_2011,brodkorb_overview_2016} which scale up particularly dense regions to reveal the network topology and are additionally able to bring subgraphs geographically closer, distorting geography.

\vspace{-0.5em}
\subsection{Dynamic Geospatial Networks}

Studying \textbf{dynamic networks}, i.e., networks that change over time as nodes and links appear or disappear or link weight changes over time, has led to many techniques, nicely summarized by Beck~et~al.~\cite{beck_state_2014}.
Geospatial networks can involve a range of additional types of changes such as nodes changing positions and geography changing structure (e.g., when countries merge or split). While visualizations for geo-temporal (non-network) data have been summarized by Bach~et~al.~\cite{bach_descriptive_2017}, little is known about visualizing dynamic geospatial networks.

One common approach is to use animation to show changes. 
Juxtaposition of small multiples, one for every time step, is another straightforward method explored and evaluated by Boyandin~et~al.~\cite{boyandin_qualitative_2012}. Their study, comparing animation with small multiples, found that animation led participants to more findings with local events and changes between subsequent years. However, small multiples led to more findings concerning longer time periods. Animation can also be used with other types of visual representations, for example animating deformations of a 3D globe to show how strongly linked different locations are~\cite{alper_dynamic_2007}. 

For smaller networks with a moderate amount of time steps, where only the weights are dynamic, the \textit{Flowstrates} technique~\cite{boyandin_flowstrates:_2011} can be used. This technique juxtaposes two maps and a custom time series visualization such that the time series of each link is shown in the center, connected to start and end location on the two maps respectively.
For some networks, using the third dimension to represent time, i.e., displaying the network in a space-time cube, may also be a solution~\cite{kapler_geotime_2005}. For exploratory visualization, Hadlak~et~al.~\cite{hadlak_situ_2011} propose the concept of \textit{in-situ exploration} of dynamic networks, where users can display and modify a variety of additional visualizations.

In cases where it is not necessary to explicitly show changes over time in sequence, temporal abstraction or summarization can be an alternative approach. Andrienko~et~al.~\cite{andrienko_revealing_2017} propose a technique to abstract dynamic network data over space and time, showing the result in a glyph-based visualization. \textit{Shifted maps}~\cite{otten_shifted_2018} is a technique intended for personal movement data which includes network layouts based on how frequently someone travels to certain locations or how long the journey takes.

\vspace{-0.5em}
\subsection{Uncertainty}

Visualizing uncertainty in networks is an underexplored, but growing, field of research. In the context of geospatial networks, uncertainty can occur in different ways: 
Firstly, there is the issue of \textbf{uncertain network topology}, i.e., uncertainty about whether a certain node or link exists or what direction a directed link has. 
Secondly, there can be \textbf{uncertain locations}, i.e., where exactly a node is located, or in the case of geolocated links, what exactly their trajectories are. 
Lastly, nodes and links can have \textbf{uncertain additional attributes}.
Uncertainty can be seen as ranging on a spectrum from \textit{exactly known} via \textit{uncertain} to \textit{unknown}, and the grade of uncertainty will affect how it can best be addressed in a visualization.

We did not find any concrete techniques specifically for visualizing geospatial networks with uncertain topology or geolocations. 
However, Von~Landesberger~et~al.~\cite{landesberger_typology_2017} have proposed a typology for uncertainty in geospatial graphs, and identified many approaches to visualizing uncertainty that have not been explored or tested yet. They focus on modifying geolocated node-link diagrams to visualize uncertain graphs and thus do not consider solutions outside of what we call \textit{explicit} geography and network representations. 
Also, techniques with an \textit{abstract} geography representation, e.g., by using spatially ordered chord diagrams~\cite{hennemann_information-rich_2013, abel_quantifying_2014}, \textit{Kriskograms}~\cite{xiao_visualizing_2009}, or matrices~\cite{bach_ontotrix_2011}, can be a solution for uncertain locations, since these techniques do not rely on precise geospatial locations to position elements of the visualization. The \textit{probabilistic graph layout} introduced by Schulz~et~al.~\cite{schulz_probabilistic_2017} is a technique to visualize networks with uncertain link weights or other numerical link attributes. The technique creates a `blurred' node-link diagram by decomposing the uncertain graph into its possible instances, then visually recombining them using node splatting and edge bundling. 

Overall, there is a clear need for a larger variety of techniques for more different use cases here. In particular, there is a lack of techniques usable for data exploration. Uncertainty in dynamic geospatial networks is another area that has not been addressed.


\vspace{-0.5em}
\section{Discussion}
\label{sec:discussion}

This survey identified \ncount\ papers presenting techniques for geospatial network visualization, to which a structured coding methodology was applied.
In this section, we discuss our design space, how to make trade-offs between techniques along our dimensions, and which issues remain future work. 

\begin{figure*}[t!]
    \centering
    \includegraphics[width=\textwidth]{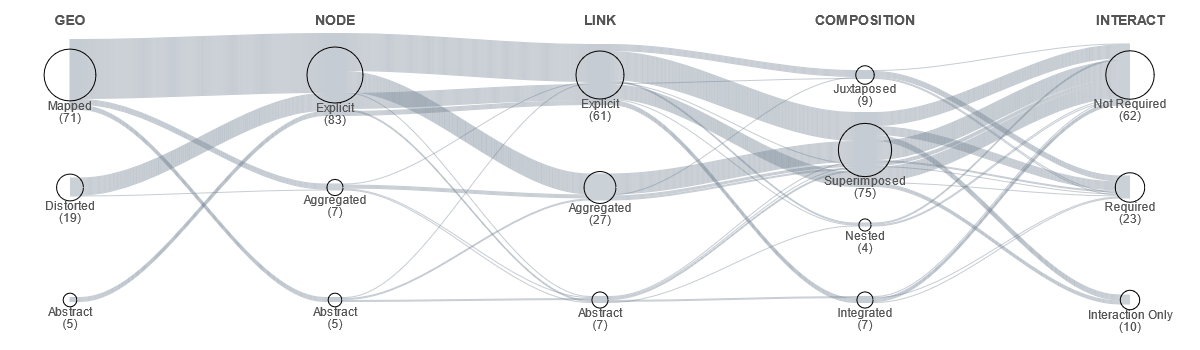}
    \caption{Overview of techniques classified in our design space. This visualization illustrates the trade-off effect between \fgeo, \fnode, and \flink: there is little correlation between the \textit{abstract} ends of these dimensions, i.e., most techniques abstract only one dimension.}
    \label{fig:parallelcoord}
\end{figure*}

\vspace{-0.5em}
\subsection{Design Space}

Our design space (Sections \ref{sec:d-geography}--\ref{sec:d-interactivity}) provides an overview of various techniques' designs with the goal of providing a conceptual understanding for possible design solutions. 
Following several iterative steps to classify techniques according to different schemata (Section~\ref{sec:designspace}), we eventually defined five design dimensions \fgeo, \ftopo:\fnode, \ftopo:\flink, \fcomp, and \finteract. 
This design space was created on the premise that each visualization technique has to solve a trade-off between showing different levels of detail and precision for different types of information in order to deal with the complexity of data in geospatial networks.

To that end, each of the dimensions \fgeo, \fnode, and \flink\ describes a spectrum to categorize techniques according to how much information they show explicitly and how much information about the data gets abstracted and aggregated. 
We classified techniques from each paper into rough categories along these dimensions. 
This allows us to capture essential steps along the dimensions and their individual characteristics. 
Our dimensions \fcomp{} and \finteract{} exhibit more discrete categories along their respective dimensions---\textit{loose} to \textit{tight integration} for \fcomp{} and \textit{not required} to \textit{interaction only} for \finteract{} since these two dimensions naturally fall into more discrete steps. 

We want to emphasize that all our dimensions are to be understood as a continuum between two respective poles.
Such a continuum has three benefits: 
\begin{itemize}
    \item \textbf{Our design space is high-level and expresses ideas and trade-offs, rather than just specific existing solutions.} This implies that there is a huge set of design solutions and techniques to explore the trade-off between explicit and abstract representations and we believe such a spectrum is a powerful thinking tool when discussing existing and designing new techniques. For example, when searching for a given technique to apply to a specific application problem, the designer or analyst can use each dimension to assess the relative importance and of that information (network topology with nodes and links, or geography). Then, they could look for specific techniques or inform their own design trying to include as much information from each dimension as they need.
    \item \textbf{Thinking of our dimensions as continuous makes them less rigid and open to capturing new techniques in the future.} A new technique can be placed onto these dimensions, potentially creating a new category, or requiring the refinement or splitting of an existing category. Our categories represent groups of techniques with common characteristics along these dimensions. They help capturing these high-level characteristics (e.g., \textit{distortion}) and to orient the user of our survey and design space.  
    \item \textbf{Continuous dimensions allow techniques to float along the continuum and exploit a richer set of designs.} For example, \textit{edge bundling} captures many different individual techniques, each of which potentially floats along the \flink-dimension. For example, one edge bundling technique can show explicit links if the bundling parameter is low, but aggregate individual links and make them harder to trace under another parameter setting. 
    Consequently, a single technique (edge bundling) can occupy an entire segment in our dimension.
\end{itemize}

In that sense, our design space is complete for the papers we could find, and we believe it is open enough to capture a range of future techniques.

\vspace{-0.5em}
\subsection{Design Space Coverage and Open Designs}

Our design space allows us to discuss existing and common designs (Figure~\ref{fig:parallelcoord}), less common and underexplored designs, drawbacks of individual designs, trade-offs between designs along the same dimension, as well as to negotiate trade-offs between design solutions \textit{across} dimensions. 

Common combinations in our design space include \textit{Mapped $\times$ Explicit Nodes $\times$ Superimposed} (\perc{51}) as well as \textit{Aggregated Links $\times$ Superimposed} (\perc{26}). The latter contains nearly all edge bundling techniques as well as flow map layout techniques. In contrast, other combinations have only a few associated techniques, such as \textit{Abstract Nodes $\times$ Abstract Links} (\perc{2}). We believe that many of the rarer combinations are underexplored but possible; 
for example, we can imagine \textit{abstract} nodes \textit{juxtaposed} with a \textit{mapped} or \textit{distorted} geography representation. We also believe there is a large unexplored design space of \textit{juxtaposed} and linked visualizations where the network can be represented in a clearer way while the geography is \textit{mapped}.  \textit{MapTrix}~\cite{yang_many--many_2017} and \textit{Flowstrates}~\cite{boyandin_flowstrates:_2011} are examples of techniques in this space. 

However, our design space is \textit{descriptive}, rather than \textit{prescriptive}. In other words, we can point to holes in the design space, e.g., we could not find any techniques for a combination of \textit{juxtaposition} and \textit{abstract geography}. However, we cannot prescribe \textit{how} a new technique should be designed. 
This is different from prescriptive design spaces which can be used to create new techniques by combining a set of options (e.g., layout, interaction technique, color encoding, etc.). 
The dimension with the most prescriptive power is \fcomp{} as our categories (\textit{juxtaposed}, \textit{superimposed}, \textit{nested}, and \textit{integrated}) are quite distinct and generic design solutions. However, the categories for \textit{nesting} and \textit{integration}, for example, both capture a range of possible techniques and are hence less useful to describe a specific visual design. Likewise, while \textit{juxtaposition} and \textit{superimposition} present very specific design solutions, their individual implementation offers a wide range of design parameters. For example, a juxtaposition needs to clarify the size of each view (e.g., large geographic view and small network topology if the network is rather small but spatially unequally distributed), as well as mechanisms for connecting information across both views (e.g., brushing and linking, visual links, or consistent visual encodings).

\vspace{-0.75em}
\subsection{Discussing Techniques} 

Our design space makes it straightforward to discuss conceptual advantages and drawbacks of specific designs. Techniques on the explicit end of the dimensions show \textit{more explicit} and, as a consequence, often \textit{more detailed} information. Generally, techniques on the abstract end show less detail and aggregate and abstract information appropriate to specific tasks. Less information makes tasks related to individual nodes, links, and geographic locations harder or impossible. Generally, abstracting information is useful in two scenarios: \textit{i)} support tasks that require abstractions (e.g., analyze nodes and connectivity per geographic region) and \textit{ii)} hide information less important to a task and that would otherwise clutter the interface.

\textbf{\fgeo:} \textit{Mapped} techniques such as a 3D globe or 2D projected maps offer high precision regarding geography but pose constraints on the position of nodes---if used in a superimposed or integrated way---or require linking information between views if presented in a side-by-side view. 3D globes also require navigation. In general, distortion is an inevitable consequence of transforming a sphere or an ellipsoid onto a plane. It is impossible to unfold the Earth onto a planar map without distortion~\cite{snyder_map_1987,jenny_guide_2017}. This can result in exaggerated and unequal link lengths for superimposed node-link diagrams, but also can result in a false perception of where a link is located, if that matters. 
Using three-dimensional globes~\cite{cox_3d_1996,kaya_multi-resolution_2016,yang_maps_2018,yang_origin-destination_2019} partly addresses this issue, but virtual reality is not always accessible.

\textbf{\fnode\ and \flink:} Explicit network topology (nodes and links) combined with explicit geography is required in tasks that require parallel understanding of network topology (node connectivity, path following, clusters, identifying network motifs~\cite{lee_task_2006}) and precise geography at the same time. It can also precisely preserve the geography. However, the resulting views are usually visually cluttered. 

\vspace{-0.25em}
\textbf{\fcomp:} A loose integration (\textit{juxtaposition}) is flexible, if not agnostic, to the respective representations of geography and network topology. 
Both facets can employ a visualization technique that best supports the information shown and the task at hand. 
As mentioned earlier, \textit{juxtaposition} requires mechanisms to relate information from both views. 
\textit{Superimposition} represents a stronger integration as network and geographical information are displayed in the \textit{same} visual space. 
Most commonly, nodes are placed at their associated geographic positions. 
This might support tasks related to nodes and their positions as well as questions of local connectivity.
On the downside, \textit{superimposition} comes with the common issues that motivated this survey: node overlap, long links, ambiguous links, visual clutter, and the general difficulties to understand network topology. 
A visual solution includes moving nodes away from the map, presenting a transition to a \textit{juxtaposed} view~\cite{speckmann_necklace_2010}. 
Interactive techniques (e.g., \textit{EdgeLens}~\cite{wong_edgelens:_2003}, \textit{Bring \& Go}
~\cite{moscovich_topology-aware_2009} or node circles~\cite{bach_networkcube:_2015}) can further alleviate these issues.
Some network representations, e.g., adjacency matrices, are hard to superimpose since node positions are constrained to vertical and horizontal rows and columns. Hybrid techniques such as \textit{NodeTrix}~\cite{henry_nodetrix_2007} superimposed on a map could provide a solution for analyzing local clusters and inter-regional connections. 
\textit{Nested} and \textit{integrated} techniques provide for a tighter connection of topological and geographic information, but the tight integration also reduces flexibility in choosing geography and network representation separately.

\vspace{-0.25em}
\textbf{\finteract:} Without interaction, a visualization needs to be very careful about clutter reduction and can present only a single visual representation. Interaction could allow for `moving along' the dimensions in our design space, e.g., by morphing from a \textit{mapped} via a \textit{distorted} to an \textit{abstract} geographic representation. A range of interaction techniques have been created to address specific issues. 
Interaction for exploring geospatial networks is a largely underexplored space in which we see great potential for novel techniques.

\vspace{-0.5em}
\subsection{Negotiating Trade-Offs} 

Knowing our design space and the individual drawbacks of each category along the dimensions helps negotiate trade-offs between each category as well as think about hybrid techniques. 
For example, one possible way to address the trade-off between geography and network topology is to morph between geographic and force-directed layouts, or to restrict geographic information to the node neighborhood~\cite{otten_shifted_2018}. 
Another possible trade-off is geographic distortion~\cite{bouts_visual_2016, alper_dynamic_2007} whereby through sophisticated geometric transformations geographic locations are relaxed and distorted to bring nodes closer to their positions in a force-directed layout.  
Hybrid approaches in terms of positioning, such as the approach taken in metro maps, are another way of giving more importance to the network topology. 

\vspace{-0.5em}
\subsection{Limitations and Possible Extensions} 

Some techniques in our collection caused more discussion when classifying than others. While \textit{mapped} techniques with \textit{explicit} nodes and links are easy to classify---perhaps due to the tradition our community has with node-link diagrams---techniques aggregating nodes and links were less common and consequently caused us more discussion in understanding and capturing their idea. Other ambiguous cases are found between distorted and abstract geography, as some \textit{abstract} designs, like spatially ordered nodes on a circular chord diagram, could also be seen as extreme \textit{distortions}. 
Still, we see the flexibility of our design space as a strength. 

Future analyses of our collection of techniques may classify them differently. For example, in Version \#2 of our taxonomy (Section 4.2), we tried to classify techniques by visualization types for networks (e.g., matrix, arc diagram). While we found that such a categorization does not capture the information necessary for our discussion, we believe such a categorization might still be useful to inform novel techniques in a more prescriptive way. However, care must be taken to not restrict creativity and to capture all possible groupings and combinations of geography and network topology. 

\vspace{-0.5em}
\subsection{Addressing Open Challenges}

Section~\ref{sec:challenges} provided an overview of how specific challenges, grouped by data type, are addressed in current techniques. Despite our survey including techniques from \ncount\ papers, there is a large array of unsolved challenges in visualizing geospatial networks. For example, among the challenges most under-addressed is uncertainty. While uncertainty in networks is generally under-addressed, we could not find solutions for visualizing multiple positions, missing node positions, missing locations on links (trajectories) and the range of uncertainties present in geographic visualization. 

We also found few techniques directly addressing problems in node overlap and resolving ambiguity when links overlap. Routing seems to be promising here but future routing algorithms could try to take geography more into account, to route links, e.g., through empty space or along semantic trajectories such as rivers or frontiers. 
Finally, techniques for dynamic geospatial networks, including nodes moving between locations, a changing network topology as well as other changing node attributes are under-explored. Approaches from spatio-temporal visualization~\cite{bach_descriptive_2017} could help finding appropriate solutions.

\vspace{-0.5em}
\subsection{Towards a Task Taxonomy} 

Much of our discussion in this survey is making reference to \textit{tasks}. Tasks are a powerful concept in visualization that inform design and evaluation and have been formulated for networks in general~\cite{lee_task_2006} and dynamic networks~\cite{bach_graphdiaries_2014,kerracher_constructing_2017} as well as for geographic visualization~\cite{roth_empirically-derived_2013}. The latter taxonomy included objectives (\textit{identify, compare, rank, associate, delineate, procure, predict, \& prescribe}) and seventeen operators (\textit{import, export, save, edit, annotate, re-express, arrange, sequence, re-symbolize, overlay, pan, zoom, re-project, search, filter, retrieve, \& calculate}). However, such tasks are similar to generic taxonomies in visualization
~\cite{amar_knowledge_2004} and do not provide the expressiveness necessary for geospatial networks.

More specifically, Andrienko et al.~\cite{andrienko_exploratory_2006} formalized tasks for spatio-temporal visualizations, considering space, time and objects as fundamental elements and tasks as queries for the information associated with them. 
Yang and Goodwin~\cite{yang_what-why_2019} interviewed domain experts who analyze geospatial networks in their professional work and identified three analytical targets: single flow (i.e., a flow between two geographic locations), total flow (i.e., all flows linked to a given location) and regional flows (i.e., flows between locations within a geographic area).
However, these works only focus on single perspectives of our design space.

To the best of our knowledge, a structured task taxonomy for geospatial networks does not exist yet. Such a task taxonomy could help to standardize evaluation and allow for comparison across technique papers, and additionally to characterize systems in terms of task support, thus allowing for more informed technique selections for applications. 

\vspace{-0.5em}
\subsection{Empirical Evidence} 

Empirical evidence on how specific geospatial network visualization techniques perform is sparse (\perc{7} in our collection). Existing papers include mostly case studies, sometimes small quantitative user studies. The tested tasks are predominantly rather basic, with more complex tasks rarely being evaluated. Even when there is a stronger focus on evaluation, comparison across publications is practically impossible due to the lack of standardization. Our design space might provide some scaffolding here. For example, open questions that should be addressed by empirical evaluations include the effectiveness of geographic distortion and the impact and readability of abstract geographic representations.


\vspace{-0.5em}
\section{Conclusion}
\label{sec:conclusion}

This survey presents a structured collection of \ncount\ visualization and interaction techniques for geospatial networks. 
We explored various ways of categorizing our techniques and ended up with five dimensions. 
Each technique can be described along each of these dimensions and compared to other techniques. 
At the same time, these dimensions provide for a design space to inspire future techniques. 
We also discussed common challenges in visualizing geospatial networks, and whether our collected techniques are suitable for addressing these challenges.
We concluded with a discussion of the design space, techniques, and a list of directions for future research and hope this survey will be an inspiration for visualization designers to find novel and creative techniques to solve the many open questions as well as a useful guide for analysts, scientists, and students (yet) outside the field of visualization and geography.

\section*{Acknowledgements}
We would like to thank the anonymous reviewers for their valuable comments.
Yalong Yang is supported by a Harvard Physical Sciences and Engineering Accelerator Award.


\vspace{-0.5em}
\printbibliography                

\end{document}